\documentclass{aastex62}

\received{}
\revised{}
\accepted{}
\submitjournal{ApJ}

\usepackage{floatrow}
\usepackage{CJKutf8}
\usepackage{graphicx}
\usepackage{subfigure}
\usepackage{epsfig}
\usepackage[perpage,symbol*]{footmisc}
\usepackage{natbib}
\usepackage{dcolumn}
\usepackage{bm}
\usepackage{longtable}
\usepackage[figureright]{rotating}
\usepackage{ragged2e}

\def\lapp{\ifmmode\stackrel{<}{_{\sim}}\else$\stackrel{<}{_{\sim}}$\fi}
\def\gapp{\ifmmode\stackrel{>}{_{\sim}}\else$\stackrel{<}{_{\sim}}$\fi}

\shorttitle{5~GHz TMRT observations of 71 pulsars}
\shortauthors{R.S. Zhao et al.}

\begin{document}

\title{5.0~GHz TMRT observations of 71 pulsars}

\author{Ru-Shuang~Zhao~\begin{CJK}{UTF8}{gbsn}(赵汝双)\end{CJK}}
\affiliation{Shanghai Astronomical Observatory, Chinese Academy of Sciences\\
Shanghai 200030, China}
\affiliation{University of Chinese Academy of Sciences\\
Beijing 100049, China}

\author{Zhen~Yan~\begin{CJK}{UTF8}{gbsn}(闫振)\end{CJK}}
\affiliation{Shanghai Astronomical Observatory, Chinese Academy of Sciences\\
Shanghai 200030, China}
\affiliation{Key Laboratory of Radio Astronomy, Chinese Academy of Sciences\\
Nanjing 210008, China}

\author{Xin-Ji~Wu~\begin{CJK}{UTF8}{gbsn}(吴鑫基)\end{CJK}}
\affiliation{School of Physics, Peking University\\
Beijing 100871, China}

\author{Zhi-Qiang~Shen~\begin{CJK}{UTF8}{gbsn}(沈志强)\end{CJK}}
\affiliation{Shanghai Astronomical Observatory, Chinese Academy of Sciences\\
Shanghai 200030, China}
\affiliation{Key Laboratory of Radio Astronomy, Chinese Academy of Sciences\\
Nanjing 210008, China}

\author{R.~N.~Manchester}
\affiliation{CSIRO Astronomy and Space Science\\
P.O. Box 76, Epping, NSW 1710, Australia}

\author{Jie~Liu~\begin{CJK}{UTF8}{gbsn}(刘杰)\end{CJK}}
\affiliation{Shanghai Astronomical Observatory, Chinese Academy of Sciences\\
Shanghai 200030, China}
\affiliation{Shanghai Tech University\\
Shanghai 201210, China}

\author{Guo-Jun~QIAO~\begin{CJK}{UTF8}{gbsn}(乔国俊)\end{CJK}}
\affiliation{School of Physics, Peking University\\
Beijing 100871, China}

\author{Ren-Xin~XU~\begin{CJK}{UTF8}{gbsn}(徐仁新)\end{CJK}}
\affiliation{School of Physics, Peking University\\
Beijing 100871, China}

\author{Ke-Jia~LEE~\begin{CJK}{UTF8}{gbsn}(李柯伽)\end{CJK}}
\affiliation{Kavil Institute for Astronomy and Astrophysics, Peking University\\
Beijing 100871, China}

\correspondingauthor{Zhi-Qiang~Shen~}
\email{zshen@shao.ac.cn}

\begin{abstract}

We present integrated pulse profiles at 5~GHz for 71 pulsars, including 
eight millisecond pulsars (MSPs), obtained using the Shanghai Tian 
Ma Radio Telescope (TMRT). Mean flux densities and pulse widths 
are measured. For 19 normal pulsars and one MSP, these are the 
first detections at 5~GHz and for a further 19, including five MPSs, the 
profiles have a better signal-to-noise ratio than 
previous observations. Mean flux density spectra between 400~MHz 
and 9~GHz are presented for 27 pulsars and correlations of power-law 
spectral index are found with characteristic age, radio pseudo-luminosity 
and spin-down luminosity. Mode changing was detected in five pulsars.
The separation between the main pulse and interpulse is shown to be 
frequency independent for six pulsars but a frequency dependence 
of the relative intensity of the main pulse
and interpulse is found. The frequency dependence of component
separations is investigated for 20 pulsars and three groups are found: 
in seven cases the separation between the outmost leading and trailing 
components decreases with frequency, roughly in agreement with 
radius-to-frequency mapping; in eleven cases the separation is nearly 
constant; in the remain two cases the separation between the outmost 
components increases with frequency. We obtain the correlations 
of pulse widths with pulsar period and estimate the core widths of 
23 multi-component profiles and conal widths of 17 multi-component 
profiles at 5.0~GHz using Gaussian fitting and discuss the width-period
relationship at 5~GHz compared with the results at at 1.0~GHz and 8.6~GHz.

\end{abstract}


\keywords{pulsars: general, radiation mechanisms: nonthermal}


\section{Introduction}\label{sec:intro}
The mechanism of pulsar radio emission has been a long-outstanding
question. A widely accepted theory is that the radio emitting regions
are confined to the open polar cap inside the light-cylinder radius
\citep{rs75}. Two different ways can be used to study the
pulsar radio emission. Firstly, study of individual pulses at
different radio frequencies helps us understand the instantaneous,
fluctuating properties of the pulsar emission region \citep{mbm+16}.
Individual pulse studies reveal a rich set of phenomena such as 
nulling, mode changing, subpulse drifting and quasi-periodic intensity
modulations. Since mode changing was first found in PSR B1237+25 
\citep{bac70a}, dozens of pulsars showing this phenomenon have been 
revealed. Many of these pulsars exhibit drifting subpulses and nulling
\citep{jv04,rwr05}. Weak emission within the `null' intervals was found 
in PSR B0826$-$34, suggesting a mode changing behaviour in this pulsar 
\citep{elg+05}. \citet{wmj07} explored the relations between 
nulling and mode changing for six pulsars in Parkes observations.
Simultaneous 328~MHz and 4.85~GHz obsrevations were performed for 
the subpulse drifting in PSR B0031$-$07 \citep{smk+05}. Mode changing 
properties of PSR B0329+54 were detected in simultaneous 2.3 and 
8.6~GHz single-pulse observations \citep{ysm+18}.

Secondly, integrated profile studies can help us investigate the radio
emission geometry and the stable structure and properties of the
magnetic field. The profile morphology at different frequencies has
been studied for many years
\citep[e.g.,][]{ran83,lm88,ran92,mr02a,jkmg08,hr10}. Despite
the variability of individual pulses, integrated profiles are
observed to be highly stable, indicating the global properties of
the pulsar magnetosphere. Various morphologies of integrated
profiles in different pulsars suggest that the observed variations
might be related to different ways of cutting across the radiated
beam and the asymmetric and irregular location of the individual
pulsar beam components \citep[e.g.,][]{lm88, ran93, md99}. The
frequency dependence of integrated pulse profiles is a key diagnostic
of the emission mechanism. Measurements of pulsar flux densities 
are of great importance for deriving the pulsar luminosity
function and investigating the population of radio pulsars at large.
Studying the radio spectra of pulsars helps us to understand the
emission mechanism \citep[e.g.][]{lylg95, kvl+15, jvk+18}. In some
pulsars at least, the frequency dependence of profile width reflects
radius-to-frequency mapping \citep{cor78} with radio emission at
different frequencies at different heights \citep[e.g.][]{mr02a}.
Characteristics of the frequency dependence include: (i) steep
intensity spectral index $\alpha \sim -1.8$ ($S_\nu \propto
\nu^{\alpha}$) \citep{sie73,lylg95} with a range of about 0.0 to
$-4.0$ \citep{mkkw00a}, (ii) radio pulsars have a diverse frequency
dependence of component separation \citep{cw14}. Many early studies 
showed that the component separation decreased with increasing 
frequency but sometimes showed a break between two power laws 
\citep[e.g.][] {cc68, srw75, ran83}. \citet{mr02a} studied a
sample of 10 pulsars with conal components suggesting three kinds of
beam-radius-to-frequency behaviors: a decreasing frequency
dependence of pulse width, a decreasing trend but approaching a
constant pulse width at higher frequency \citep[cf.][]{tho91a}, and
a nearly constant trend of pulse width. (iii) According to
the magnetic-pole model, the profile is interpreted as the
combination of central core and outer conal emission. The core
components of multi-component pulse profiles generally have a
steeper spectrum than conal components \citep{bac76, ran83, lm88}.
(iv) The linear polarization level generally decreases with
increasing frequency \citep{mth73,jkmg08}.

Because of their generally steep radio spectra, pulsars are more
difficult to study at higher frequencies, say above a few GHz. Several
high-frequency observations have been performed around 5.0~GHz
\citep{sgg+95,mj95,kkwj97,hx97,kkwj98,jnk98,hkk98,kll+99}. There are 
159 pulsars that have been observed in the frequency range from 4~GHz to
5~GHz. The observed profiles of these pulsars provide important
information on the emission mechanism. Furthermore, the high-frequency
flux densities are important in constraining the spectral behaviour of
pulsar radio emission. Some pulsars have a spectral break and some
have a spectral peak in the GHz region \citep{mgj+94,xkj+96,klm+11}.
Thus, the study of pulse intensities and pulse profiles over a wide 
radio-frequency range (from handreds of megahertz to tens of gigahertz) 
is very important for the understanding of the pulsar radiation mechanism 
and the geometry of pulsar radio emission region.

The Shanghai 65-m Tian Ma Radio Telescope (TMRT) is located at
a relatively low latitude and is suited for observing the pulsars
located north of $-45\degr$ declination. Using the TMRT we have made
5~GHz observations of a sample of 71 pulsars which were expected to
be detectable with good signal-to-noise ratio (S/N) at this
frequency. A description of the TMRT observations is given in
\S\ref{sec:obs}. \S\ref{sec:results} gives the results for individual
pulsars. These results are discussed in the context of other
observations in \S\ref{sec:discn} and \S\ref{sec:concl} gives a
concluding summary.

\section{Observations}\label{sec:obs}

Observations of a sample of 71 pulsars at 5~GHz were performed with
the TMRT between 2015 May (MJD 57155) and 2017 December (MJD
58114). The sample was chosen from a list of pulsars with 1.4
GHz flux densities $\gtrsim$ 4~mJy, corresponding to flux densities
$\gtrsim$ 0.4~mJy at 5.0~GHz (assuming a spectral index of
$-1.8$). Table~\ref{tb:fluxpsr} lists the observational parameters
for the 71 pulsars.  Initially a central frequency of 5.12~GHz was
used, but it was changed to 4.92~GHz and then to 4.82~GHz in late 2015
to avoid radio-frequency interference. The observations were made
using the FPGA-based spectrometer digital backend system (DIBAS)
\citep{ysw+15,ysw+18} to provide incoherent dedispersion and on-line
folding. The total recording bandwidth of 1~GHz was subdivided into
512 frequency channels and each pulsar period was divided into 1024
phase bins. Table~\ref{tb:fluxpsr} lists the total integration time
(T$_{\rm on}$) for each observation. A subintegration time of 30~s was
used. Observational data were written out in 8-bit {\sc psrfits}
format and {\sc psrchive} programs \citep{hvm04} were used for data
editing and processing to produce the integrated pulse profiles.
Pulsar parameters for on-line folding were obtained from the ATNF
Pulsar Catalogue.\footnote{http://www.atnf.csiro.au/research/pulsar/psrcat
\citep{mhth05}}

\section{Results}\label{sec:results}

Table~\ref{tb:fluxpsr} lists measured flux densities for the 71
observed pulsars, including eight millisecond pulsars (MSPs). For 30 
pulsars, the flux densities were calibrated
using the calibration sources: 3C48, 3C123, 3C196 or 3C295
\citep{pb13,pb17}. The pulsars calibrated in this way are marked
by a * in Table~\ref{tb:fluxpsr}. For the other 41 pulsars which have
no real-time flux density calibration, we estimated the mean flux
density using the system equivalent flux density \citep[cf.,][]{zwy+17}, 
which is $\geq$ 26 Jy at 5.0~GHz \citep{wzy+15}. From multiple
observations of 3C123 we estimate a 20\% calibration uncertainty. This
was assumed for pulsars that were observed only once. For pulsars that
were observed several times, we selected all observations with S/N higher 
than 5 and gave the mean flux density and its rms uncertainty in 
Table~\ref{tb:fluxpsr}. For comparison, flux densities around 5~GHz 
measured by others are listed in the last column in Table~\ref{tb:fluxpsr}.

Also listed in Table~\ref{tb:fluxpsr} are $\Delta\phi(N)$, where
$\Delta\phi$ is the separation of the outer-most profile components
and $N$ is the number of identifiable components. Pulse widths at 50\%
and 10\% of the profile maximum, $W_{50}$ and $W_{10}$ respectively,
are also given in Table~\ref{tb:fluxpsr}. Component separations and
pulse widths were computed from fits of Gaussian profiles to
components \citep[cf.,][]{zwy+17}. As for flux densities, for
pulsars that were observed several times, mean widths and their
uncertainties are given in Table~\ref{tb:fluxpsr}.

Mean pulse profiles for the 71 observed pulsars are given in
Fig.~\ref{fg:psrprf}. For 19 pulsars (PSRs J0454+5543, J0738$-$4042,
J0835$-$4510, J1012+5307, J1509+5531, J1518+4904, J1643$-$1224,
J1713+0747, J1740$-$3015, J1744$-$1134, J1745$-$3040, J1752$-$2806,
J1807$-$0847, J1829$-$1751, J1833$-$0827, J1848$-$0123, J1935+1616,
J1937+2544 and J1955+5059), these profiles have a much better S/N than
previous observations at frequencies around 5~GHz \citep{sgg+95, hx97,
  hkk98, kkwj98, jnk98, hoe99, kll+99}. For PSRs J0738$-$4042 and
J0835$-$4510, our profiles with 1024 bins have better time-resolution
compared to those with 128 bins in \citet{jnk98}. The S/Ns of profiles
in six pulsars (PSRs J1012+5307, J1509+5531, J1518+4904, J1643$-$4042,
J1744$-$1134, J1833$-$0827) are in the range from 10 to 20 in our
observations, while in the previous observations, the S/Ns were lower
than 10. For PSRs J1745$-$3040, J1752$-$2806 and J1848$-$0123, the
profile S/Ns are between 20 to 40 in our observations, which is higher
than those in the previous observations. For the rest, in the previous
observations, the S/Ns were lower than 33, while in our observations,
they are higher than 40. Among the profiles given in
Fig.~\ref{fg:psrprf}, there are 20 where there are no previously
published 5~GHz profiles. These, along with other MSPs (PSRs
J1012+5307, J1518+4904, J1643-1224, J1713+0747 and J1744$-$1134) in
the sample, are discussed in \S\ref{sec:normal} and \S\ref{sec:msp}
below.

TMRT 5~GHz pulse profiles are compared with profiles at other
frequencies obtained from the EPN profile
database\footnote{http://www.epta.eu.org/epndb} and ATNF observation
database\footnote{https://datanet.csiro.au/dap/public/atnf/pulsarSearch.zul}
in Fig.~\ref{fg:mul-freq}.  Corresponding components are connected by
dashed lines assuming a power-law frequency dependence of component 
separation as discussed in \S\ref{sec:Freq-W}. Spectral plots for 27 
pulsars are given in Fig.~\ref{fg:spec} and discussed in \S\ref{sec:spec}.

\subsection{Normal pulsars}\label{sec:normal}
In this section we discuss the 19 pulsars for which there are
no previously published 5~GHz pulse profiles. 

{\bf PSR J0248+6021.} With 12 years of radio timing data from Nan\c{c}ay
Radio Telescope observations, \citet{tpc+11} published the basic
parameters of PSR J0248+6021, which is a relatively young pulsar
(characteristic age $\tau_c \sim 62$~kyr) with a spin period of 217 ms
and a large dispersion measure (DM = 370 pc~cm$^{-3}$).  At 1.4~GHz,
profile broadening due to interstellar scattering is evident but at
2.05~GHz and 2.68~GHz the profile has a single narrow component
\citep{tpc+11}.  Fig.~\ref{fg:psrprf} shows that at 5.0~GHz there is a
single dominant component similar to that at 2.05~GHz and 2.68~GHz,
but the profile also has extended wings on either side of this
dominant component. These are almost certainly conal emission which
has flatter spectrum than the dominant component which can be
identified as core emission. The flux density of this pulsar is 3.1
mJy as listed in Table~\ref{tb:fluxpsr} which, within the uncertainties, is consistent
with the value of 3.15 mJy obtained by extrapolating with the spectral
index of $-1.19$ given by \citet{tpc+11}.

{\bf PSR J0837$-$4135 (B0835$-$41).} This pulsar was discovered at
408~MHz by \citet{lvw68}. Published 1400~MHz flux densities
\citep{tml93, hfs+04,jk18} are variable as are the 5.0~GHz flux densities
given in Table~\ref{tb:fluxpsr}, suggesting that the observed flux
densities are affected by scintillation. However, even with this
uncertainty, Fig.~\ref{fg:spec} suggests a break frequency around 400~MHz. 
The 5.0~GHz profile for this pulsar (Fig.~\ref{fg:psrprf})
has a strong central (core) component with two equal outlier (conal)
components on each side, similar to that at 8.6~GHz \citep{zwy+17}.
At 1.4 and 3.1~GHz the central core component completely dominates the
profile \citep{kj06}, consistent with the normally steeper spectral
index for core emission. The separation of the components in longitude
is nearly constant with frequency.

{\bf PSR J1559$-$4438 (B1556$-$44).} Flux densities of this pulsar at
lower frequencies were measured by \citet{mlt+78}, \citet{fgl+92} and
\citet{mhq98}. The spectrum from 400 to 5000~MHz shown in
Fig.~\ref{fg:spec} suggests a broken power-law form. The 5~GHz
profile has a basically double shape, but with an unresolved
asymmetric central component preceding the trailing component by about
$3\degr$ of longitude. Fig.~\ref{fg:mul-freq} shows that this offset
central component has a much steeper spectrum than the outer
components showing that it can be identified as core-type emission
despite its offset position. The outer components are clearly
conal. These properties are consistent with a gradual change in
emission properties across the beam \citep{lm88} rather than a
distinct emission process for core emission \citep{ran83}.

{\bf PSR J1651$-$4246 (B1648$-$42).} The 5.0~GHz integrated pulse
profile given in Fig.~\ref{fg:psrprf} has a basically double pulse
profile. At lower frequencies, this pulsar has a similar but wider
profile, and the components become more equal in amplitude with
decreasing frequency \citep{jk18}.

{\bf PSR J1703$-$3241 (B1700$-$32).} Observations of this relatively
long-period pulsar at frequencies between 0.6~GHz and 1.4~GHz
\citep[e.g.,][]{gl98,jk18} show that it has a relatively flat-topped
profile probably consisting of three overlapping components of similar
strength. The 5.0~GHz pulse profile shown in Fig.~\ref{fg:psrprf} is
consistent with this, perhaps indicating a somewhat weaker and
steeper-spectrum central core component. Fig.~\ref{fg:spec} shows a
relatively flat spectrum for this pulsar.

{\bf PSR J1705$-$3423.} This pulsar was discovered in the Parkes
Southern Pulsar Survey by \citet{mld+96}. It has an approximately
gaussian-shaped profile at lower frequencies, perhaps with some
structure about the profile peak \citep{jk18}. At 5.0~GHz, the profile
(Fig.~\ref{fg:psrprf}) is more asymmetric with two identifiable
components, the leading one about twice the strength of the trailing
one. The spectral behaviour is very similar to that of PSR
J1703$-$3241.

{\bf PSR J1707$-$4053 (B1703$-$40).} At 1.4~GHz the pulse profile for
this relatively high-DM pulsar (DM $\sim 360$~cm$^{-3}$~pc) is dominated by 
interstellar scattering with no clear intrinsic profile structure
\citep{jk18}. At 3.1~GHz, there are two obvious 
peaks with a relatively weak component at the 
trailing edge \footnote{https://datanet.csiro.au/dap/public/atnf/pulsarSearch.zul}.
As shown in Fig.~\ref{fg:psrprf}, at 5.0~GHz, the trailing component 
is much stronger but the components at the leading part of profile 
can not be resolved clearly because of the low S/N. 
The spectrum shown in Fig.~\ref{fg:spec} has a turn-over around
1.4~GHz.

{\bf PSR J1709$-$4429 (B1706$-$44).} This young Vela-like pulsar has
been extensively observed at radio frequencies between 450~MHz
\citep{jnk98} and 32~GHz \citep{kxj+97} and has been detected as a
$\gamma$-ray source with {\it Fermi} \citep{aaa+13} and {\it HESS}
\citep{hok+09}. It has a simple highly polarised radio profile with one
dominant component \citep[e.g.,][]{jk18}. The
spectrum is relatively flat as shown in Fig.~\ref{fg:spec} with a
spectral index of $-$0.62. Our 5.0~GHz pulse profile
(Fig.~\ref{fg:psrprf}) shows a single component of 50\% width about
$19\degr$ of longitude. This is narrower than the 50\%
width (~$20\degr$) at 1.4~GHz \citep{jk18} and broader than that
(~$16\degr$) at 8.6~GHz \citep{zwy+17}.

{\bf PSR J1721$-$3532 (B1718$-$35).} This pulsar with a short pulse
period of 0.28~s has a very high dispersion measure about 496 pc $\rm
cm^{-3}$ \citep{hlk+04}.  At 1.4 GHz the profile is highly scattered
\citep{jk18} with a scattering time scale of about 30~ms. At 5.0~GHz,
Fig.~\ref{fg:psrprf} shows a simple asymmetric profile
with a slow rising edge and a steeper falling edge, which is similar
to the higher-frequency results from \citet{jkw06} and
\citet{zwy+17}.  Given the flux densities at from 1.4~GHz to 17~GHz,
Fig.~\ref{fg:spec} shows a relatively flat spectrum.

{\bf PSR J1730$-$3350 (B1727$-$33).} This pulsar is relatively close
both on the sky and in distance to PSR J1721$-$3532 and is similarly
highly scattered at 1.4~GHz \citep{jk18}. At 5.0~GHz,
Fig.~\ref{fg:psrprf} shows that this pulsar has a simple symmetric
profile. Fig.~\ref{fg:spec} suggests that the spectrum is a
  broken power law which is flatter at high freqencies.

{\bf PSR J1739$-$2903 (B1736$-$29).} This pulsar has a relatively
strong interpulse separated from the main pulse by very close to
$180\degr$. At 0.6~GHz, the ratio of the peak flux of the interpulse
to the peak flux of the main pulse is about 42\% \citep{gl98} and at
1.4 GHz, slightly less, about 40\% \citep{jk18}.  In our observation
at 5.0~GHz, the flux density ratio between interpulse and main pulse
is about 17\%. The longitude separation of the pulse centroid is very
close to $180\degr$ at all frequencies. Fig.~\ref{fg:spec} shows a 
relatively flat spectrum between 400 and 8400~MHz with a spectral 
index of $-0.88$.

{\bf PSR J1809$-$1917.} This is a young energetic pulsar with pulse
period 82.7~ms, age 5.1~kyr and spin-down luminosity $\sim 1.8\times
10^{36}$~erg~s$^{-1}$. The pulse period is very similar to that of the
Vela pulsar, but PSR J1809$-$1917 is about five times older and five times less
energetic than Vela. X-ray observations show that there is an
associated pulsar wind nebula \citep{kp07}. At 5.0~GHz, this profile has two 
widely spaced but overlapping components separated by about $40\degr$
of longitude, with the leading component being about half as strong as
the trailing one. At 1.4 GHz \citep{jk18}, the component separation is
larger, about $54\degr$, the components do not overlap and the leading
component is a little stronger, about 60\% of the trailing one. 

{\bf PSR J1835$-$1106.} This relatively young pulsar ($\tau_c \sim
100$~kyr) has a highly polarised symmetric single-component pulse
profile at 1.4~GHz \citep{jk18}. At lower frequencies, the profile is
affected by interstellar scattering \citep[e.g.,][]{dsb+98}.  The
5.0~GHz pulse profile shown in Fig.~\ref{fg:psrprf} appears to have
some asymmetry. Gaussian fitting shows the profile to have three
components with a partially resolved leading component and a weaker
trailing one.

{\bf PSR J1844+00} This is a distant, high-dispersion pulsar
discovered at 0.4~GHz with a highly scattered profile at that
frequency \citep{cnst96}. At 1.4 GHz, the scattering is about two
orders of magnitude less and the profile is seen to have two
components, the trailing one about 30\% as strong as the leading one,
with a component separation of about $15\degr$ \citep{hfs+04}. At 5.0~GHz,
Fig.~\ref{fg:psrprf} shows a similar pulse profile with similar
component separation, but with the trailing component a little stronger, 
about 40\% of the leading component. The spectrum of this pulsar has 
a turn-over around 1~GHz (see Fig.~\ref{fg:spec}).

{\bf PSR J1853+0545.} This middle-aged ($\sim 3.3$ Myr) pulsar was
discovered at 1.4 GHz in the Parkes Multibeam Survey and shows an
apparently scattered profile (observed scattering time about 8~ms or
$24\degr$ of longitude) at this frequency \citep{kbm+03}.  At 5.0~GHz,
Fig.~\ref{fg:psrprf} shows that the profile has a dominant leading
component with a steep rising edge and a partially resolved trailing
component about 30\% as strong as the leading component and separated
from it by $\sim 6\degr$ of longitude. While there are only five
measurements of flux density at different frequencies, Fig.~\ref{fg:spec}
suggests that the pulsar has a relatively flat spectral index.

{\bf PSR J1900$-$2600 (B1857-26).} \textbf{At low frequencies, this
  well-known pulsar has a mutiple-component profile with a strong
  central or core component. Five components are distinguished clearly
  in the profile at frequencies around 300~MHz \citep{mbm+16}. The
  second and central components started merging with each other at
  410~MHz and 610~MHz \citep{gl98, mbm+16}. At frequencies above
  1~GHz, all components are merged and have about the same amplitude, giving the
  profile a flat-topped appearance \citep[e.g.,][]{jk18}. As
  Fig.~\ref{fg:psrprf} shows, this remains true at 5~GHz.}
Table~\ref{tb:fluxpsr} shows that the observed 5-GHz flux density is
quite stable and Fig.~\ref{fg:spec} shows that the spectrum between
200~MHz and 5~GHz is a broken power law.

{\bf PSR J1909+1102 (B1907+10).} This pulsar has a 
relatively short period ($\sim 0.284$~s) but a relatively weak
magnetic field giving it a characteristic age of about 1.7~Myr. At frequencies 
lower than 1~GHz, the pulse profile is dominated by a single 
component. A leading component with peak flux density about 25\% of the
main component can be seen at 1.4~GHz 
\citep{wcl+99}. Fig.~\ref{fg:psrprf} shows that at 5.0~GHz 
the leading component has a stronger peak flux 
density than the trailing component. The steeper spectrum of the
trailing component is confirmed in Fig.~\ref{fg:mul-freq} which also
shows that the component separation is essentially 
independent of frequency.

{\bf PSR J2257+5909 (B2255+58).} At low frequencies, below about
1.4~GHz, this pulsar shows two main components with the trailing
component dominant \citep{gl98}. At 1.4~GHz, these two components are
of equal strength and there is a hint of a third component trailing
the main components by about $15\degr$ \citep{gl98}. Fig.~\ref{fg:psrprf} 
shows that at 5.0~GHz the leading component is the stronger of the two
main components. Fig.~\ref{fg:mul-freq} shows that the separation of
these two components is almost constant with frequency. The spectrum
shown in Fig.~\ref{fg:spec} suggests a low-frequency turn-over below
about 1~GHz.

{\bf PSR J2330$-$2005 (B2327-20).} At 410~MHz, three components are
visible for this pulsar, with the leading one the strongest and a weak
overlapping third component \citep{gl98}. The observed pulse
profile at 5~GHz has rather low S/N (Fig.~\ref{fg:psrprf}) but within
the uncertainties has a similar pulse profile to that at lower
frequencies. Fig.~\ref{fg:mul-freq} shows that the central component
becomes somewhat stronger with decreasing frequency, consistent with a
core origin. The component separation appears to be independent of
frequency. Fig.~\ref{fg:spec} shows a good fit to
a relatively steep power-law spectrum over the frequency range 400~MHz
to 5~GHz.

\subsection{Millisecond pulsars}\label{sec:msp} 

MSPs form a separate group of pulsars
characterised by their short spin period ($\lapp 30$~ms) and very low
spin-down rate ($\dot P \lapp 10^{-18}$). These properties are
attributed to a different evolutionary history involving spin-up, or
``recycling'', of an old neutron star by accretion from a binary
companion \citep[e.g.,][]{bv91}. Although implied surface dipole
magnetic field strengths are much smaller for MSPs (typically $10^8$
-- $10^{10}$~G) compared to normal pulsars ($10^{11}$ -- $10^{13}$~G),
the properties of the observed radio emission, for example, radio
spectral indices and polarisation properties, are remarkably similar
to those of radio emission from normal pulsars
\citep{kll+99,dhm+15}. Although radio luminosities are generally
smaller for MSPs, there is a large overlap in the distribution
\citep{kxl+98}. On the other hand, MSP pulse widths are generally much
larger, often with significant emission over much of the pulse period,
and the spacing of profile components is generally independent of
frequency. Observations at high radio frequencies are important in
quantifying these trends \citep{kkwj97}. In our 5~GHz observations, we
obtained pulse profiles for eight MSPs (PSRs J1012+5307, J1518+4904,
J1600$-$3053, J1643$-$1224, PSR J1713+0747, J1744$-$1134, J1939+2134 
and J2145$-$0750) (Fig.~\ref{fg:psrprf}). Of these, there is 
no previous published 5~GHz profile for PSR
J1600$-$3053, for four pulsars (PSRs J1012+5307, J1518+4904,
J1643$-$1224 and J1744$-$1134) our 5~GHz observations have higher 
S/N than those in \citet{kll+99}, and the profile for PSR J1713+0747 
at 5~GHz has comparable S/N with the one in \citet{kll+99}. 
Descriptions of these six MSPs are given below.

{\bf PSR J1012+5307.} As shown by \citet{kxl+98}, at 1.4~GHz the pulse
profile is complex with a broad main pulse having several identifiable
components and two interpulses trailing the main peak by about
$110\degr$ and $160\degr$ respectively. Polarisation data at 610~MHz
\citet{stc99} show a nearly flat polarization position angle (PA) through 
all pulse components, suggesting this pulsar might be an aligned rotator. 
 
As shown in Fig.~\ref{fg:psrprf}, our 5~GHz profile has rather low S/N
and only two pulse components can be identified. The observed
separation of $115\degr \pm 5\degr$ suggests that we are seeing the
trailing components of the main pulse and first interpulse seen at
1.4~GHz. At 5~GHz the interpulse is about 50\% as strong as the main
pulse, whereas at 1.4 GHz the interpulse is a little weaker, about
40\% of the main peak. At 610~MHz it is only about 20\% as strong. The
separation of these components is the same at all observed frequencies
and Fig.~\ref{fg:spec} shows that, overall, the pulsar has a relatively
steep power-law spectrum.

{\bf PSR J1518+4904.} This pulsar has a 40-ms spin period, which is
typical of a mildly recycled pulsar. At most frequencies, the pulse
profile has two identifiable components spaced by about $4\degr$ of
longitude with the leading component the stronger of the two
\citep{kll+99}. However, at 370~MHz there is a weak third component
trailing the main component by about $20\degr$ \citep{nst96}, also
seen by \citet{kxl+98} at 1.4 GHz. Our
5-GHz observation, shown in Fig.~\ref{fg:psrprf}, which has a S/N
comparable to that of the profile shown by \citet{kll+99}, shows an
approximately Gaussian profile of 50\% width $12.5\degr$
(Table~\ref{tb:fluxpsr}). Given the relatively low S/N, it is not
possible to distinguish individual components, but the width indicates
a blend of the two main components seen at lower frequencies. 

{\bf PSR J1600$-$3053.} The 5-GHz profile shown in Fig.~\ref{fg:psrprf}
has two identifiable components, spaced by about $12\degr$. In the Parkes
three-band observations \citep{dhm+15}, shown in Fig.~\ref{fg:mul-freq},
these two components are clearly seen at 3.1~GHz, but the leading
component becomes weaker and more extended with decreasing
frequency. This pulsar may be an exception to the general rule that
MSP component spacings are independent of frequency. Alternatively, there may
additional blended components at the leading edge of the profile
that become stronger at lower frequencies. 

{\bf PSR J1643$-$1224.} This millisecond pulsar is the third most 
luminous millisecond pulsar after PSRs J0437$-$4715 
and B1937+21. Four components can be identified in the integrated profile at 1.4~GHz 
\citep{kxl+98}, with the leading component having a flatter spectrum
than the others \citep{dhm+15}. At 5.0~GHz, the leading  
components is more distinguishable, consistent with this trend. 

{\bf PSR J1713+0747.} Parkes observations of this well-known 4.57~ms
pulsar at 0.7, 1.4 and 3.1~GHz show at least four components with the
third becoming more prominent and narrower with increasing frequency
\citep{dhm+15}. The leading component has a relatively flat spectrum,
whereas the trailing component has a steeper spectrum
\citep{dhm+15}. At 5~GHz, the four components are clearly
distinguishable. The component separation is constant with frequency.

{\bf PSR J1744$-$1134.} At lower frequencies, this isolated MSP has a 
main pulse consisting of two overlapping components and a double and 
much weaker precursor pulse leading the main pulse by about $120\degr$
\citep{dhm+15}. Fig.~\ref{fg:psrprf} shows that at 5.0~GHz 
there is a partially resolved leading component, consistent with the results 
of \citet{dhm+15}.

\section{Discussion}\label{sec:discn}

\subsection{Spectra of 27 pulsars}\label{sec:spec}

Based on available multi-frequency data between 400~MHz and 9000~MHz,
flux-density spectral indices for 27 pulsars are listed in
Table~\ref{tb:spec} along with the pulsar characteristic parameters
and references. The corresponding spectra are shown in
Fig.~\ref{fg:spec}. For most of the pulsars in our sample, the spectra
can be well described by a simple power law ($S_\nu \propto
\nu^{\alpha}$). For seven pulsars, PSRs J0837-4135, J1559$-$4438,
J1707$-$4053, J1730$-$3350, J1844+00, J1900-2600, and J2257+5909, the
observed spectrum is better fitted by two power laws
\citep[cf.,][]{mkkw00,jvk+18}, although for three of these, PSRs
J0837-4135, J1559-4438 and J1844+00, the interpretation relies on just
one discrepant point. Note that for PSR J1730$-$3350, the spectrum is
evidently flatter at high frequencies. \citet{kbl+17} and \citet{brkl18}
identified PSR J1809$-$1917 as a ``gighertz-peaked spectrum'' pulsar, 
with a peak flux density around 1~GHz. \citet{brkl18} fitted the 
spectrum with a thermal absorption plus power-law model, giving a 
high-frequency spectral index of $-1.00^{+0.57}_{-0.98}$. \citet{jvk+18} 
fitted a power-law spectrum to data between 1 GHz and 7~GHz, obtaining a
spectral index of $-0.4\pm 0.2$. \textbf{In September 2018 (MJD 58370),
we re-observed this pulsar at 2.25~GHz with the total recording bandwidth 
of 100~MHz for 30~min, to give a calibrated flux density of $1.1\pm0.1$ 
mJy.} As Fig.~\ref{fg:spec} shows, we similarly fitted a power law to 
the high-frequency data and obtained a flatter spectral index of 
$-0.11\pm0.26$ (Table~\ref{tb:spec}). For PSR J0837$-$4135,
\citet{jvk+18} identified the spectrum as broken power-law, with a break
frequency around 700~MHz, but the spectrum in Fig.~\ref{fg:spec} suggests 
a break frequency around 400~MHz. For PSR J1703$-$3241, \citet{jvk+18} 
suggested a weak log-parabolic spectrum, but also fitted a power law, 
giving a spectral index of $-1.5\pm0.2$. Our data, shown in 
Fig.~\ref{fg:spec}, supports the power-law interpretation and gave
essentially the same spectral index as that obtained by \citet{jvk+18}.

To study the spectal index correlations we chose the 19 pulsars that
have simple power law spectra plus four more that have well-defined
power-law spectra above a break (viz., PSRs J0837$-$4135,
J1707$-$4053, J1900$-$2600 and J2257+5909). PSR J0248+6021 is excluded
since there are just two measurements of flux
density. Fig.~\ref{fg:spec-corr} shows the Spearman rank correlations
of radio spectral index $\alpha$ with pulsar period $\rm P$, period
derivative $\rm \dot{P}$, characteristic age $\tau$, the 1400-MHz
psuedo-luminosity $\rm L_{1400} = S_{1400}d^2$, the surface magnetic
field $\rm B_{surf}$ and the spin-down luminosity $\rm
\dot{E}$.\footnote{Pulsar parameters are from the ATNF Pulsar
  Catalogue.} These correlations were also fitted by weighted power
laws giving the results shown in Table~\ref{tb:spec-fit}. The
correlation coefficients for all pulsars in our sample, normal pulsars
and MSPs are listed in Table~\ref{tb:spec-corr}, along with previous
results from \citet{jvk+18, hwxh16,lylg95}. \citet{jvk+18} observed
441 pulsars, including nine MSPs, at frequencies between 730 MHz and
3100 MHz. \textbf{For the 276 pulsars with simple power-law spectra,
  they obtained Spearman rank correlations of spectral index with
  various pulsar parameters.} They found very significant correlations
with $\dot\nu$, $\rm B_{\rm LC}$ and $\rm \dot{E}$ for normal pulsars,
where $\rm B_{\rm LC}$ is the magnetic field strength at the light
cylinder. \citet{hwxh16} obtained Parkes observations at 1.4~GHz for
224 normal pulsars and combined these data with previously published
results for 372 pulsars to give a total sample of 572 normal
pulsars. They found a weak correlation of spectral index with
spin-down luminosity $\rm \dot{E}$. \citet{lylg95} studied a set of
343 pulsars, including 20 MSPs, for correlations between $\alpha$ and
P, $\rm \dot{P}$, $\tau$, $\rm B_{\rm surf}$, $\rm \dot{E}$ and pulse
width. For normal pulsars, they found two significant correlations:
$\alpha$ with P and $\alpha$ with $\tau$. For MSPs they found a strong
positive correlation of spectral index with P.

For all pulsars in our sample, we find that the correlations of 
spectral index with $\tau$ and $\rm \dot{E}$ are relatively
strong, whereas we find no correlation with $\rm B_{surf}$ and only
weak correlation with the other parameters. The stronger
correlations are in the sense of steeper spectra for old pulsars and
flatter spectra for pulsars with high radio luminosity and high $\rm
\dot{E}$. In contrast to \citet{lylg95}, we find no strong
correlation of MSP spectral indices with pulse period. The most
consistent correlation across the different samples is the positive
correlation with spin-down luminosity $\rm \dot{E}$ for normal
pulsars. MSPs do not show any strong correlations and results from
different samples are different but this may be largely due to the
small sample sizes.

\subsection{Pulsar profiles with mode changing}\label{sec:md}

Mode changing is a phenomenon, first observed by \citet{bac70a}, in
which the pulse profile abruptly changes from a given state to another
state. Usually there are just two quasi-stable states or modes
\citep[e.g.,][]{wmj07,ran17}, \textbf{but in a few pulsars three or 
more modes are observed, for example, in PSRs J1822-2256, J1946+1805 
and J2321+6024 \citep{wf81,dchr86,bm18}}. Mode changing is closely
related to pulse nulling, where the observable emission abruptly 
ceases for some interval and then resumes equally abruptly 
\citep[e.g.,][]{wmj07}. Both mode changing and nulling
evidently result from an abrupt change in the magnetospheric current
distribution, leading to a different radiation beam pattern which,
in some cases, has little or no emission directed toward us 
\citep{klo+06,tim10}. Simultaneous mode changing 
in the radio and X-ray emission was detected by \citet{hhk+13},
suggesting a rapid and global magnetospheric state change.

In our 5-GHz observations, mode changes have been detected in five
pulsars: PSRs J0332+5434, J0538+2817, J0742$-$2822, J1825$-$0935 and
J2022+2854. We discuss each in turn.

PSR J0332+5434 (B0329+54) is a bright pulsar which has
frequent mode changing and is therefore an ideal candidate for
studying mode changing at different frequencies. \citet{lyn71a}
first found that the ratio of the amplitude of the outer components
to that of the central component changed simultaneously at
408~MHz. \citet{bmsh82} found evidence for several different
abnormal modes in this pulsar. They found that the mode changing
occured simultaneously at 1.4 and 9.0~GHz, and concluded that
abnormal modes lasted from several periods to hours.  \citet{sp02}
showed that the profile shape changes within one pulse
period. High-frequency observations \citep{kxj+96,kxj+97} showed
that the mode changing was clearly visible at 32 and 43~GHz. Long
term observations performed by \citet{cww+11} found that this pulsar
spent 84.9\% of the total time in the normal mode, and 15.1\% in the
abnormal mode. Six 10-hour observing sessions of this pulsar were
performed by \citet{blk+18} with LOFAR at 140~MHz, they found that
12.6\% of profiles are abnormal-mode profile which was lower than
that in \citet{cww+11}. Simultaneous 13cm/3cm observations for this
pulsar was performed by \citet{ysm+18}, and 13\% of time for PSR
J0332+5434 was in the abnormal mode. Besides, bright narrow pulses,
whose peak flux densities are 10 times of those of integrated
profile, were observed at both frequencies.

Several observations at 5.0~GHz for this pulsar were performed
between MJD 57721 and MJD 57887 with total observation time of 180
mins. Sub-integrations of 300 pulse periods were formed to get
relatively stable intergrated profiles. We found that 12.5\% of
profiles were in abnormal mode (see Fig.~\ref{fg:J0332_md}), in
agreement with the results of \citet{blk+18} and \citet{ysm+18}, 
but we found no evidence for multiple abnormal modes. \textbf{Note
  that, because of our 300-period sub-integrations, some short
  bursts of abnormal mode may have been missed. } The left 
panel of Fig.~\ref{fg:J0332_md} shows the comparison between the 
normal and abnormal modes at 5~GHz. In both modes, there are three 
main components connected by bridges. In the abnormal mode, the 
leading and trailing components are of larger amplitude relative 
to the central component and the trailing component appears earlier 
by about $1\degr$ of longitude \citep[cf.][]{ysm+18}. The right 
panel in Fig.~\ref{fg:J0332_md} shows that the ratio of leading and
trailing components to the central one ($\rm R_{12}$ and $\rm R_{32}$) 
change simultaneouly at mode switching. The duration of the abnormal 
mode is between 3 and 8~min, which is limited by the sub-integration 
time and observing time in our sample.

PSR J0538+2817 is a young pulsar associated with the supernova remnant
Sim 147. In the discovery paper \citet{acj+96} found that the pulsar
had two emission modes with the trailing component being relatively
weaker in Mode B. Observations at 430~MHz and 1.4~GHz showed that the
trailing component has a flatter spectrum and that the mode changes are
more pronounced at the higher frequency. There are only two TMRT 
observations for this pulsar, at MJD 58102 and MJD 58114 (see 
Table~\ref{tb:fluxpsr}). Fig.~\ref{fg:J0538_md} shows that the spectral
evolution of the modes continues to higher frequencies, with the
trailing component dominating the profile in Mode A at 5~GHz. We also 
find that the profile components are wider and overlap more in Mode A.

For PSR J0742$-$2822 (B0740$-$28), in observations at 1.4~GHz and
3.1~GHz, \citet{ksj13} found interesting mode change behaviour with
complex changes in the pulse profile between two modes which they
labelled as Mode I and Mode II. Generally the trailing half of the
profile is relatively stronger in Mode II. They also found that Mode
II was associated with a higher spindown rate and that this
correlation was stronger after a glitch and remained so for about 3
years. In our 5.0~GHz observations, this pulsar was observed on a
number of occasions between MJD 57291 to MJD 57887. The left
panel in Fig.~\ref{fg:J0742_md} shows that the mode-changing 
behaviour at 5.0~GHz is similar to that at lower frequencies
although the trailing components in both modes are more prominent at
5.0~GHz. In Mode II, the trailing component is broader, leading to
and increased overall pulse width in this mode. Observed mode
changes at frequencies around 8 GHz \citep{jkw06,zwy+17} have
similar properties. For this pulsar, stable intergrated profiles can
be obtained by averaging pulses over 1.5~min (about 550 rotation
periods). We find that the ratio between leading and trailing
components ($\rm R_{12}$) varies frequently between 0.45 to 0.98 at
5.0~GHz.  The right panel of Fig.~\ref{fg:J0742_md} shows the
distribution of $\rm R_{12}$ fitted with two gaussian components
following the method of \citet{wwy+16} to distinguish the different
modes. Based on these fits, the pulsar is in Mode I for 77\% and
Mode II for 23\% of the total observation time of 4 hours, with
durations of Mode I and Mode II in the ranges 10 to 40 minutes and
10 to 20 minutes, respectively.

PSR J1825$-$0935 (B1822$-$09) has been observed many times and shown
to have complex mode changes and related drifting subpulse behaviour
\citep[e.g.,][]{fwm81}. In the so-called ``quiescent'' or Q-mode, the
``precursor'' component which leads the main pulse by about $15\degr$
of longitude is weak and the main component shows strong drifting
subpulses. In the ``burst'' or B-mode, the precursor is much stronger
and the drifting behaviour in the main pulse disappears. Even more
interestingly, \citet{gjk+94} showed that the interpulse, which is
about $180\degr$ from the main pulse, becomes relatively much stronger
in the Q-mode. This reveals a strong connection between the
precursor/main-pulse emission and the interpulse emission which
challenges standard orthogonal two-pole models for
main-pulse/interpulse structure. \citet{dzg05} addressed this issue
with a one-pole bi-directional model, in which both the main pulse and
the interpulse originate from the same magnetic pole with oppositely
directed emission beams forming the precursor and interpulse
respectively. \citet{lmr12} found there was a strong modulation with 
$P_3 = 46.6$ rotation periods in Q mode. A weak modulation with 70 
rotation periods was found in B mode. They also found a triple structure 
of the main pulse in both modes and in Q mode that the leading and 
trailing regions of the main pulse correlated both with each other 
and with the interpulse. In a 55-hour radio observation \citet{hkh+17} 
found that 63.9\% of pulses were in Q mode.

PSR J1825$-$0935 was observed several times between MJD 57378 and MJD
57825 with a total observation time of about 110~mins. Observed durations 
of B-mode emission were from 1.5 to 15~mins and for Q-mode the durations 
were from 4 to 14~mins. The left panel in Fig.~\ref{fg:J1825_md} shows 
a comparison of the B-mode and Q-mode integrated profiles at 5~GHz. The
behaviour is similar to that at lower frequencies with a strong 
anti-correlation between the intensities of the precursor and interpulse. 
The 5~GHz observations show that a feature on the trailing side of the
main pulse becomes stronger and more distinct in Q-mode. For each
30~s sub-integration, we measured the S/N of the main pulse,
precursor and interpulse as shown in the right panel of
Fig.~\ref{fg:J1825_md}. We find that setting an S/N threshold ($\sim 3$) 
on the precursor emission distinguishes the different modes clearly. 
This plot also shows that the precursor is anti-correlated with the 
interpulse as found at lower frequencies.

Mode changing for PSR J2022+2854 was first discovered at 1.4~GHz
\citep{wwy+16}, and consists of a change of relative intensity 
between the leading and trailing profile components. Two-component
gaussian fitting to the profile was used to obtain the ratio of leading 
component to trailing one ($\rm R_{12}$) confirming that there are two 
emission modes (normal and abnormal mode). In their 8~hours 
quasi-contiuous observation, only 11\% of profiles averaged over 30 
seconds are in abnormal mode.

In our 5.0~GHz observations, this pulsar was observed between MJD
57292 and MJD 57844 and the total observation time was about
96~mins. We also averaged profiles over 30~s. The left panel
of Fig.~\ref{fg:J2022_md} shows the two emission modes at 5~GHz. The
right panel shows the distribution of $\rm R_{12}$ with the normal
mode having the smaller values and the abnormal mode represented by
the small tail at higher $\rm R_{12}$. Of the 167 sub-integrations, 
about 6\% are in abnormal mode in our observations, smaller than 
the 11\% observed at 1.4~GHz \citep{wwy+16}.

\textbf{For another five pulsars (PSRs J0358+5413, J0826+2637,
  J1239+2453, J1946+1805 and J2321+6024), mode changing has been
  previously reported, but was not detected in our observations. In
  most cases, the reason for our non-detection of mode changing is
  that our observations were of short duration compared to the typical
  time in ``normal'' mode. For example, for PSR J0358+5413 (PSR
  B0355+54) \citet{msfb80} found one mode change in over 3 hours which
  lasted about 5\% of the total observation time. For PSR J1239+2453
  (PSR B1237+25), \citet{bac70a} found typical normal mode durations
  of several hours. For PSR J1946+1805 (PSR B1944+17), the profile has
  relatively low S/N in our 15~min observation, so we cannot 
  distinguish the different modes \citep{dchr86,kr10}. Similarly, in 
  PSR J0826+2637 (PSR B0823+26) \citep{syh+15} and PSR J2321+6024 
  (PSR B2319+60) \citep{wf81} it is likely that our observations had
  insufficient sensitivity to detect the emission in the weaker mode.}

\subsection{Pulsar profiles with interpulse emission }\label{sec:IP}

Interpulses have been found in many pulsars and are especially common
in young pulsars \citep{wj08} and millisecond pulsars
\citep{dhm+15}. Often the interpulse (IP) lies close to $180\degr$
from the main pulse (MP) and for this reason interpulses are commonly
assumed to originate near the second pole of a dipolar magnetic field
with the magnetic axis approximately perpendicular to the rotation
axis. For some pulsars the observed position angle variations are
consistent with this idea \citep{kjwk10} whereas in others they are
not very consistent \citep[e.g.,][]{ew01,mn13}. Often in these latter
cases the separation of the MP and IP is less than $180\degr$ and
there is a bridge of emission between the two components
\citep{hf86}. These and similar observations led to models in which
the MP and IP originate from a single magnetic polar region, possibly
with the magnetic axis close to alignment with the rotation axis
\citep[e.g.,][]{ml77,hc81}, or with bi-directional emission from a
single source region \citep[e.g.,][]{dzg05}. It has been suggested, 
most recently by \citet{bmr15}, that these wide profiles may be related 
to the precursor or postcursor components that are seen in some pulsars, 
e.g., PSR B0823+26, as they generally share some properties such as 
frequency-independent separation from the main pulse and high linear
polarisation.

Six pulsars with interpulses (PSRs J0953+0755, J1012+5307, J1705$-$1906,
J1739$-$2903, J1825$-$0935 and J1939+2134) are detected in our 5.0~GHz
observations (see Fig.~\ref{fg:psrprf} and Fig.~\ref{fg:mp_ip}). All
data at other frequencies are from ATNF and EPN pulsar databases.  The
frequency dependence of MP--IP separations for these six pulsars is
shown in Fig.~\ref{fg:mp_ip} along with weighted power-law fits to the
data. Absolute values for spectral indices are all less than 0.01
and consistent with zero, i.e., the MP--IP separations are
independent of frequency. This is consistent with previous studies
\citep[e.g.,][]{hf86}.

For PSRs J1705$-$1906, J1739$-$2903 and J1939+2134, the longitude
separations between the IP and MP are $179\degr \pm 4\degr$, $182\degr
\pm 1\degr$ and $173\degr \pm 1\degr$, respectively. It is likely that
a two-pole model is applicable to these three cases
\citep{stc99,wws07,kjwk10,mn13}. However, there are
complications. \citet{dhm+15} show that for PSR J1939+2134 the
polarisation position angle variations are inconsistent with the
rotating-vector model (RVM) at each pole, and that there is low-level
emission between the two components, similar to that observed in the
Crab pulsar \citep{hje15}. Furthermore, for PSR J1705$-$1906, there
are in-phase modulations of the IP and MP which are difficult to
explain in the two-pole model \citep{wws07}. Similarly, as discussed
in \S\ref{sec:md}, the emission of MP and IP of Q-mode of PSR
J1825$-$0935 might be from the same pole \citep{gjk+94,dzg05,mn13},
even though the separation between the MP and IP is $174\degr \pm
3\degr$. There is clearly a relationship between the interpulse and
precursor in this pulsar \citep[cf.][]{bmr15} but, as
Fig.~\ref{fg:J1825_md} shows, the trailing edge of the main pulse also
participates in the mode change, so the situation is complicated.

The discussion of PSR J1012+5307 in \S\ref{sec:msp} shows that this 
pulsar might be a nearly aligned rotator with a relatively wide beam
\citep{kxl+98,stc99} having three main components, although only two
(with a separation of $\sim 115\degr$) are visible at 5~GHz. 
In PSR J0953+0755, the separation between IP and MP is $156\degr \pm7 
\degr$ and is basically frequency independent, the inner edges
of the MP and IP are flatter than the outer edges and there is a
low-level bridge of emission joining them \citep{hf86}. \citet{hc81}
showed that the polarisation position angle rotates continuously from
the MP through the bridge to the IP. These observations all suggest
that the emission in pulsars with component separations similar to or
greater than that of PSR J0953+0755 may come from a single wide
emission cone, possibly close to alignment with the rotation axis
\citep{ml77,nv83a,mn13}. The pulse emission from this pulsar may be a
wide-separation example of the precursor class as suggested by \citet{bmr15}.

Fig.~\ref{fg:mp_ip}b shows us that for two pulsars (PSRs J1739$-$2903
and J1939+2134), the ratio of the peak flux of the IP to the peak
flux of MP decreases with increasing frequency, and this behavior can
be described by a simple power law. For PSR J0953+0755, we find that
the IP/MP ratio increases with frequency, in agreement with the
results of \citet{hf86}. Also in the case of PSR J1012+5307, the behavior of 
IP/MP ratio with frequency is the same as that in PSR J0953+0755, 
which is consistent with the low-frequency results of \citet{nll+95}. 
For PSRs J1705$-$1906 and J1825$-$0935, there appears to be a maximum IP/MP 
ratio at about 1.5~GHz, with a decreasing power law at higher frequencies 
and possibly an increasing one at lower frequencies.

\subsection{Frequency dependence of profile width}\label{sec:Freq-W}

The frequency dependence of pulse width is a very useful tool for
studying the geometry and location of radio emission in the pulsar
magnetosphere.  Previous studies showed that the component separation
$\Delta\phi$ decreases with increasing frequency with a typical
power-law index $\beta \sim -0.25$, where $\Delta\phi \sim \nu^\beta$
\citep{srw75,mt77,cor78}.  By studying the pulsar widths at 333 and
618~MHz \citet{mbm+16} found that $\beta \sim -0.19\pm0.1$.  But for
some pulsars, two power laws with a break frequency separating the
high- and low-frequency regimes better describes the frequency
dependence of profile width \citep{ran83a, sba87}. \citet{tho91a}
analysed this dependency and concluded that a break freqency is not
necessary and that a single power law combined with a minimum pulse
width can fit the data at all frequencies. Low-frequency observations
for 100 radio pulsars with LOFAR were analysed by \citet{phs+16}. They
found a preference for profile widening at low frequencies, as
expected for radius-to-frequency mapping (RFM) \citep[e.g.,][]{cor78}, but
with a wide spread of $\beta$ values, including about 15\% with $\beta
> 0$. In some cases this apparent broadening at high frequencies
resulted from the appearance of new components at the edge of the
profile. 

From our observations, 20 pulsars with multiple components are
selected to study the frequency dependence of pulsar
width. Table~\ref{tb:fluxpsr} lists the separations of the outer
components, $\Delta\phi$, and the number of discernable components for
these pulsars. Including data from previous work
\citep{dhm+15,gl98,wcl+99,sgg+95} and dividing the 20 pulsars into
three categories: seven with $\beta < 0$, eleven with $\beta \sim 0$
and two with $\beta > 0$, parameters from weighted power-law fits to
the frequency dependence are listed in Tables~\ref{tb:FW_ng},
\ref{tb:FW_fl} and \ref{tb:FW_ps}, respectively, and plotted in
Fig.~\ref{fg:sp_freq}. The pulse profiles at different frequencies are
given in Fig.~\ref{fg:mul-freq}, along with the component phase at
each frequency based on the derived power-law fits.

For PSR J1740$-$3015, the multi-frequency 
profile comparison for this pulsar was excluded in our earlier work 
\citep{zwy+17} because that the components cannot be resolved clearly at 
lower frequencies. However, the TMRT profiles at three frequencies
(2.3, 5.0 and 8.6~GHz) and the Parkes profile at 1.4~GHz shown in 
Fig.~\ref{fg:mul-freq} clearly show that there are two components with the
trailing one becoming more obvious at higher frequencies.

In the first group, all but one of the frequency index values
are between $-0.26$ and $-0.11$ (Table~\ref{tb:FW_ng}), similar to
previously reported values.  The exception is PSR J2354+6155 which
has a much steeper dependence, with index $-0.58$. 
Fig.~\ref{fg:mul-freq} shows that there is little ambiguity
about the component identification and consequently that the rapid
frequency evolution of component spacing is likely to be real. We
find no evidence for a minimum pulse width at higher frequencies
\citep[cf.][]{tho91a}

Pulsars with component separation index $<0.075$ (but
excluding interpulse pulsars and MSPs) are listed in 
Table~\ref{tb:FW_fl}. The reason(s) why these pulsars evidently do 
not have RFM is not understood. Some, e.g., PSR J1825-0935, may be 
in the precursor group which may have a distinct emission mechanism
\citep{bmr15}.

There are two pulsars, listed in Table~\ref{tb:FW_ps}, where
the components evidently increase in separation with increasing
frequency. For PSR J1954+2923, there are two prominent profile
components.  Fig.~\ref{fg:mul-freq} shows that the Tian Ma 5~GHz
profile clearly has a wider component separation compared to the
lower-frequency profiles. For PSR J1946+1805, the identification of
the different components at different frequencies is less clear. In
Fig.~\ref{fg:J1946_G}, we show the results of fitting Gaussian
components to the profiles at the four frequencies. At the lowest
frequency only three components are needed but at higher frequencies
four are required. These results confirm that the separation of the
outermost components increases with frequency (see also
Fig.~\ref{fg:sp_freq}) contrary to the standard RFM. We also checked
on the effect of the observed mode changing at lower frequencies
\citep{kr10} and found that this has no effect on this
conclusion. Whether the increasing separation with frequency results
from a different magnetospheric configuration, different emission-region
location or different viewing angle is not clear.

\subsection{Period dependence of pulse widths}\label{sec:W-P}

The inverse dependence of pulse widths with pulsar period is a
well-established phenomenon \citep{lm88,ran90,ran93}. \citet{ran90} found 
the lower bound for $W_{50}$ of core components, normalised to a
radio frequency of 1~GHz, is $2\fdg45 P^{-0.5}$. Widths of interpulse pulsars
where the magnetic axis was believed to be orthogonal to the rotation
axis followed this relation. Wider observed widths
are naturally interpreted as pulsars with more aligned magnetic axes,
giving a pulse-width relation for core components:

\begin{equation}\label{eq:core1}
W_{50}^{core,1.0} = 2\fdg45 P^{-1/2}/sin\alpha_B.
\end{equation}
where $\alpha_{B}$ is the magnetic inclination angle. \citet{mg11}
analysed a much larger sample of interpulse pulsars and obtained the
relation:
\begin{equation}\label{eq:core2}
W_{50}^{core,1.0} = (2\fdg51 \pm 0\fdg08)  P^{-0.5\pm 0.02}.
\end{equation}
\citet{mg11} and \citet{mgm12} examined the half-power pulse widths,
$W_{50}$, of a large sample (about 1500) pulsars and showed that a
similar lower bound applied.  \citet{mbm+16} found the lower bound at
618~MHz to be $2\fdg7 P^{-0.5}$ for $W_{50}$ and $5\fdg7 P^{-0.5}$ for
$W_{10}$. \citet{phs+16} used the interpulse pulsars in their 150~MHz
sample and obtained the the relations ($3\fdg5 \pm 0\fdg6) P^{-0.5}$
for $W_{50}$ and ($10\degr \pm 4\degr) P^{-0.5}$ for $W_{10}$. 
\citet{sbm+18} studied the profile component widths of 123 normal pulsars
at 333 and 618~MHz and found that the core and conal component widths 
versus period had a lower boundary line following the $P^{-0.5}$ relation.

As in \citet{zwy+17}, we take triple or multiple components pulsars
whose core components can be confirmed at 5.0~GHz to establish a
relation between the pulsar period and the core widths.  We used
gaussian fitting to measure half-power widths of core components at
5.0~GHz, $W_{50}^{core,5.0}$ and these are given in
Table~\ref{tb:core-width}. Also listed in Table~\ref{tb:core-width}
are the core widths at 1.0~GHz of these
pulsars taken directly from \citet{ran90},
\citep[cf.][]{mg11}. The average ratio of 5.0~GHz width to 1.0~GHz
width for these pulsars is about 0.93, so we adopt a scale factor for
the 5.0~GHz intrinsic core widths of $2\fdg29$:

\begin{equation}\label{eq:core3}
W_{50}^{core,5.0} = 2\fdg29 P^{-1/2}/sin\alpha_B.
\end{equation}

Fig.~\ref{fg:w-p} shows the observed 5~GHz pulse widths, $W_{50}$ and
$W_{10}$, as a function of pulse period. As in \citet{sbm+18},
  we estimate the lower bound for $W_{50}$ and $W_{10}$ using quantile
  regression with the quantile of 0.1:

\begin{equation}\label{eq:w50_QG}
W_{50} = (2\fdg19^{+0\fdg63}_{-0\fdg49}) P^{-(0.50\pm 0.20)}
\end{equation}
\begin{equation}\label{eq:w10_QG}
W_{10} = (7\fdg08^{+1\fdg04}_{-0\fdg91}) P^{-(0.32\pm 0.10)}.
\end{equation}

For comparison, the estimated relations, $(2\fdg37 \pm
0\fdg13) P^{-0.51\pm0.07}$ and $(2\fdg16 \pm 0\fdg14)
P^{-0.49\pm0.08}$, for $W_{50}$ at 333 and 618~MHz from
\citet{sbm+18} are also shown in Fig.~\ref{fg:w-p},
respectively. The lower boundary lines for $W_{50}$ at different
frequencies are very close and follow the $P^{-0.5}$ relation. For
$W_{10}$, the lower boundary line is flatter than that for
$W_{50}$ and is in fact essentially identical to the $P^{-1/3}$ relation
obtained by \citet{lm88}.

Power-law fits to the period dependence of $W_{50}$ and $W_{10}$ gave
the results:

\begin{equation}\label{eq:w50}
W_{50} = (6\fdg04\pm 0\fdg06) P^{-(0.37\pm 0.11)}
\end{equation}
\begin{equation}\label{eq:w10}
W_{10} = (13\fdg15\pm 0\fdg05) P^{-(0.37\pm 0.09)}.
\end{equation}

These fitted lines are shown on Fig.~\ref{fg:w-p}. We can see that the
relation of $W_{50}$ versus period ($P^{-0.37}$) is different to the 
lower bound relation ($P^{-0.50}$) although within the margin of 
uncertainty, they are consistent. For $W_{10}$ the lower-bound and fit 
indices are essentially the same and both close to $-1/3$.

Fig.~\ref{fg:Wc-P} shows the relation between the observed 5.0~GHz
core widths $W_{50}^{core,5.0}$ and conal widths $W_{50}^{conal,5.0}$
versus period. In the left panel, we show the estimated
  minimum boundary scaled to 5~GHz given by Equation~\ref{eq:core3}
  and the median regression relation:
\begin{equation}\label{eq:wcor50QG}
W_{50}^{core,5.0} = (2\fdg75^{+1\fdg05}_{-0\fdg75}) P^{-(0.50\pm 0.26)}.
\end{equation} 
Also shown are the minimum width-period relations obtained by 
\citet{ran90} and \citet{zwy+17}. The right panel shows the conal
component widths obtained by the multi-gaussian fitting, as listed
in Table~\ref{tb:core-width}. The lower bound for $W_{50}^{conal,5.0}$ is 
estimated using quantile regression with the quantile of 0.1:
\begin{equation}\label{eq:wcone50QG}
W_{50}^{conal,5.0} = (1\fdg32^{+0\fdg26}_{-0\fdg23}) P^{-(0.60\pm 0.14)}.
\end{equation} 
Although different samples give somewhat different relations,
overall, we find that the lower boundary lines for conal and core 
component widths are consistent with a $P^{-0.5}$ relation within the
uncertainties.

We then use Equation~\ref{eq:core3} to compute the magnetic
inclination angles $\alpha_B^{5.0}$ as listed in the third-last column
of Table~\ref{tb:core-width}. Corresponding $\alpha_B$ values for
8.6~GHz from \citet{zwy+17} and 1~GHz from \citet{ran90} are given in
the second-last and last columns, respectively. For some pulsars, the
estimated inclination angles at different frequencies are similar to each
other, but for others they are not. For example, PSR J0454+5543 was
classified as triple pulsar at 1.0~GHz \citep{ran90}, but the central
component is split into two at higher frequencies (see
Fig.~\ref{fg:mul-freq}), which might account for the discrepant
$\alpha_B$ estimates for this pulsar. For some other pulsars
(e.g. PSR J1935+1616), the core emission shows bifurcation behaviour
at higher frequencies, perhaps related to orthogonal polarization
modes \citep{mra16}, PSR J1644$-$4559, another case of discrepant
$\alpha_B$ estimates, shows no sign of either bifurcation or
orthogonal modes in the core component, even at 8.5~GHz
\citep{jkw06}. \citet{kjwk10} applied the rotating vector model
\citep{rc69a} to fit the MP and IP inclination angles in PSR
J1739$-$2903 ($\alpha_B \sim 84\degr$ for MP, $\sim 96\degr$ for IP), which
are different to the values we obtained
(Table~\ref{tb:core-width}). PSR J1955+5059 was classified as single
core component in \citet{ran90}, but may not in fact be a ``core''
component. Similarly, fitting of polarization position angles
at different frequencies \citep{bcw91,hx97a,ml04,ew01} give
different values of $\alpha_B$ with a large number of possibilities
\citep{ml04,ew01}. The different results obtained for $\alpha_B$
using different methods for a given pulsar illustrates the fact that
all of these methods rely on different assumptions, some of which may
not be valid. There is no clear and unambiguous way of determining the
magnetic inclination in the general case.

\section{Conclusions}\label{sec:concl}

Integrated pulse profiles of 71 pulsars, including eight MSPs, have
been obtained at 5~GHz with the Shanghai Tian Ma Radio Telescope
(TMRT). For 20 pulsars these are the first high-frequency observations
and for a further 19 pulsars our results have a higher signal-to-noise
ratio (S/N) than the previous observations around this frequency. By
combining the 5-GHz data with published data at frequencies between
400~MHz and 9~GHz, radio-frequency spectra are derived for 27
pulsars. For most pulsars the spectra are close to power-law, but for
seven, a broken power law is indicated. A significant correlation
between power-law slope and characteristic age, 1400~MHz
pseudo-luminosity and spin-down luminosity is observed. Mode changing
at 5.0~GHz was detected in five pulsars. For six pulsars with
interpulses (IP), the separation of the IP and main pulse (MP) was
shown to frequency independent, but the ratio of IP to MP intensity is
frequency dependent. Comparison of 5.0~GHz profiles with those at
other frequencies for 20 pulsars shows three distinct behaviours in
the profile width.  For seven pulsars, the component separation
decreases with frequency; there are eleven pulsars with no significant
change in separation between 1~GHz and 5~GHz; for the remaining two
pulsars, the separation between the outmost components increases with
frequency. The inverse dependence of pulse widths with period is shown at
5.0~GHz and the lower bound at high frequency for pulse widths is
estimated. The core half-power widths of 23 pulsars and conal 
half-power widths of 17 pulsars are measured around 5.0~GHz 
and a modified width-period relation is given for 5.0~GHz data. 
Magnetic inclination angles estimated from observed pulse widths
are often inconsistent with other estimates, showing that care must 
be taken when estimating the pulsar geometry.

\section*{Acknowledgements}
This work was supported in part by the National Natural Science Foundation of China (grants U1631122, 11173046, 11403073, 11633007 and 11673002), the strategic Priority Research Program of Chinese Academy of Sciences, Grant No. XDB23010200, the Knowledge Innovation Program of the Chinese Academy of Sciences (Grant No. KJCX1-YW-18), and the scientific Program of Shanghai Municipality (08DZ1160100)

\label{lastpage}


\begin{thebibliography}{}
\expandafter\ifx\csname natexlab\endcsname\relax\def\natexlab#1{#1}\fi
\providecommand{\url}[1]{\href{#1}{#1}}

\bibitem[{{Abdo} {et~al.}(2013){Abdo}, {Ajello}, {Allafort}, \&
  et~al.}]{aaa+13}
{Abdo}, A.~A., {Ajello}, M., {Allafort}, A., \& et~al. 2013, ArXiv:1305.4385,
  arXiv:1305.4385

\bibitem[{Anderson {et~al.}(1996)Anderson, Cadwell, Jacoby, Wolszczan, Foster,
  \& Kramer}]{acj+96}
Anderson, S., Cadwell, B.~J., Jacoby, B.~A., {et~al.} 1996, ApJ, 468, L55

\bibitem[{Backer(1970)}]{bac70a}
Backer, D.~C. 1970, Nature, 228, 1297

\bibitem[{Backer(1976)}]{bac76}
---. 1976, ApJ, 209, 895

\bibitem[{Bailes {et~al.}(1997)Bailes, Johnston, Bell, Lorimer, Stappers,
  Manchester, Lyne, D'Amico, \& Gaensler}]{bjb+97}
Bailes, M., Johnston, S., Bell, J.~F., {et~al.} 1997, ApJ, 481, 386

\bibitem[{Bartel {et~al.}(1982)Bartel, Morris, Sieber, \& Hankins}]{bmsh82}
Bartel, N., Morris, D., Sieber, W., \& Hankins, T.~H. 1982, ApJ, 258, 776

\bibitem[{{Basu} \& {Mitra}(2018)}]{bm18}
{Basu}, R., \& {Mitra}, D. 2018, \mnras, 476, 1345

\bibitem[{{Basu} {et~al.}(2015){Basu}, {Mitra}, \& {Rankin}}]{bmr15}
{Basu}, R., {Mitra}, D., \& {Rankin}, J.~M. 2015, \apj, 798, 105

\bibitem[{{Basu} {et~al.}(2018){Basu}, {Ro{\.Z}ko}, {Kijak}, \&
  {Lewandowski}}]{brkl18}
{Basu}, R., {Ro{\.Z}ko}, K., {Kijak}, J., \& {Lewandowski}, W. 2018, MNRAS,
  475, 1469

\bibitem[{Bhattacharya \& {van den Heuvel}(1991)}]{bv91}
Bhattacharya, D., \& {van den Heuvel}, E. P.~J. 1991, Phys. Rep., 203, 1

\bibitem[{{Bia{\l}kowski} {et~al.}(2018){Bia{\l}kowski}, {Lewandowski},
  {Kijak}, {B{\l}aszkiewicz}, {Krankowski}, \& {Os{\l}owski}}]{blk+18}
{Bia{\l}kowski}, S., {Lewandowski}, W., {Kijak}, J., {et~al.} 2018, \apss, 363,
  110

\bibitem[{Blaskiewicz {et~al.}(1991)Blaskiewicz, Cordes, \& Wasserman}]{bcw91}
Blaskiewicz, M., Cordes, J.~M., \& Wasserman, I. 1991, ApJ, 370, 643

\bibitem[{Camilo {et~al.}(1996)Camilo, Nice, Shrauner, \& Taylor}]{cnst96}
Camilo, F., Nice, D.~J., Shrauner, J.~A., \& Taylor, J.~H. 1996, ApJ, 469, 819

\bibitem[{{Chen} \& {Wang}(2014)}]{cw14}
{Chen}, J.~L., \& {Wang}, H.~G. 2014, ApJS, 215, 11

\bibitem[{{Chen} {et~al.}(2011){Chen}, {Wang}, {Wang}, {Lyne}, {Liu},
  {Jessner}, {Yuan}, \& {Kramer}}]{cww+11}
{Chen}, J.~L., {Wang}, H.~G., {Wang}, N., {et~al.} 2011, ApJ, 741, 48

\bibitem[{Cordes(1978)}]{cor78}
Cordes, J.~M. 1978, ApJ, 222, 1006

\bibitem[{Craft \& Comella(1968)}]{cc68}
Craft, H.~D., \& Comella, J.~M. 1968, Nature, 220, 676

\bibitem[{{Dai} {et~al.}(2015){Dai}, {Hobbs}, {Manchester}, {Kerr}, {Shannon},
  {van Straten}, {Mata}, {Bailes}, {Bhat}, {Burke-Spolaor}, {Coles},
  {Johnston}, {Keith}, {Levin}, {Os{\l}owski}, {Reardon}, {Ravi}, {Sarkissian},
  {Tiburzi}, {Toomey}, {Wang}, {Wang}, {Wen}, {Xu}, {Yan}, \& {Zhu}}]{dhm+15}
{Dai}, S., {Hobbs}, G., {Manchester}, R.~N., {et~al.} 2015, MNRAS, 449, 3223

\bibitem[{D'Amico {et~al.}(1998)D'Amico, Stappers, Bailes, Martin, Bell, Lyne,
  \& Manchester}]{dsb+98}
D'Amico, N., Stappers, B.~W., Bailes, M., {et~al.} 1998, MNRAS, 297, 28

\bibitem[{Deich {et~al.}(1986)Deich, Cordes, Hankins, \& Rankin}]{dchr86}
Deich, W. T.~S., Cordes, J.~M., Hankins, T.~H., \& Rankin, J.~M. 1986, ApJ,
  300, 540

\bibitem[{Dyks {et~al.}(2005)Dyks, Zhang, \& Gil}]{dzg05}
Dyks, J., Zhang, B., \& Gil, J. 2005, ApJ, 626, L45

\bibitem[{{Esamdin} {et~al.}(2005){Esamdin}, {Lyne}, {Graham-Smith}, {Kramer},
  {Manchester}, \& {Wu}}]{elg+05}
{Esamdin}, A., {Lyne}, A.~G., {Graham-Smith}, F., {et~al.} 2005, MNRAS, 356, 59

\bibitem[{{Everett} \& {Weisberg}(2001)}]{ew01}
{Everett}, J.~E., \& {Weisberg}, J.~M. 2001, ApJ, 553, 341

\bibitem[{Fomalont {et~al.}(1992)Fomalont, Goss, Lyne, Manchester, \&
  Justtanont}]{fgl+92}
Fomalont, E.~B., Goss, W.~M., Lyne, A.~G., Manchester, R.~N., \& Justtanont, K.
  1992, MNRAS, 258, 497

\bibitem[{Fowler {et~al.}(1981)Fowler, Wright, \& Morris}]{fwm81}
Fowler, L.~A., Wright, G. A.~E., \& Morris, D. 1981, A\&A, 93, 54

\bibitem[{Gil {et~al.}(1994)Gil, Jessner, Kijak, Kramer, Malofeev, Malov,
  Seiradakis, Sieber, \& Wielebinski}]{gjk+94}
Gil, J.~A., Jessner, A., Kijak, J., {et~al.} 1994, A\&A, 282, 45

\bibitem[{Gould \& Lyne(1998)}]{gl98}
Gould, D.~M., \& Lyne, A.~G. 1998, MNRAS, 301, 235

\bibitem[{{Han} {et~al.}(2016){Han}, {Wang}, {Xu}, \& {Han}}]{hwxh16}
{Han}, J., {Wang}, C., {Xu}, J., \& {Han}, J.-L. 2016, Res. Astron. Astrophys.,
  16, 159

\bibitem[{Hankins \& Cordes(1981)}]{hc81}
Hankins, T.~H., \& Cordes, J.~M. 1981, ApJ, 249, 241

\bibitem[{Hankins \& Fowler(1986)}]{hf86}
Hankins, T.~H., \& Fowler, L.~A. 1986, ApJ, 304, 256

\bibitem[{{Hankins} {et~al.}(2015){Hankins}, {Jones}, \& {Eilek}}]{hje15}
{Hankins}, T.~H., {Jones}, G., \& {Eilek}, J.~A. 2015, ApJ, 802, 130

\bibitem[{{Hankins} \& {Rankin}(2010)}]{hr10}
{Hankins}, T.~H., \& {Rankin}, J.~M. 2010, AJ, 139, 168

\bibitem[{{Hermsen} {et~al.}(2013){Hermsen}, {Hessels}, {Kuiper}, {van
  Leeuwen}, {Mitra}, {de Plaa}, {Rankin}, {Stappers}, \& et~al.}]{hhk+13}
{Hermsen}, W., {Hessels}, J.~W.~T., {Kuiper}, L., {et~al.} 2013, Science, 339,
  436

\bibitem[{{Hermsen} {et~al.}(2017){Hermsen}, {Kuiper}, {Hessels}, {Mitra},
  {Rankin}, {Stappers}, {Wright}, {Basu}, {Szary}, \& {van Leeuwen}}]{hkh+17}
{Hermsen}, W., {Kuiper}, L., {Hessels}, J.~W.~T., {et~al.} 2017, \mnras, 466,
  1688

\bibitem[{Hobbs {et~al.}(2004)Hobbs, Lyne, Kramer, Martin, \& Jordan}]{hlk+04}
Hobbs, G., Lyne, A.~G., Kramer, M., Martin, C.~E., \& Jordan, C. 2004, MNRAS,
  353, 1311

\bibitem[{{Hobbs} {et~al.}(2004){Hobbs}, {Faulkner}, {Stairs}, {Camilo},
  {Manchester}, {Lyne}, {Kramer}, {D'Amico}, {Kaspi}, {Possenti}, {McLaughlin},
  {Lorimer}, {Burgay}, {Joshi}, \& {Crawford}}]{hfs+04}
{Hobbs}, G., {Faulkner}, A., {Stairs}, I.~H., {et~al.} 2004, MNRAS, 352, 1439

\bibitem[{{Hoppe} {et~al.}(2009){Hoppe}, {de O{\~n}a-Wilhemi}, {Kh{\'e}lifi},
  {Chaves}, {de Jager}, {Stegmann}, {Terrier}, \& {for the
  H.~E.~S.~S.~Collaboration}}]{hok+09}
{Hoppe}, S., {de O{\~n}a-Wilhemi}, E., {Kh{\'e}lifi}, B., {et~al.} 2009,
  ArXiv:0906.5574, arXiv:0906.5574

\bibitem[{{Hotan} {et~al.}(2004){Hotan}, {van Straten}, \&
  {Manchester}}]{hvm04}
{Hotan}, A.~W., {van Straten}, W., \& {Manchester}, R.~N. 2004, PASA, 21, 302

\bibitem[{{Jankowski} {et~al.}(2018){Jankowski}, {van Straten}, {Keane},
  {Bailes}, {Barr}, {Johnston}, \& {Kerr}}]{jvk+18}
{Jankowski}, F., {van Straten}, W., {Keane}, E.~F., {et~al.} 2018, MNRAS, 473,
  4436

\bibitem[{{Janssen} \& {van Leeuwen}(2004)}]{jv04}
{Janssen}, G.~H., \& {van Leeuwen}, J. 2004, A\&A, 425, 255

\bibitem[{{Johnston} {et~al.}(2008){Johnston}, {Karastergiou}, {Mitra}, \&
  {Gupta}}]{jkmg08}
{Johnston}, S., {Karastergiou}, A., {Mitra}, D., \& {Gupta}, Y. 2008, MNRAS,
  388, 261

\bibitem[{{Johnston} {et~al.}(2006){Johnston}, {Karastergiou}, \&
  {Willett}}]{jkw06}
{Johnston}, S., {Karastergiou}, A., \& {Willett}, K. 2006, MNRAS, 369, 1916

\bibitem[{{Johnston} \& {Kerr}(2018)}]{jk18}
{Johnston}, S., \& {Kerr}, M. 2018, MNRAS, 474, 4629

\bibitem[{{Johnston} {et~al.}(1992){Johnston}, {Lyne}, {Manchester}, {Kniffen},
  {D'Amico}, {Lim}, \& {Ashworth}}]{jlm+92}
{Johnston}, S., {Lyne}, A.~G., {Manchester}, R.~N., {et~al.} 1992, MNRAS, 255,
  401

\bibitem[{Johnston {et~al.}(1998)Johnston, Nicastro, \& Koribalski}]{jnk98}
Johnston, S., Nicastro, L., \& Koribalski, B. 1998, MNRAS, 297, 108

\bibitem[{{Karastergiou} \& {Johnston}(2006)}]{kj06}
{Karastergiou}, A., \& {Johnston}, S. 2006, MNRAS, 365, 353

\bibitem[{{Karastergiou} {et~al.}(2005){Karastergiou}, {Johnston}, \&
  {Manchester}}]{kjm05}
{Karastergiou}, A., {Johnston}, S., \& {Manchester}, R.~N. 2005, MNRAS, 359,
  481

\bibitem[{{Kargaltsev} \& {Pavlov}(2007)}]{kp07}
{Kargaltsev}, O., \& {Pavlov}, G.~G. 2007, ApJ, 670, 655

\bibitem[{{Keith} {et~al.}(2011){Keith}, {Johnston}, {Levin}, \&
  {Bailes}}]{kjlb11}
{Keith}, M.~J., {Johnston}, S., {Levin}, L., \& {Bailes}, M. 2011, MNRAS, 416,
  346

\bibitem[{{Keith} {et~al.}(2010){Keith}, {Johnston}, {Weltevrede}, \&
  {Kramer}}]{kjwk10}
{Keith}, M.~J., {Johnston}, S., {Weltevrede}, P., \& {Kramer}, M. 2010, MNRAS,
  402, 745

\bibitem[{{Keith} {et~al.}(2013){Keith}, {Shannon}, \& {Johnston}}]{ksj13}
{Keith}, M.~J., {Shannon}, R.~M., \& {Johnston}, S. 2013, MNRAS, 432, 3080

\bibitem[{{Kijak} {et~al.}(2017){Kijak}, {Basu}, {Lewandowski}, {Ro{\.z}ko}, \&
  {Dembska}}]{kbl+17}
{Kijak}, J., {Basu}, R., {Lewandowski}, W., {Ro{\.z}ko}, K., \& {Dembska}, M.
  2017, ApJ, 840, 108

\bibitem[{Kijak {et~al.}(1997)Kijak, Kramer, Wielebinski, \& Jessner}]{kkwj97}
Kijak, J., Kramer, M., Wielebinski, R., \& Jessner, A. 1997, A\&A, 318, L63

\bibitem[{Kijak {et~al.}(1998)Kijak, Kramer, Wielebinski, \& Jessner}]{kkwj98}
---. 1998, A\&AS, 127, 153

\bibitem[{{Kijak} {et~al.}(2011){Kijak}, {Lewandowski}, {Maron}, {Gupta}, \&
  {Jessner}}]{klm+11}
{Kijak}, J., {Lewandowski}, W., {Maron}, O., {Gupta}, Y., \& {Jessner}, A.
  2011, A\&A, 531, A16

\bibitem[{{Kloumann} \& {Rankin}(2010)}]{kr10}
{Kloumann}, I.~M., \& {Rankin}, J.~M. 2010, \mnras, 408, 40

\bibitem[{{Kowali{\'n}ska} {et~al.}(2012){Kowali{\'n}ska}, {Kijak}, {Maron}, \&
  {Jessner}}]{kkmj12}
{Kowali{\'n}ska}, M., {Kijak}, J., {Maron}, O., \& {Jessner}, A. 2012, in
  Astronomical Society of the Pacific Conference Series, Vol. 466,
  Electromagnetic Radiation from Pulsars and Magnetars, ed. W.~{Lewandowski},
  O.~{Maron}, \& J.~{Kijak}, 101

\bibitem[{Kramer {et~al.}(1999)Kramer, Lange, Lorimer, Backer, Xilouris,
  Jessner, \& Wielebinski}]{kll+99}
Kramer, M., Lange, C., Lorimer, D.~R., {et~al.} 1999, ApJ, 526, 957

\bibitem[{{Kramer} {et~al.}(2006){Kramer}, {Lyne}, {O'Brien}, {Jordan}, \&
  {Lorimer}}]{klo+06}
{Kramer}, M., {Lyne}, A.~G., {O'Brien}, J.~T., {Jordan}, C.~A., \& {Lorimer},
  D.~R. 2006, Science, 312, 549

\bibitem[{Kramer {et~al.}(1997)Kramer, Xilouris, Jessner, Lorimer, Wielebinski,
  \& Lyne}]{kxj+97}
Kramer, M., Xilouris, K.~M., Jessner, A., {et~al.} 1997, A\&A, 322, 846

\bibitem[{Kramer {et~al.}(1996)Kramer, Xilouris, Jessner, \&
  Wielebinski}]{kxj+96}
Kramer, M., Xilouris, K.~M., Jessner, A., \& Wielebinski, R.;~Timofeev, M.
  1996, A\&A, 306, 867

\bibitem[{Kramer {et~al.}(1998)Kramer, Xilouris, Lorimer, Doroshenko, Jessner,
  Wielebinski, Wolszczan, \& Camilo}]{kxl+98}
Kramer, M., Xilouris, K.~M., Lorimer, D.~R., {et~al.} 1998, ApJ, 501, 270

\bibitem[{{Kramer} {et~al.}(2003){Kramer}, {Bell}, {Manchester}, {Lyne},
  {Camilo}, {Stairs}, {D'Amico}, {Kaspi}, {Hobbs}, {Morris}, {Crawford},
  {Possenti}, {Joshi}, {McLaughlin}, {Lorimer}, \& {Faulkner}}]{kbm+03}
{Kramer}, M., {Bell}, J.~F., {Manchester}, R.~N., {et~al.} 2003, MNRAS, 342,
  1299

\bibitem[{{Kuniyoshi} {et~al.}(2015){Kuniyoshi}, {Verbiest}, {Lee}, {Adebahr},
  {Kramer}, \& {Noutsos}}]{kvl+15}
{Kuniyoshi}, M., {Verbiest}, J.~P.~W., {Lee}, K.~J., {et~al.} 2015, \mnras,
  453, 828

\bibitem[{Large {et~al.}(1968)Large, Vaughan, \& Wielebinski}]{lvw68}
Large, M.~I., Vaughan, A.~E., \& Wielebinski, R. 1968, Nature, 220, 753

\bibitem[{{Latham} {et~al.}(2012){Latham}, {Mitra}, \& {Rankin}}]{lmr12}
{Latham}, C., {Mitra}, D., \& {Rankin}, J. 2012, \mnras, 427, 180

\bibitem[{{Levin} {et~al.}(2016){Levin}, {McLaughlin}, {Jones}, {Cordes},
  {Stinebring}, {Chatterjee}, {Dolch}, {Lam}, {Lazio}, {Palliyaguru},
  {Arzoumanian}, {Crowter}, {Demorest}, {Ellis}, {Ferdman}, {Fonseca},
  {Gonzalez}, {Jones}, {Nice}, {Pennucci}, {Ransom}, {Stairs}, {Stovall},
  {Swiggum}, \& {Zhu}}]{lmj+16}
{Levin}, L., {McLaughlin}, M.~A., {Jones}, G., {et~al.} 2016, ApJ, 818, 166

\bibitem[{Lorimer {et~al.}(1995{\natexlab{a}})Lorimer, Yates, Lyne, \&
  Gould}]{lylg95}
Lorimer, D.~R., Yates, J.~A., Lyne, A.~G., \& Gould, D.~M. 1995{\natexlab{a}},
  MNRAS, 273, 411

\bibitem[{Lorimer {et~al.}(1995{\natexlab{b}})Lorimer, Nicastro, Lyne, Bailes,
  Manchester, Johnston, Bell, D'Amico, \& Harrison}]{lnl+95}
Lorimer, D.~R., Nicastro, L., Lyne, A.~G., {et~al.} 1995{\natexlab{b}}, ApJ,
  439, 933

\bibitem[{{Lyne}(1971)}]{lyn71a}
{Lyne}, A.~G. 1971, MNRAS, 153, 27P

\bibitem[{Lyne \& Manchester(1988)}]{lm88}
Lyne, A.~G., \& Manchester, R.~N. 1988, MNRAS, 234, 477

\bibitem[{Lyne {et~al.}(1998)Lyne, Manchester, Lorimer, Bailes, D'Amico,
  Tauris, Johnston, Bell, \& Nicastro}]{lml+98}
Lyne, A.~G., Manchester, R.~N., Lorimer, D.~R., {et~al.} 1998, MNRAS, 295, 743

\bibitem[{{Maciesiak} \& {Gil}(2011)}]{mg11}
{Maciesiak}, K., \& {Gil}, J. 2011, MNRAS, 417, 1444

\bibitem[{{Maciesiak} {et~al.}(2012){Maciesiak}, {Gil}, \& {Melikidze}}]{mgm12}
{Maciesiak}, K., {Gil}, J., \& {Melikidze}, G. 2012, MNRAS, 424, 1762

\bibitem[{Malofeev {et~al.}(1994)Malofeev, Gil, Jessner, Malov, Seiradakis,
  Sieber, \& Wielebinski}]{mgj+94}
Malofeev, V.~M., Gil, J.~A., Jessner, A., {et~al.} 1994, A\&A, 285, 201

\bibitem[{{Malov} \& {Nikitina}(2013)}]{mn13}
{Malov}, I.~F., \& {Nikitina}, E.~B. 2013, Astronomy Reports, 57, 833

\bibitem[{Manchester {et~al.}(1998)Manchester, Han, \& Qiao}]{mhq98}
Manchester, R.~N., Han, J.~L., \& Qiao, G.~J. 1998, MNRAS, 295, 280

\bibitem[{{Manchester} {et~al.}(2005){Manchester}, {Hobbs}, {Teoh}, \&
  {Hobbs}}]{mhth05}
{Manchester}, R.~N., {Hobbs}, G.~B., {Teoh}, A., \& {Hobbs}, M. 2005, AJ, 129,
  1993

\bibitem[{Manchester \& Johnston(1995)}]{mj95}
Manchester, R.~N., \& Johnston, S. 1995, ApJ, 441, L65

\bibitem[{Manchester \& Lyne(1977)}]{ml77}
Manchester, R.~N., \& Lyne, A.~G. 1977, MNRAS, 181, 761

\bibitem[{Manchester {et~al.}(1978)Manchester, Lyne, Taylor, Durdin, Large, \&
  Little}]{mlt+78}
Manchester, R.~N., Lyne, A.~G., Taylor, J.~H., {et~al.} 1978, MNRAS, 185, 409

\bibitem[{Manchester \& Taylor(1977)}]{mt77}
Manchester, R.~N., \& Taylor, J.~H. 1977, Pulsars (San Francisco: Freeman)

\bibitem[{{Manchester} {et~al.}(1973){Manchester}, {Taylor}, \&
  {Huguenin}}]{mth73}
{Manchester}, R.~N., {Taylor}, J.~H., \& {Huguenin}, G.~R. 1973, ApJ, 179, L7

\bibitem[{Manchester {et~al.}(1996)Manchester, Lyne, D'Amico, Bailes, Johnston,
  Lorimer, Harrison, Nicastro, \& Bell}]{mld+96}
Manchester, R.~N., Lyne, A.~G., D'Amico, N., {et~al.} 1996, MNRAS, 279, 1235

\bibitem[{{Manchester} {et~al.}(2013){Manchester}, {Hobbs}, {Bailes}, {Coles},
  {van Straten}, {Keith}, {Shannon}, {Bhat}, {Brown}, {Burke-Spolaor},
  {Champion}, {Chaudhary}, {Edwards}, {Hampson}, {Hotan}, {Jameson}, {Jenet},
  {Kesteven}, {Khoo}, {Kocz}, {Maciesiak}, {Oslowski}, {Ravi}, {Reynolds},
  {Sarkissian}, {Verbiest}, {Wen}, {Wilson}, {Yardley}, {Yan}, \&
  {You}}]{mhb+13}
{Manchester}, R.~N., {Hobbs}, G., {Bailes}, M., {et~al.} 2013, PASA, 30, 17

\bibitem[{{Maron} {et~al.}(2000){Maron}, {Kijak}, {Kramer}, \&
  {Wielebinski}}]{mkkw00a}
{Maron}, O., {Kijak}, J., {Kramer}, M., \& {Wielebinski}, R. 2000, A\&AS, 147,
  195

\bibitem[{Maron {et~al.}(2000)Maron, Kijak, Kramer, \& Wielebinski}]{mkkw00}
Maron, O., Kijak, J., Kramer, M., \& Wielebinski, R. 2000, in Pulsar Astronomy
  - 2000 and Beyond, {IAU} Colloquium 177, ed. M.~Kramer, N.~Wex, \&
  R.~Wielebinski (San Francisco: Astronomical Society of the Pacific), 227--228

\bibitem[{{Mitra} {et~al.}(2016{\natexlab{a}}){Mitra}, {Basu}, {Maciesiak},
  {Skrzypczak}, {Melikidze}, {Szary}, \& {Krzeszowski}}]{mbm+16}
{Mitra}, D., {Basu}, R., {Maciesiak}, K., {et~al.} 2016{\natexlab{a}}, ApJ,
  833, 28

\bibitem[{Mitra \& Deshpande(1999)}]{md99}
Mitra, D., \& Deshpande, A.~A. 1999, A\&A, 346, 906

\bibitem[{Mitra \& Li(2004)}]{ml04}
Mitra, D., \& Li, X.~H. 2004, A\&A, 421, 215

\bibitem[{{Mitra} {et~al.}(2016{\natexlab{b}}){Mitra}, {Rankin}, \&
  {Arjunwadkar}}]{mra16}
{Mitra}, D., {Rankin}, J., \& {Arjunwadkar}, M. 2016{\natexlab{b}}, \mnras,
  460, 3063

\bibitem[{Mitra \& Rankin(2002)}]{mr02a}
Mitra, D., \& Rankin, J.~M. 2002, ApJ, 322

\bibitem[{Morris {et~al.}(1980)Morris, Sieber, Ferguson, \& N.}]{msfb80}
Morris, D., Sieber, W., Ferguson, D.~C., \& N., B. 1980, A\&A, 84, 260

\bibitem[{{Morris} {et~al.}(2002){Morris}, {Hobbs}, {Lyne}, {Stairs}, {Camilo},
  {Manchester}, {Possenti}, {Bell}, {Kaspi}, {Amico}, {McKay}, {Crawford}, \&
  {Kramer}}]{mhl+02}
{Morris}, D.~J., {Hobbs}, G., {Lyne}, A.~G., {et~al.} 2002, MNRAS, 335, 275

\bibitem[{{Murphy} {et~al.}(2017){Murphy}, {Kaplan}, {Bell}, {Callingham},
  {Croft}, {Johnston}, {Dobie}, {Zic}, {Hughes}, {Lynch}, {Hancock},
  {Hurley-Walker}, {Lenc}, {Dwarakanath}, {For}, {Gaensler}, {Hindson},
  {Johnston-Hollitt}, {Kapi{\'n}ska}, {McKinley}, {Morgan}, {Offringa},
  {Procopio}, {Staveley-Smith}, {Wayth}, {Wu}, \& {Zheng}}]{mdm+17}
{Murphy}, T., {Kaplan}, D.~L., {Bell}, M.~E., {et~al.} 2017, PASA, 34, e020

\bibitem[{Narayan \& Vivekanand(1983)}]{nv83a}
Narayan, R., \& Vivekanand, M. 1983, ApJ, 274, 771

\bibitem[{Nicastro {et~al.}(1995)Nicastro, Lyne, Lorimer, Harrison, Bailes, \&
  Skidmore}]{nll+95}
Nicastro, L., Lyne, A.~G., Lorimer, D.~R., {et~al.} 1995, MNRAS, 273, L68

\bibitem[{Nice {et~al.}(1996)Nice, Sayer, \& Taylor}]{nst96}
Nice, D.~J., Sayer, R.~W., \& Taylor, J.~H. 1996, ApJ, 466, L87

\bibitem[{{Perley} \& {Butler}(2013)}]{pb13}
{Perley}, R.~A., \& {Butler}, B.~J. 2013, ApJS, 204, 19

\bibitem[{{Perley} \& {Butler}(2017)}]{pb17}
---. 2017, ApJS, 230, 7

\bibitem[{{Pilia} {et~al.}(2016){Pilia}, {Hessels}, {Stappers}, {Kondratiev},
  {Kramer}, {van Leeuwen}, {Weltevrede}, {Lyne}, \& et~al.}]{phs+16}
{Pilia}, M., {Hessels}, J.~W.~T., {Stappers}, B.~W., {et~al.} 2016, A\&A, 586,
  A92

\bibitem[{Radhakrishnan \& Cooke(1969)}]{rc69a}
Radhakrishnan, V., \& Cooke, D.~J. 1969, Astrophys. Lett., 3, 225

\bibitem[{Rankin(1983{\natexlab{a}})}]{ran83}
Rankin, J.~M. 1983{\natexlab{a}}, ApJ, 274, 333

\bibitem[{Rankin(1983{\natexlab{b}})}]{ran83a}
---. 1983{\natexlab{b}}, ApJ, 274, 359

\bibitem[{Rankin(1990)}]{ran90}
---. 1990, ApJ, 352, 247

\bibitem[{{Rankin}(1992)}]{ran92}
{Rankin}, J.~M. 1992, in IAU Colloq. 128: Magnetospheric Structure and Emission
  Mechanics of Radio Pulsars, 133

\bibitem[{Rankin(1993)}]{ran93}
---. 1993, ApJ, 405, 285

\bibitem[{{Rankin}(2017)}]{ran17}
{Rankin}, J.~M. 2017, J. Astrophys. Astr., 38, 53

\bibitem[{{Redman} {et~al.}(2005){Redman}, {Wright}, \& {Rankin}}]{rwr05}
{Redman}, S.~L., {Wright}, G.~A.~E., \& {Rankin}, J.~M. 2005, MNRAS, 357, 859

\bibitem[{Ruderman \& Sutherland(1975)}]{rs75}
Ruderman, M.~A., \& Sutherland, P.~G. 1975, ApJ, 196, 51

\bibitem[{Sayer {et~al.}(1997)Sayer, Nice, \& Taylor}]{snt97}
Sayer, R.~W., Nice, D.~J., \& Taylor, J.~H. 1997, ApJ, 474, 426

\bibitem[{Seiradakis {et~al.}(1995)Seiradakis, Gil, Graham, Jessner, Kramer,
  Malofeev, Sieber, \& Wielebinski}]{sgg+95}
Seiradakis, J.~H., Gil, J.~A., Graham, D.~A., {et~al.} 1995, A\&AS, 111, 205

\bibitem[{{Shang} {et~al.}(2017){Shang}, {Lu}, {Du}, {Hao}, {Li}, {Lee}, {Li},
  {Li}, {Qiao}, {Shen}, {Wang}, {Wang}, {Wu}, {Wu}, {Xu}, {Yue}, {Yan}, {Zhi},
  {Zhao}, \& {Zhao}}]{sld+17}
{Shang}, L.-H., {Lu}, J.-G., {Du}, Y.-J., {et~al.} 2017, MNRAS, 468, 4389

\bibitem[{Sieber(1973)}]{sie73}
Sieber, W. 1973, A\&A, 28, 237

\bibitem[{Sieber {et~al.}(1975)Sieber, Reinecke, \& Wielebinski}]{srw75}
Sieber, W., Reinecke, R., \& Wielebinski, R. 1975, A\&A, 38, 169

\bibitem[{{Skrzypczak} {et~al.}(2018){Skrzypczak}, {Basu}, {Mitra},
  {Melikidze}, {Maciesiak}, {Koralewska}, \& {Filothodoros}}]{sbm+18}
{Skrzypczak}, A., {Basu}, R., {Mitra}, D., {et~al.} 2018, \apj, 854, 162

\bibitem[{Slee {et~al.}(1987)Slee, Bobra, \& Alurkar}]{sba87}
Slee, O.~B., Bobra, A.~D., \& Alurkar, S.~K. 1987, Aust. J. Phys., 40, 557

\bibitem[{{Smits} {et~al.}(2005){Smits}, {Mitra}, \& {Kuijpers}}]{smk+05}
{Smits}, J.~M., {Mitra}, D., \& {Kuijpers}, J. 2005, A\&A, 440, 683

\bibitem[{{Sobey} {et~al.}(2015){Sobey}, {Young}, {Hessels}, {Weltevrede},
  {Noutsos}, {Stappers}, {Kramer}, {Bassa}, {Lyne}, {Kondratiev}, {Hassall},
  {Keane}, {Bilous}, {Breton}, {Grie{\ss}meier}, {Karastergiou}, {Pilia},
  {Serylak}, {Veen}, {van Leeuwen}, {Alexov}, {Anderson}, {Asgekar}, {Avruch},
  {Bell}, {Bentum}, {Bernardi}, {Best}, {B{\^i}rzan}, {Bonafede}, {Breitling},
  {Broderick}, {Br{\"u}ggen}, {Corstanje}, {Carbone}, {de Geus}, {de Vos}, {van
  Duin}, {Duscha}, {Eisl{\"o}ffel}, {Falcke}, {Fallows}, {Fender}, {Ferrari},
  {Frieswijk}, {Garrett}, {Gunst}, {Hamaker}, {Heald}, {Hoeft}, {H{\"o}randel},
  {J{\"u}tte}, {Kuper}, {Maat}, {Mann}, {Markoff}, {McFadden},
  {McKay-Bukowski}, {McKean}, {Mulcahy}, {Munk}, {Nelles}, {Norden},
  {Orr{\`u}}, {Paas}, {Pandey-Pommier}, {Pandey}, {Pietka}, {Pizzo},
  {Polatidis}, {Rafferty}, {Renting}, {R{\"o}ttgering}, {Rowlinson}, {Scaife},
  {Schwarz}, {Sluman}, {Smirnov}, {Steinmetz}, {Stewart}, {Swinbank}, {Tagger},
  {Tang}, {Tasse}, {Thoudam}, {Toribio}, {Vermeulen}, {Vocks}, {van Weeren},
  {Wijers}, {Wise}, {Wucknitz}, {Yatawatta}, \& {Zarka}}]{syh+15}
{Sobey}, C., {Young}, N.~J., {Hessels}, J.~W.~T., {et~al.} 2015, \mnras, 451,
  2493

\bibitem[{Stairs {et~al.}(1999)Stairs, Thorsett, \& Camilo}]{stc99}
Stairs, I.~H., Thorsett, S.~E., \& Camilo, F. 1999, ApJS, 123, 627

\bibitem[{{Suleimanova} \& {Pugachev}(2002)}]{sp02}
{Suleimanova}, S.~A., \& {Pugachev}, V.~D. 2002, Astronomy Reports, 46, 309

\bibitem[{Taylor {et~al.}(1993)Taylor, Manchester, \& Lyne}]{tml93}
Taylor, J.~H., Manchester, R.~N., \& Lyne, A.~G. 1993, ApJS, 88, 529

\bibitem[{{Theureau} {et~al.}(2011){Theureau}, {Parent}, {Cognard},
  {Desvignes}, {Smith}, {Casandjian}, {Cheung}, {Craig}, {Donato}, {Foster},
  {Guillemot}, {Harding}, {Lestrade}, {Ray}, {Romani}, {Thompson}, {Tian}, \&
  {Watters}}]{tpc+11}
{Theureau}, G., {Parent}, D., {Cognard}, I., {et~al.} 2011, A\&A, 525, A94+

\bibitem[{Thorsett(1991)}]{tho91a}
Thorsett, S.~E. 1991, ApJ, 377, 263

\bibitem[{{Timokhin}(2010)}]{tim10}
{Timokhin}, A.~N. 2010, MNRAS, 408, L41

\bibitem[{Toscano {et~al.}(1998)Toscano, Bailes, Manchester, \&
  Sandhu}]{tbms98}
Toscano, M., Bailes, M., Manchester, R., \& Sandhu, J. 1998, ApJ, 506, 863

\bibitem[{van Ommen {et~al.}(1997)van Ommen, D'Alesssandro, Hamilton, \&
  McCulloch}]{vdhm97}
van Ommen, T.~D., D'Alesssandro, F.~D., Hamilton, P.~A., \& McCulloch, P.~M.
  1997, MNRAS, 287, 307

\bibitem[{von Hoensbroech(1999)}]{hoe99}
von Hoensbroech, A. 1999, PhD thesis, University of Bonn

\bibitem[{von Hoensbroech {et~al.}(1998)von Hoensbroech, Kijak, \&
  Krawczyk}]{hkk98}
von Hoensbroech, A., Kijak, J., \& Krawczyk, A. 1998, A\&A, 334, 571

\bibitem[{von Hoensbroech \& Xilouris(1997{\natexlab{a}})}]{hx97}
von Hoensbroech, A., \& Xilouris, K.~M. 1997{\natexlab{a}}, A\&AS, 126, 121

\bibitem[{von Hoensbroech \& Xilouris(1997{\natexlab{b}})}]{hx97a}
---. 1997{\natexlab{b}}, A\&A, 324, 981

\bibitem[{{Wang} {et~al.}(2015){Wang}, {Zhao}, {Yu}, {Yin}, {Lao}, {Wu}, {Li},
  {Dong}, {Jiang}, {Xia}, {Zuo}, {Gou}, {Guo}, {Wu}, {Lu}, {Liu}, {Fan},
  {Jiang}, \& {Qian}}]{wzy+15}
{Wang}, J.~Q., {Zhao}, R.~B., {Yu}, L.~F., {et~al.} 2015, Acta Astronomica
  Sinica, 56, 278

\bibitem[{{Wang} {et~al.}(2007){Wang}, {Manchester}, \& {Johnston}}]{wmj07}
{Wang}, N., {Manchester}, R.~N., \& {Johnston}, S. 2007, MNRAS, 377, 1383

\bibitem[{Weisberg {et~al.}(1999)Weisberg, Cordes, Lundgren, Dawson, Despotes,
  Morgan, Weitz, Zink, \& Backer}]{wcl+99}
Weisberg, J.~M., Cordes, J.~M., Lundgren, S.~C., {et~al.} 1999, ApJS, 121, 171

\bibitem[{{Weltevrede} \& {Johnston}(2008)}]{wj08}
{Weltevrede}, P., \& {Johnston}, S. 2008, MNRAS, 391, 1210

\bibitem[{Weltevrede {et~al.}(2007)Weltevrede, Wright, \& Stappers}]{wws07}
Weltevrede, P., Wright, G.~A.~E., \& Stappers, B.~W. 2007, A\&A, 467, 1163

\bibitem[{{Wen} {et~al.}(2016){Wen}, {Wang}, {Yan}, {Yuan}, {Liu}, {Chen}, \&
  {Chen}}]{wwy+16}
{Wen}, Z.~G., {Wang}, N., {Yan}, W.~M., {et~al.} 2016, \apss, 361, 261

\bibitem[{Wright \& Fowler(1981)}]{wf81}
Wright, G.~A., \& Fowler, L.~A. 1981, A\&A, 101, 356

\bibitem[{Xilouris {et~al.}(1996)Xilouris, Kramer, Jessner, Wielebinski, \&
  Timofeev}]{xkj+96}
Xilouris, K.~M., Kramer, M., Jessner, A., Wielebinski, R., \& Timofeev, M.
  1996, A\&A, 309, 481

\bibitem[{{Yan} {et~al.}(2015){Yan}, {Shen}, {Wu}, {Manchester}, {Weltevrede},
  {Wu}, {Zhao}, {Yuan}, {Lee}, {Fan}, {Hong}, {Jiang}, {Li}, {Liang}, {Ling},
  {Liu}, {Qian}, {Zhang}, {Zhong}, \& {Ye}}]{ysw+15}
{Yan}, Z., {Shen}, Z.-Q., {Wu}, X.-J., {et~al.} 2015, ApJ, 814, 5

\bibitem[{{Yan} {et~al.}(2018{\natexlab{a}}){Yan}, {Shen}, {Manchester}, {Ng},
  {Weltevrede}, {Wang}, {Wu}, {Yuan}, {Wu}, {Zhao}, {Liu}, {Zhao}, \&
  {Liu}}]{ysm+18}
{Yan}, Z., {Shen}, Z.-Q., {Manchester}, R.~N., {et~al.} 2018{\natexlab{a}},
  \apj, 856, 55

\bibitem[{{Yan} {et~al.}(2018{\natexlab{b}}){Yan}, {Shen}, {Wu}, {Zhao},
  {Zhao}, {Liu}, {Huang}, {Liu}, \& {Wu}}]{ysw+18}
{Yan}, Z., {Shen}, Z.-Q., {Wu}, Y.-J., {et~al.} 2018{\natexlab{b}}, Radio
  Science Bulletin, 366, 10

\bibitem[{{Zhao} {et~al.}(2017){Zhao}, {Wu}, {Yan}, {Shen}, {Manchester},
  {Qiao}, {Xu}, {Wu}, {Zhao}, {Li}, {Du}, {Lee}, {Hao}, {Liu}, {Lu}, {Shang},
  {Wang}, {Wang}, {Yuan}, {Zhi}, \& {Zhong}}]{zwy+17}
{Zhao}, R.-S., {Wu}, X.-J., {Yan}, Z., {et~al.} 2017, ApJ, 845, 156

\end{thebibliography}

\clearpage
\startlongtable 
\begin{deluxetable}{c c c c c c c c c c c c c}
\tabletypesize{\footnotesize}
\tablecaption{Parameters for the 71 pulsars observed with the TMRT\label{tb:fluxpsr}}
\tablehead{
\colhead{PSR J2000}  & \colhead{MJD} & \colhead{$\rm Freq$}& \colhead{$\rm T_{on}$} & \colhead{$\rm S \pm \sigma_{\rm b}$}& \colhead{$\Delta\phi(N)$} & \colhead{$\rm W_{50}$} & \colhead{$\rm W_{10}$} & \colhead{$\rm S^\ddag$}\\
\colhead{Name} & \colhead{(d)}  & \colhead{(MHz)}& \colhead{(min)} & \colhead{(mJy)} & \colhead{(deg)} & \colhead{(deg)} & \colhead{(deg)} & \colhead{(mJy)}
}
\startdata
J0147+5922 & 57721$^\dag$ & 4820.0 &	40 & $\,\ 2.49\pm0.13^*$ & 6.7(3) & 2.9 & 4.1 \\
                 & 57245$\ $ & 4920.0 & 30 & $\,0.91\pm0.03$ & 6.0(3) & 2.2 & 4.3 \\
             & & & & $\,1.70\pm0.80$ & $6.3\pm0.4$ & $2.5\pm0.3$ & $4.2\pm0.1$ & $0.35\pm0.11^1$\\
J0248+6021 & 57845$^\dag$ & 4820.0 & 17 & $\,\ 3.10\pm0.21^*$	& -- & $7.6\pm0.2$ & $23.2\pm1.0$ & \\
J0332+5434 & 57721$^\dag$ & 4820.0 & 30 & $\,\ 7.13\pm0.03^*$	& 20.2(3) & 4.4 & 22.4 \\
		   & 57378$\ $ & 4820.0 & 8 & $\,3.37\pm0.07$ & 20.1(3) & 4.5 & 22.6 \\
		   & 57414$\ $ & 4820.0 & 8 & $\,13.41\pm0.07$ & 20.2(3) & 4.5 & 22.5 \\
		   & 57443$\ $ & 4820.0 & 8 & $\,14.06\pm0.08$ & 20.3(3) & 4.3 & 21.7 \\
		   & 57466$\ $ & 4820.0 & 8 & $\,19.15\pm0.08$ & 20.1(3) & 4.6 & 22.6 \\
		   & 57536$\ $ & 4820.0 & 8 & $\,11.47\pm0.07$ & 18.7(3) & 12.3 & 23.0 \\
		   & 57571$\ $ & 4820.0 & 8 & $\,8.11\pm0.07$ & 20.2(3) & 4.4 & 22.5 \\
		   & 57575$\ $ & 4820.0 & 15 & $\,7.61\pm0.07$ & 20.2(3) & 4.6 & 22.5 \\
		   & 57740$\ $ & 4820.0 & 8 & $\,5.41\pm0.06$ & 20.4(3) & 5.0 & 22.7 \\
		   & 57760$\ $ & 4820.0 & 8 & $\,7.03\pm0.07$ & 18.4(3) & 12.5 & 23.7 \\
		   & 57802$\ $ & 4820.0 & 8 & $\,13.22\pm0.07$ & 20.3(3) & 4.4 & 23.0 \\
		   & 57806$\ $ & 4820.0 & 8 & $\,6.27\pm0.07$ & 19.5(3) & 4.3 & 21.7 \\
		   & 57827$\ $ & 4820.0 & 8 & $\,6.46\pm0.07$ & 18.5(3) & 12.3 & 23.0 \\
		   & 57844$\ $ & 4820.0 & 8  & $\,\ 5.11\pm0.42^*$	& 19.7(3) &	4.0	& 22.2 \\
		   & 57872$\ $ & 4820.0 & 30 & $\,13.42\pm0.03^*$ & 20.1(3) & 3.8 & 22.3 \\
		   & 57887$\ $ & 4820.0 & 8 & $\,18.81\pm0.07$ & 20.0(3) & 3.8 & 22.3 \\
		   & & & & $\,10.00\pm4.87$& $19.8\pm0.7$ & $5.9\pm3.2$ & $22.5\pm0.5$ & $5.64^2$,$20.5^3$\\
J0358+5413 & 57378$\ $ & 4820.0 & 8 & $\,6.23\pm0.07$ & 14.4(3) & 5.1 & 25.5 \\
		   & 57443$\ $ & 4820.0 & 15 & $\,6.86\pm0.05$ & 17.3(3) & 5.5 & 27.9 \\
		   & 57536$\ $ & 4820.0 &	15 & $\,5.89\pm0.05$ & 18.8(3) & 5.1 & 25.8 \\
		   & 57549$^\dag$ & 4820.0 &	15 & $\,7.76\pm0.05$ & 17.8(3) & 5.9 & 28.7 \\
		   & 57576$\ $ & 4820.0 & 20 & $\,11.38\pm0.04$ & 14.8(3) & 8.8 & 31.7 \\
		   & 57771$\ $ & 4820.0 & 15 & $\,6.73\pm0.05$ & 15.7(3) & 6.1 & 28.8 \\
		   & 57887$\ $ & 4820.0 & 10 & $\,5.67\pm0.06$ & 18.2(3) & 7.5 & 30.5 \\
		   & & & & $\,7.22\pm1.97$ & $16.7\pm1.7$ & $6.3\pm1.4$ & $28.4\pm2.3$ & $9.10^2$,$3.36^3$\\
J0454+5543 & 57414$^\dag$ & 4820.0 & 10 & $\,1.97\pm0.06$ & 13.1(3) & 18.6 & 30.0 \\
		   & 57443$\ $ & 4820.0 & 10 & $\,1.75\pm0.06$ & 13.9(3) & 18.9 & 31.4 \\
		   & 57466$^\dag$ & 4820.0 &	10 & $\,1.47\pm0.05$ & 13.3(3) & 18.0 & 28.0 \\
		   & 57571$\ $ & 4820.0 &	10 & $\,1.57\pm0.06$ & 13.8(3) & 17.8 & 29.7 \\
		   & 57802$\ $ & 4820.0 & 10 & $\,\ 3.24\pm0.09^*$ & 13.3(3) & 18.5 & 28.0 \\
		   & & & & $\,2.00\pm0.73$	& $13.5\pm0.3$ & $18.4\pm0.5$ & $29.5\pm1.5$ & $0.07^2$,$1.78^3$\\
J0528+2200  & 57575$^\dag$ & 4820.0 &	60 & $\,0.27\pm0.02$	& 11.3(2) &	12.9 & 15.4 \\ 
			& 57845$\ $ & 4820.0 &	30 & $\,\ 0.26\pm0.12^*$	& 11.1(2) & 12.5 & 16.3	\\ 
		 	& 57887$\ $ & 4820.0 &	30 & $\,0.30\pm0.03$	& 11.4(2) &	13.2 & 16.5 \\ 
		    & & & & $\,0.28\pm0.13$	& $11.3\pm0.2$ & $12.9\pm0.4$ & $16.1\pm0.6$ & $0.46^2$,$0.10^3$\\ 
J0538+2817 & 58102$^\dag$ & 4820.0 &	24 & $\ \,0.90\pm0.18^*$	& 10.8(2) &	$20.9\pm1.4$ & $37.2\pm2.2$ \\ 
	       & 58114$^\dag$ & 4820.0 &	60 & $\ \,0.42\pm0.09^*$	& 12.5(2) &	$18.2\pm1.0$ & $28.1\pm2.4$ \\
	 &  &  &	 & $\,0.67\pm0.31$	& $11.7\pm0.8$ & $19.5\pm2.2$ & $32.7\pm5.6$ \\ 
J0543+2329	 & 57536$\ $ & 4820.0 &	30 & $\,1.83\pm0.03$	& -- & 10.8 & 25.9 \\
			 & 57579$\ $ & 4820.0 &	30 & $\,1.67\pm0.03$	& -- & 12.8 & 25.8 \\
			 & 57845$^\dag$ & 4820.0 &	30 & $\,\ 1.96\pm0.16^*$	& -- & 11.0 & 26.3 \\
			 & & & & $\,1.82\pm0.22$	& -- & $11.5\pm1.1$ & $26.0\pm0.3$ & $2.79^2$,$2.19^3$\\
J0738$-$4042 & 57845$^\dag$ & 4820.0 &	15 & $\,\ 8.06\pm0.66^*$	& 11.3(3) &	$16.1\pm0.3$ & $30.1\pm0.9$ & \\
J0742$-$2822 & 57291$\ $ & 4820.0 & 10 & $\,1.16\pm0.06$ & 5.6(2) & 11.0 & 17.3 \\
			 & 57349$\ $ & 4820.0 & 20 & $\,0.59\pm0.04$ & 6.2(2) & 9.5 & 16.9 \\
			 & 57378$\ $ & 4820.0 & 10 & $\,3.11\pm0.06$ & 5.5(2) & 8.4 & 17.6 \\
			 & 57414$\ $ & 4820.0 & 10 & $\,2.50\pm0.06$ & 4.8(2) & 10.2 & 18.4 \\
			 & 57466$\ $ & 4820.0 & 10 & $\,3.36\pm0.06$ & 5.6(2) & 9.0 & 16.3 \\
			 & 57536$\ $ & 4820.0 & 10 & $\,1.64\pm0.06$ & 5.1(2) & 8.4 & 17.5 \\
			 & 57549$\ $ & 4820.0 & 30 & $\,3.88\pm0.04$ & 4.8(2) & 8.6 & 17.5 \\
			 & 57550$\ $ & 4820.0 & 40 & $\,3.50\pm0.03$ & 5.3(2) & 8.8 & 17.4 \\
			 & 57571$\ $ & 4820.0 & 10 & $\,2.48\pm0.06$ & 5.7(2) & 8.8 & 17.0 \\
			 & 57572$\ $ & 4820.0 & 10 & $\,2.68\pm0.06$ & 5.6(2) & 8.8 & 19.4 \\
			 & 57579$\ $ & 4820.0 & 10 & $\,2.62\pm0.06$ & 5.5(2) & 9.1 & 19.7 \\
			 & 57740$\ $ & 4820.0 & 20 & $\,3.36\pm0.04$ & 5.4(2) & 8.4 & 17.3 \\
			 & 57760$\ $ & 4820.0 & 20 & $\,5.78\pm0.04$ & 5.5(2) & 9.7 & 17.3 \\
			 & 57845$\ $ & 4820.0 & 15 & $\,\ 1.65\pm0.13^*$ & 5.3(2) & 10.8 & 20.0 \\
			 & 57872$^\dag$ & 4820.0 &	10 & $\,\ 2.07\pm0.30^*$ & 5.9(2) & 8.7 & 16.9 \\
			 & 57887$\ $ & 4820.0 & 15 & $\,1.95\pm0.05$ & 4.7(2) & 11.4 & 20.3 \\
			 & & & & $\,2.65\pm1.29$ & $5.4\pm0.4$ & $9.4\pm1.0$ & $17.9\pm1.2$ & $1.73^2$,$2.47^3$ \\
J0820$-$1350 & 57350$\ $ & 4820.0 & 30 & $0.23\pm0.04$ & 2.7(2) & 5.0 & 9.1 \\
			 & 57806$^\dag$ & 4820.0 & 30 & $0.74\pm0.03$ & 2.8(2)	& 6.1 & 10.1 \\
			 & 57845$\ $ & 4820.0 & 30 & $0.41\pm0.04$ & 2.2(2) & 5.6 & 10.3 \\
			 & & & & $\,0.47\pm0.27$ & $2.6\pm0.3$ & $5.6\pm0.6$ & $9.8\pm0.6$ & $0.33\pm0.03$\\
J0826+2637 & 57350$\ $ & 4820.0 &	80 & $0.41\pm0.02$ & --	& 3.0 & 6.5	\\
		   & 57802$\ $ & 4820.0 &	30 & $\,\ 0.59\pm0.03^*$ & --	& 2.7 &	5.5 \\
		   & 57806$\ $ & 4820.0 &	30 & $1.62\pm0.03$ & --	& 2.7 & 6.4	\\
		   & 57825$\ $ & 4820.0 &	30 & $0.95\pm0.03$ & --	& 2.7 & 6.3	\\
		   & 57887$^\dag$ & 4820.0 & 30 & $2.21\pm0.04$ & --	& 2.8 & 6.0 \\
		   & & & & $\,1.15\pm0.75$ & -- & $2.8\pm0.1$ & $6.1\pm0.4$ & $0.71^2$,$2.63^3$\\
J0835$-$4510 & 57291$\ $& 4820.0 & 8 & $28.61\pm0.11\,\ $ &	6.9(3) & 13.4 &	23.4 \\
			 & 57349$\ $& 4820.0 & 8 & $24.61\pm0.11\,\ $ &	6.5(3) & 13.3 &	22.5 \\
			 & 57378$\ $& 4820.0 & 8 & $116.15\pm0.11\,\ $ & 7.1(3) & 13.8 & 23.1 \\
			 & 57414$\ $& 4820.0 & 8 & $79.42\pm0.11\,\ $ &	6.8(3) & 14.3 &	27.4 \\
			 & 57443$\ $& 4820.0 & 8 & $93.93\pm0.11\,\ $ &	6.6(3) & 13.5 & 22.7 \\
			 & 57466$^\dag$	& 4820.0 & 8 & $95.42\pm0.11\,\ $ &	7.0(3) & 13.6 &	23.2 \\
			 & & & & $73.02\pm37.83\,\ $&	$6.8\pm0.2$ & $13.7\pm0.4$ & $23.7\pm1.8$ \\
J0837$-$4135 & 57414$\ $ & 4820.0 & 10 & $2.73\pm0.09$ & 10.7(3) & 3.7 & 12.8 \\
			 & 57443$\ $ & 4820.0 & 10 & $2.30\pm0.09$ & 10.5(3) & 3.0 & 12.6 \\
			 & 57466$\ $ & 4820.0 & 10 & $2.80\pm0.09$ & 10.6(3) & 3.1 & 12.5 \\
			 & 57571$^\dag$ & 4820.0 & 10 & $6.55\pm0.09$ & 10.7(3) & 3.0 & 13.1 \\
			 & 57760$\ $ & 4820.0 & 10 & $3.46\pm0.09$ & 10.5(3) & 3.2 & 12.6 \\
			 & & & & $\,3.57\pm1.73$ & $10.6\pm0.1$	& $3.2\pm0.3$ & $12.7\pm0.2$ \\
J0953+0755 & 57771$^\dag$ & 4820.0 & 110 & $12.76\pm2.55\,\ $ & -- & $11.8\pm0.1$ & $30.1\pm0.4$ & $5.27^2$,$41.3^3$\\
J0953+0755i& 57771$\ $ & 4820.0 & 110 & $0.35\pm0.07$ & -- & $13.5$ & -- \\
J1012+5307 & 57549$^\dag$ & 4820.0 & 140 & $0.17\pm0.04$ & -- & $14.5\pm1.8$ & -- & $0.2\pm0.1^4$\\
J1012+5307i& 57549$\ $ & 4820.0 & 140 & $0.05\pm0.04$ & -- & $11.6\pm2.4$ & -- \\
J1136+1559 & 57466 & 4820.0 & 8 & $1.44\pm0.07\,\ $ & 5.0(2) & 1.7 & 7.8 \\
		   & 57570 & 4820.0 & 8 & $1.16\pm0.07\,\ $ & 5.3(2) & 1.4 & 8.3 \\
		   & 57721$^\dag$ & 4820.0 & 40 & $\,\ 2.89\pm0.14^*$ & 5.2(2) & 1.4 & 8.1 \\
		   & 57772 & 4820.0 & 15 & $3.80\pm0.05\,\ $ & 5.0(2) & 1.4 & 8.0 \\
		   & 57844$\ $ & 4820.0 & 30	& $\,\ 1.48\pm0.14^*$ & 5.3(2) & 1.3 & 8.7 \\
		   & 57912$\ $ & 4820.0 & 30 & $\,\ 2.38\pm0.07^*$ & 4.9(2) & 1.5 & 8.1 \\
		   & & & & $\,2.19\pm1.05$ & $5.1\pm0.2$ & $1.5\pm0.1$ & $8.2\pm0.3$ & $0.92^2$,$4.38^3$\\
J1239+2453 & 57245$\ $ & 4920.0 & 13	& $\,1.01\pm0.05$ & 9.1(5) & 10.5 & 12.6 \\
		   & 57721$^\dag$ & 4820.0 & 40 & $\,\ 1.03\pm0.07^*$ & 8.9(5) & 10.5 & 12.9 \\
		   & 57827$\ $ & 4820.0 & 30 & $\,0.54\pm0.03$ & 8.3(5) & 10.6 & 13.1 \\
		   & 57844$\ $ & 4820.0 & 30	& $\,\ 0.71\pm0.08^*$ & 8.8(5)& 10.4 & 12.7 \\
 		   & & & & $\,0.82\pm0.27$ & $8.8\pm0.3$ & $10.5\pm0.1$ & $12.8\pm0.2$ & $1.50^2$,$0.79^3$\\
J1509+5531 & 57721$^\dag$ & 4820.0 &	30 & $\,\ 0.54\pm0.07^*$	& 5.9(2) & $8.5\pm0.2$ & $12.4\pm0.7$ & $0.39^2$\\
J1518+4904 & 57844$^\dag$ & 4820.0 &	30 & $\,\ 0.24\pm0.04^*$	& -- & 12.5 & 20.2 \\
 		   & 58114$\ $ & 4820.0 &	60 & $\,\ 0.36\pm0.13^*$	& -- & 12.0 & 21.3 \\
 		   & & & & $\,0.30\pm0.16$	& -- & $12.3\pm0.4$ & $20.8\pm0.8$ & $0.24\pm0.07^4$ \\
J1559$-$4438 & 57155$^\dag$ & 5124.0 &	30 & $1.59\pm0.32$ & 17.5(3) & $24.3\pm0.6$ & $30.7\pm1.9$ \\
J1600$-$3053 & 57844$^\dag$ & 4820.0 &	30 & $\,\ 0.27\pm0.04^*$	& 12.7(2) & $27.0\pm1.4$ & $41.6\pm3.5$ \\
J1643$-$1224 & 57825$^\dag$ & 4820.0 &	30 & $0.68\pm0.14$ & 28(2) & $18.9\pm2.0$ & $56.1\pm6.0$ & $0.4\pm0.2^4$\\
J1644$-$4559 & 57155$^\dag$ & 5124.0 &	30 & $30.00\pm6.00\,\ $	& 9.6(2) & $5.48\pm0.01$ & $10.54\pm0.02$ \\
J1645$-$0317 & 57331$\ $ & 4820.0 & 23 & $2.34\pm0.04$ & 10.9(3) & 3.9 & 17.6 \\
			 & 57466$\ $ & 4820.0 &	8 & $1.57\pm0.06$ & 11.5(3) & 3.4 & 16.8 \\
			 & 57536$^\dag$ & 4820.0 & 8 & $1.28\pm0.07$ & 11.4(3) & 3.6 & 15.9 \\
			 & 57844$\ $ & 4820.0 &	30 & $1.34\pm0.05$ & 11.3(3) & 3.2 & 16.5 \\
			 & & & & $\,1.63\pm0.50$ & $11.3\pm0.3$ & $3.5\pm0.3$ & $16.7\pm0.7$ & $2.35^2$,$1.10^3$ \\
J1651$-$4246 & 57155$^\dag$ & 5124.0 & 30 & $1.15\pm0.23$ & 35.5(2) & $16.1\pm1.0$ & $58.0\pm7.7$ \\
J1703$-$3241 & 57155$^\dag$ & 5124.0 & 30 & $0.72\pm0.14$ & 7.1(3) & $10.3\pm0.7$	& $12.8\pm1.2$ \\
J1705$-$1906 & 57155$^\dag$ & 5124.0 & 30 & $2.43\pm0.50$ & 4.7(2) & $10.7\pm1.7$ & $15.2\pm1.9$ & $1.10\pm0.11^1$\\
J1705$-$1906i& 57155$\ $ & 5124.0 & 30 & $0.25\pm0.05$ & -- & $6.0\pm1.0$ & $12.0\pm1.3$ & $0.14\pm0.11^1$\\ 
J1705$-$3423 & 57245$^\dag$ & 4920.0 & 30 & $1.01\pm0.20$ & 8.5(2) & $13.7\pm0.8$ & $25.4\pm2.0$	\\
J1707$-$4053 & 57155$^\dag$ & 5124.0 & 30 & $0.55\pm0.11$ & -- & $11.0\pm1.2$ & $16.1\pm2.9$ \\
J1709$-$4429 & 57155$^\dag$ & 5124.0 & 30 & $3.51\pm0.70$ & -- & $19.4\pm0.8$ & $35.2\pm2.4$ \\
J1713+0747  & 57379$^\dag$ & 4820.0 & 10 & $\,1.42\pm0.06$ &	74.6(4) & 9.8 & 82.2 \\
			& 57570$\ $ & 4820.0 &	15 & $\,1.97\pm0.03$ & 76.6(4) & 10.1 & 88.3 \\
			& 57721$\ $ & 4820.0 &	26 & $\,\ 0.52\pm0.03^*$ & 75.8(4) & 10.9 & 87.9 \\
			& 57772$\ $ & 4820.0 &	26 & $\,2.03\pm0.04$ & 73.0(4) & 10.5 & 83.5 \\
			& 57844$\ $ & 4820.0 & 15 & $\,\ 1.19\pm0.11^*$ & 77.0(4) & 10.5 & 79.3 \\
			& & & & $\,1.42\pm0.64$	&$75.4\pm1.6$&$10.4\pm0.4$&$84.2\pm3.8$&$0.80\pm0.04^1$,$0.8\pm0.2^4$\\
J1721$-$3532 & 57157$^\dag$ & 5124.0 &	30 & $\,6.03\pm1.20$ & -- & $10.5\pm0.2$ & $19.9\pm0.2$ \\
			 & 57844$\ $ & 4820.0 &	9 & $\,12.25\pm1.34^*$ & -- & $11.6\pm0.5$ & $23.2\pm1.6$ \\
			 & & & & $\,9.14\pm3.59$	& -- & $11.1\pm0.8$ & $21.5\pm2.3$&	\\
J1730$-$3350 & 57245$^\dag$ & 4920.0 & 30 & $0.83\pm0.17$ & -- & $10.7\pm0.6$ & $20.5\pm1.4$ \\
J1739$-$2903 & 57245$^\dag$ & 4920.0 & 30 & $1.22\pm0.24$ & -- & $6.5\pm0.2$ & $11.4\pm1.4$ \\
J1739$-$2903i& 57245$\ $ & 4920.0 & 30 & $0.26\pm0.05$ & -- & $7.2\pm0.7$& $15.6\pm1.6$ \\
J1740$-$3015 & 57157$^\dag$ & 5124.0 &	30 & $2.29\pm0.46$ & 1.4(2)	& $2.92\pm0.02$	& $6.13\pm0.03$ \\
J1744$-$1134 & 57802$^\dag$ & 4820.0	& 30 & $\,\ 0.36\pm0.02^*$ & -- & $7.9\pm0.5$ & $22.2\pm2.0$ & $0.19\pm0.06^4$\\
J1745$-$3040 & 57292$\ $ & 4820.0 & 10 & $\,3.91\pm0.07$	& 26.4(3) & 9.5 & 24.1 \\
			 & 57379$\ $ & 4820.0 &	10 & $\,\ 2.34\pm0.06$	& 21.7(3) & 9.8 & 24.4 \\
             & 57572$^\dag$ & 4820.0 &	10 & $\,\ 2.42\pm0.06^*$	& 24.8(3) & 10.0 & 26.0 \\
             & 57802$\ $ & 4820.0 & 10 & $\,3.46\pm0.06$	& 24.4(3) & 9.3 & 24.1 \\
			 & 57844$\ $ & 4820.0 &	10 & $\,\ 1.96\pm0.18^*$	& 24.0(3) & 9.1 & 23.2 \\
             & & & & $\,2.82\pm0.0.86$	& $24.3\pm1.7$ & $9.5\pm0.4$ & $24.4\pm1.0$ & $1.63^2$\\
J1752$-$2806 & 57292$\ $ & 4820.0 & 10 & $\,1.86\pm0.07$ & 5.2(2) & 2.9 & 8.9 \\
			 & 57320$\ $ & 4820.0 & 25 & $\,0.83\pm0.05$ & 5.1(2) & 2.9 & 8.1 \\
			 & 57379$^\dag$ & 4820.0 & 10 & $\,\ 1.58\pm0.07^*$	& 5.9(2) & 2.8 & 9.1 \\
			 & 57444$\ $ & 4820.0 & 10 & $\,2.74\pm0.07$ & 4.9(2) & 3.2 & 8.8 \\
			 & 57466$\ $ & 4820.0 & 10 & $\,0.94\pm0.07$ & 5.5(2) & 2.7 & 8.3 \\
			 & 57570$\ $ & 4820.0 & 10 & $\,2.76\pm0.07$ & 5.1(2) & 2.9 & 8.8 \\
			 & 57572$\ $ & 4820.0 & 10 & $\,1.97\pm0.07$ & 5.5(2) & 2.9 & 9.2 \\
			 & 57844$\ $ & 4820.0 & 10 & $\,\ 1.37\pm0.15^*$ & 5.4(2) & 2.4 & 8.3 \\
			 & & & & $\,1.76\pm0.77$	& $5.3\pm0.34$ & $2.8\pm0.2$ & $8.7\pm0.4$ & $0.44^2$\\
J1803$-$2137 & 57245$^\dag$ & 4920.0 &	30 & $8.62\pm1.72$ & 46.3(2) & $19.4\pm0.3$	& $88.1\pm2.2$ & $8.04^2$\\
J1807$-$0847 & 57721$^\dag$ & 4820.0 & 30 & $\,\ 1.90\pm0.09^*$	& 16.5(3) & $20.9\pm0.3$ & $26.5\pm0.6$ & $2.30^2$\\
J1809$-$1917 & 57245$^\dag$ & 4920.0 & 30 & $3.56\pm0.71$ & 39.7(2) & $57.6\pm1.0$ & $84.3\pm1.8$ \\
J1818$-$1422 & 57155$^\dag$ & 5124.0 & 30 & $0.79\pm0.16$ & 6.4(2) & $4.8\pm0.1$ & $15.1\pm0.4$ & $0.51\pm0.06^1$\\
J1820$-$0427 & 57245$^\dag$ & 4920.0 & 15 & $0.57\pm0.11$ & 7.2(2) & $3.8\pm0.2$ & $13.7\pm0.9$ & $0.43\pm0.04^1$ \\
J1825$-$0935m& 57415$\ $ & 4820.0 & 10 & $2.69\pm0.06$	& 13.3(2) &	3.2 & 21.4 \\
			 & 57444$\ $ & 4820.0 & 10 & $1.05\pm0.06$	& 13.6(2) &	2.9 & 21.2 \\
			 & 57467$\ $ & 4820.0 & 10 & $2.46\pm0.06$	& 11.9(2) &	2.7 & 18.5 \\
			 & 57536$\ $ & 4820.0 & 10 & $3.11\pm0.06$	& 14.3(2) &	3.9 & 21.5 \\
			 & 57570$\ $ & 4820.0 & 10 & $3.32\pm0.06$	& 13.8(2) &	3.4 & 21.9 \\
			 & 57572$\ $ & 4820.0 & 10 & $2.53\pm0.06$	& 12.8(2) &	3.3 & 21.0 \\
			 & 57721$^\dag$ & 4820.0 & 20 & $\,\ 1.59\pm0.05^*$ & 13.3(2) & 3.3 & 21.4 \\
			 & 57913$\ $ & 4820.0 & 30 & $\,\ 1.20\pm0.05^*$	& 14.1(2) & 3.8 & 21.4 \\
 			 & & & & $\,2.24\pm0.88$ & $13.4\pm0.8$ & $3.3\pm0.5$ & $21.0\pm1.1$ & $6.52^2$,$4.76^3$\\
J1825$-$0935i& 57467$\ $ & 4820.0 & 10 & $0.32\pm0.06$	& -- &	2.1 & -- \\
			 & 57572$\ $ & 4820.0 & 10 & $0.15\pm0.06$	& -- &	4.2 & -- \\
			 & & & & $\,0.23\pm0.15$ & -- & $3.2\pm1.5$ & -- & \\
J1826$-$1334 & 57245$^\dag$ & 4920.0 & 30 & $4.59\pm0.92$ & 57.9(2) & $14.2\pm0.3$ & $96.9\pm1.9$ & $2.60\pm0.26^1$\\
J1829$-$1751 & 57155$^\dag$ & 5124.0 & 30 & $1.63\pm0.33$ & 12.0(2) & $2.45\pm0.03$ & $16.9\pm0.1$ & $1.23^2$\\
J1830$-$1059 & 57245$^\dag$ & 4920.0 & 30 & $0.56\pm0.11$ & -- & $3.8\pm0.1$ & $7.6\pm0.2$ \\
J1833$-$0827 & 57245$^\dag$ & 4920.0 & 30 & $2.02\pm0.40$ & 43.7(3) & $54.6\pm2.5$ & $71.7\pm4.8$ & $0.90\pm0.08^1$\\
J1835$-$1106 & 57245$^\dag$ & 4920.0 & 30 & $0.38\pm0.08$ & 8.5(3) & $6.9\pm0.4$ & $16.5\pm1.8$ \\
J1844+00$\ \ \ $& 57155$^\dag$ & 5124.0 & 30 & $0.40\pm0.03$ & 17.7(2)	& 20.6 & 32.7 \\
 			 & 58090$^\dag$ & 4820.0 & 50 & $0.46\pm0.03$ & 18.2(2) & 20.1 & 37.2\\
 			 & & & & $\,0.43\pm0.05$ & $17.9\pm0.4$ &$20.4\pm0.4$ & $35.0\pm3.2$ &  \\
J1848$-$0123 & 57155$^\dag$ & 5124.0 & 30 & $1.24\pm0.25$ & 12.8(2) & $16.7\pm0.2$ & $23.4\pm0.5$ & $1.07^2$\\
J1853+0545 & 57245$^\dag$ & 4920.0 & 30 & $0.97\pm0.19$ & 7.3(2) & $3.2\pm0.1$ & $13.9\pm0.4$ \\
J1900$-$2600  & 57570$^\dag$ & 4820.0 & 10 & $1.03\pm0.07$ & 23.6(3) & 27.8 & 37.3 \\
        	  & 57536$^\dag$ & 4820.0 & 10 & $0.86\pm0.07$ & 25.0(3) & 29.4 & 39.9 \\
		 	  & 57292$^\dag$ & 4820.0 & 10 & $0.87\pm0.07$ & 19.9(3) & 24.1 & 33.5 \\
 			  & & & & $\,0.89\pm0.16$ & $22.8\pm2.6$ & $27.1\pm2.7$ & $36.9\pm3.2$ & $0.75^2$ \\
J1909+1102 & 57245$^\dag$ & 4920.0 & 30 & $0.53\pm0.11$ & 7.1(2) & $12.7\pm0.5$ & $20.7\pm1.4$ \\
J1932+1059 & 57721$^\dag$ & 4820.0 & 30 & $\,\ 7.42\pm0.35^*$ & 2.1(2) & 7.0 & 15.9 \\
		   & 57845$\ $ & 4820.0 & 30 & $\,\ 9.96\pm0.80^*$	& 2.4(2) & 7.1 & 15.6 \\
		   & 57825$\ $ & 4820.0 & 10 & $\ 6.34\pm0.05$    & 2.4(2) & 7.1 & 15.3 \\
		   & & & & $\,7.91\pm2.06$	& $2.3\pm0.2$ & $7.1\pm0.1$ & $15.6\pm0.3$ & $9.08^2$,$6.8^3$\\
J1935+1616 & 57292$\ $ & 4820.0 & 8 & $\,3.23\pm0.07$ & 13.9(3) & 10.5 & 19.4 \\
		   & 57320$\ $ & 4820.0 & 20 & $\,2.57\pm0.05$ & 13.8(3) & 10.3 & 20.1 \\
		   & 57414$\ $ & 4820.0 & 8 & $\,3.36\pm0.07$ & 13.7(3) & 10.1 & 19.5 \\
		   & 57443$\ $ & 4820.0 & 8 &	$\,3.73\pm0.09$	& 13.5(3) & 10.6 & 20.6 \\
		   & 57466$\ $ & 4820.0 & 8 & $\,1.90\pm0.07$ & 13.9(3) & 10.5 & 19.7 \\
		   & 57570$\ $ & 4820.0 & 8 & $\,2.62\pm0.07$ & 13.6(3) & 10.4 & 20.0 \\
		   & 57825$^\dag$ & 4820.0 &	8 &	$\,3.93\pm0.07$	& 13.6(3) & 10.8 & 22.1 \\
		   & & & & $\,3.05\pm0.74$ & $13.7\pm0.2$ & $10.5\pm0.2$ & $20.2\pm0.9$ & $3.99^2$ \\
J1937+2544 & 57245$^\dag$ & 4920.0 & 30 & $3.58\pm0.72$ & 16.5(2) & $23.2\pm0.1$ & $30.8\pm0.4$ \\
J1939+2134 & 57721$^\dag$ & 4820.0 & 10 & $\,\ 0.48\pm0.02^*$ & -- & $11.6\pm	0.8$ & $20.8\pm1.7$ & $1.0\pm0.2^4$\\
J1939+2134i& 57721$\ $ & 4820.0 & 10 & $\,\ 0.07\pm0.02^*$ & -- & $7.3\pm4.0$ & $16.5\pm7.0$ \\
J1946+1805 & 57245$^\dag$ & 4920.0 & 15 & $2.58\pm0.52$ & 29.7(3) & $37.8\pm1.1$ & $45.9\pm2.2$ & $1.71^2$ \\
J1948+3540 & 57155$^\dag$ & 5124.0 & 30 & $0.40\pm0.08$ & 12.4(3) & $14.6\pm0.1$ & $16.4\pm0.4$ & $0.59^2$,$0.5^3$\\
J1954+2923 & 57245$^\dag$ & 4920.0 & 15 & $1.46\pm0.29$ & 16.8(2) & $18.1\pm0.1$ & $22.8\pm0.5$ & $0.32^2$ \\
J1955+5059 & 57245$^\dag$ & 4920.0 & 25 & $\,0.75\pm0.15$ & -- & $3.4\pm0.2$ & $7.6\pm0.7$ & $0.51^2$ \\
		   & 57721$\ $ & 4820.0 & 25 & $\,\ 0.74\pm0.25^*$ & -- & $3.5\pm0.5$ & $6.2\pm1.3$ &  \\
		   & & & & $\,0.745\pm0.292$ & -- & $3.45\pm0.55$ & $6.9\pm1.6$ &  \\
J2018+2839 & 57740$\ $ & 4820.0 & 8  & $\,1.26\pm0.07$ & -- & 7.2 & 15.7 \\
  		   & 57466$\ $ & 4820.0 & 8  & $\,1.37\pm0.07$ & -- & 6.8 & 16.2 \\
  		   & 57740$\ $ & 4820.0 & 8  & $\,1.48\pm0.07$ & -- & 7.7 & 15.0 \\
  		   & 57844$^\dag$ & 4820.0 & 8  & $\,\ 1.78\pm0.15^*$ & -- & 6.4 & 15.7 \\
  		   & & & & $\,1.47\pm0.29$ & -- & $7.0\pm0.6$ & $15.6\pm0.5$ & $0.83\pm0.09^1$,$1.44^2$,$0.85^3$\\
J2022+2854 & 57292$\ $ & 4820.0 & 8  & $\,1.72\pm0.07$ & 9.1(2) & 11.6 & 15.0 \\
           & 57351$\ $ & 4820.0 & 8  & $\,1.23\pm0.07$ & 9.1(2) & 11.8 & 15.1 \\
           & 57379$\ $ & 4820.0 & 8  & $\,3.91\pm0.07$ & 9.5(2) & 11.5 & 14.4 \\
           & 57414$\ $ & 4820.0 & 8  & $\,0.85\pm0.07$ & 9.4(2) & 11.6 & 15.1 \\
           & 57466$\ $ & 4820.0 & 8  & $\,1.82\pm0.07$ & 9.4(2) & 11.6 & 14.4 \\
  	       & 57536$\ $ & 4820.0 & 8  & $\,2.78\pm0.07$ & 9.4(2) & 11.5 & 14.8 \\
           & 57570$\ $ & 4820.0 & 8  & $\,1.82\pm0.07$ & 9.5(2) & 11.7 & 15.3 \\
           & 57572$\ $ & 4820.0 & 8  & $\,3.86\pm0.07$ & 9.4(2) & 12.0 & 15.0 \\
           & 57740$\ $ & 4820.0 & 8  & $\,1.99\pm0.07$ & 9.2(2) & 11.5 & 14.6 \\
           & 57802$\ $ & 4820.0 & 8  & $\,0.86\pm0.07$ & 9.2(2) & 11.6 & 14.5 \\
           & 57825$\ $ & 4820.0 & 8  & $\,0.62\pm0.07$ & 9.0(2) & 11.6 & 13.9 \\
           & 57844$^\dag$ & 4820.0 & 8  & $\,\ 1.89\pm0.16^*$ & 9.2(2) & 11.9 & 14.8 \\
           & & & & $\,1.94\pm1.16$ & $9.3\pm0.2$ & $11.7\pm0.2$ & $14.7\pm0.4$ & $1.72^2$,$1.64^3$\\
J2022+5154 & 57292$\ $ & 4820.0 & 8 & $\,3.98\pm0.07$ & -- & 7.8 & 15.2 \\
		   & 57330$\ $ & 4820.0 & 25 & $\,6.15\pm0.05$ & -- & 7.6 & 15.9 \\
		   & 57414$\ $ & 4820.0 & 8 & $\,5.61\pm0.07$ & -- & 7.9 & 14.9 \\
		   & 57536$\ $ & 4820.0 & 8 & $\,4.16\pm0.07$ & -- & 7.9 & 15.2 \\
		   & 57572$\ $ & 4820.0 & 8 & $\,5.29\pm0.07$ & -- & 8.4 & 15.3 \\
		   & 57721$\ $ & 4820.0 & 30 & $\,6.61\pm0.04$ & -- & 7.9 & 15.0 \\
		   & 57740$\ $ & 4820.0 & 10 & $\,4.87\pm0.06$ & -- & 8.3 & 15.7 \\
		   & 57844$^\dag$ & 4820.0 & 10 & $\,\ 6.84\pm0.55^*$ & -- & 7.9 & 15.1 \\
		   & & & & $\,5.44\pm1.21$ & -- & $8.0\pm0.3$ & $15.3\pm0.3$ & $3.01^2$,$9.73^3$\\
J2145$-$0750 & 57156$^\dag$	& 5124.0 & 30 & $1.77\pm0.35$ & -- & $9.2\pm0.4$ & $66.6\pm3.3$ & $0.44\pm0.03^1$,$0.4\pm0.1^4$\\
J2257+5909 & 57721$^\dag$ & 4820.0 & 30 & $\,\ 0.45\pm0.03^*$ & 6.1(2) & $9.9\pm0.4$ & $16.2\pm2.4$ & $0.38^2$\\
J2321+6024 & 57845$^\dag$ & 4820.0 & 30 & $\,\ 2.46\pm0.22^*$ & 11.4(2) & 6.1 & 18.9 \\
		   & 57536$\ $ & 4820.0 & 10 & $2.66\pm0.06$ & 11.3(2) & 6.2 & 18.8  \\
		   & 57572$\ $ & 4820.0 & 10 & $1.46\pm0.06$ &  11.7(2) & 7.4 & 20.0 \\
		   & & & & $\,2.19\pm0.64$ & $11.5\pm0.2$ & $6.6\pm0.7$ & $19.2\pm0.7$ & $1.75^2$,$0.24^3$\\
J2330$-$2005 & 57245$^\dag$ & 4920.0 & 30 & $0.15\pm0.04$ & 3.2(2) & $1.7\pm0.1$ & $8.0\pm1.0$ \\
J2354+6155 & 57245$^\dag$ & 4920.0 & 30 & $1.15\pm0.23$ & 1.2(2) & $5.7\pm0.1$ & $9.1\pm0.2$ & $0.26^2$,$0.08^3$\\
\enddata
\tablenotetext{}{Notes: The observational parameters for 71 pulsars were derived from our measurements. Col.(2)--(4) give the MJD, central frequency and integration time for each observation; the flux densities, separation of the outer-most profile components, pulse widths at 50\% and 10\% of the profile maximum are given in col.(5)--(8), and the corresponding mean values are measured for pulsars that observed several times. $\rm N$ in col.(6) is the number of identifiable components. Col.(9) gives previously published 5~GHz flux densities; references are: 1 \citet{kkwj98}; 2 \citet{sgg+95}; 3 \citet{hx97}; 4 \citet{kll+99}.\\
$^*$ Flux densities calibrated using the calibrators (3C48, 3C123, 3C196 or 3C295).\\
$^\dag$ MJDs corresponding to the profiles in Fig.\ref{fg:psrprf}. For three pulsars (PSRs J0538+2817, J1844+00 and J1900$-$2600) two or three observations are summed to form the profile in Fig.\ref{fg:psrprf}.\\
}
\end{deluxetable}

\clearpage

\begin{table}[htp]
 \caption{Characteristic parameters: $\rm P$, $\rm \dot{P}$, $\tau$, $\rm L_{1400}$, 
 $\rm B_{surf}$ and $\rm \dot{E}$ and power-law ($\nu^{\alpha}$) indices for flux 
 densities of 27 pulsars.}\label{tb:spec}
\begin{tabular}{ccccccccccr}
\hline
\hline
PSR J2000 & P  &  $\rm \dot{P}$ & $\tau$ & $\rm L_{1400}$ & $\rm B_{surf}$ &$\rm \dot{E}$ & Freq & $\alpha$ & Ref. \\ 
 Name & (ms) &  &(Myr) &($\rm mJy\ kpc^2)$ & (G) & (ergs/s) & (MHz) &  &  \\ 
\hline 
J0248$+$6021  &217.1  &5.5E-14 &0.0624  &54.80  &3.5e+12  &2.1e+35 & 1400,5000   & $-1.17$           &   24  \\ 
J0837$-$4135  &751.6  &3.5E-15 &3.36    &36.00  &1.65e+12 &3.3e+32 & 200,400     & $1.05$            & 4,12,20,21,30,31,32,33    \\ 
              &       &                 &       &         &        &             & 400-8600    & $-1.64\pm0.12$    &           \\
J1012$+$5307  &5.3    &1.7E-20 &4860    &1.57   &3.04e+08 &4.7e+33 & 400-5000    & $-2.04\pm0.15$    & 7,13,29                   \\ 
J1518$+$4904  &40.9   &2.7E-20 &23900   &3.69   &1.07e+09 &1.6e+31 & 400-8350    & $-1.56\pm0.21$    & 11,13,25                  \\ 
J1559$-$4438  &257.1  &1.0E-15 &4.00    &211.60 &5.18e+11 &2.4e+33 & 400-728     & $0.32\pm0.75$     & 1,2,12,15,32              \\ 
              &       &                 &       &         &        &             & 728-5000          & $-2.33\pm0.24$    &        \\ 
J1600$-$3053  &3.6    &9.5E-21 &6000    &8.10   &1.87e+08 &8.1e+33 & 400-5000    & $-1.24\pm0.27$    & 26,27                     \\ 
J1643$-$1224  &4.6    &1.8E-20 &3970    &2.63   &2.96e+08 &7.4e+33 & 200-5000    & $-1.78\pm0.08$    & 6,15,16,17,26,27,30       \\ 
J1651$-$4246  &844.1  &4.8E-15 &2.78    &432.64 &2.04e+12 &3.2e+32 & 200-5000    & $-2.03\pm0.10$    & 3,4,12,21,28,30,32,33     \\ 
J1703$-$3241  &1211.8 &6.6E-16 &29.1    &76.37  &9.05e+11 &1.5e+31 & 400-5000    & $-1.49\pm0.18$    & 4,5,12,21,32,33           \\ 
J1705$-$3423  &255.4  &1.1E-15 &3.76    &60.46  &5.31e+11 &2.5e+33 & 400-5000    & $-1.40\pm0.08$    & 9,21,32,33                \\ 
J1707$-$4053  &581.0  &1.9E-15 &4.78    &115.20 &1.07e+12 &3.9e+32 & 400,640     & $0.08$            & 3,4,21,22,28,32,33        \\ 
              &       &        &        &       &         &        & 728-5000    & $-2.01\pm0.11$    &                           \\   
J1709$-$4429  &102.5  &9.3E-14 &0.0175  &49.35  &3.12e+12 &3.4e+36 & 400-8600    & $-0.62\pm0.10$    & 3,4,14,21,22,31,32,33     \\ 
J1713$+$0747  &4.6    &8.5E-21 &8490    &15.55  &2e+08    &3.5e+33 & 1400-8350   & $-2.40\pm0.40$    & 13,25,27                  \\ 
J1721$-$3532  &280.4  &2.5E-14 &0.176   &232.76 &2.69e+12 &4.5e+34 & 950-17000   & $-0.78\pm0.13$    & 3,4,12,21,23,32,33        \\ 
J1730$-$3350  &139.5  &8.5E-14 &0.0226  &38.98  &3.48e+12 &1.2e+36 & 400-3100    & $-1.13\pm0.16$    & 3,4,5,18,21,22,32,33      \\ 
              &       &        &        &       &         &        &  3100-8400  & $-0.12\pm0.03$    &                           \\   
J1739$-$2903  &322.9  &7.9E-15 &0.649   &16.94  &1.61e+12 &9.2e+33 & 400-5000    & $-0.88\pm0.13$    & 4,15,21,32,33             \\ 
J1744$-$1134  &4.1    &8.9E-21 &7230    &0.48   &1.93e+08 &5.2e+33 & 400-5000    & $-1.71\pm0.22$    & 10,13,16,17,26,27         \\ 
J1809$-$1917  &82.7   &2.6E-14 &0.0513  &26.73  &1.47e+12 &1.8e+36 & 1170-6500   & $-0.11\pm0.26$    & 19,32,33                  \\ 
J1820$-$0427  &598.1  &6.3E-15 &1.5     &0.55   &1.97e+12 &1.2e+33 & 200-5000    & $-2.15\pm0.06$    & 4,5,26,30,32              \\ 
J1833$-$0827  &85.3   &9.2E-15 &0.147   &72.90  &8.95e+11 &5.8e+35 & 606-5000    & $-1.06\pm0.17$    & 5,18,21,32,33             \\ 
J1835$-$1106  &165.9  &2.1E-14 &0.128   &21.97  &1.87e+12 &1.8e+35 & 400-5000    & $-1.77\pm0.11$    & 9,21,28,32,33             \\ 
J1844$+$00$\ \ \ $&460.5   &$--$    &$--$   &402.36   &$--$    &$--$ & 400,1400  & $0.50        $    & 8,21                      \\ 
              &       &    &        &   &       &         & 1400,5000   & $-2.35       $    &                           \\   
J1853$+$0545  &126.4  &6.1E-16 &3.27    &68.43  &2.82e+11 &1.2e+34 & 1278,5000   & $-0.83\pm0.19$    & 20,32,33                  \\ 
J1900$-$2600  &612.2  &2.0E-16 &47.4    &6.37   &3.58e+11 &3.5e+31 & 200-800     & $-1.21\pm0.26$    & 4,5,12,15,18,30,32,33     \\ 
              &       &        &        &       &         &        & 800-10700   & $-2.54\pm0.12$    &                           \\
J1909$+$1102  &283.6  &2.6E-15 &1.7     &43.78  &8.76e+11 &4.6e+33 & 100-5000    & $-1.89\pm0.14$    & 4,5,18,21,32,33           \\ 
J2257$+$5909  &368.2  &5.8E-15 &1.01    &82.80  &1.47e+12 &4.5e+33 & 100-600     & $0.30\pm0.13$     & 5,18                      \\ 
              &       &        &        &       &         &        & 600-5000    & $-2.24\pm0.25$    &                           \\
J2330$-$2005  &1643.6 &4.6E-15 &5.62    &2.22   &2.79e+12 &4.1e+31 & 90-5000     & $-2.14\pm0.12$    & 4,5,12,18,32              \\
\hline
\end{tabular}\\
\justifying 
Notes: Col.(2)--(7) give the period, period derivative, characteristic age, 1400-MHz psuedo-luminosity, surface magnetic field and spin-down luminosity of each pulsar. The frequency range of spectrum and corresponding spectral index are listed in col.(8) and (9). References for published data are given in col.(10).\\
References: 1 \citet{mlt+78}; 2 \citet{fgl+92}; 3 \citet{jlm+92}; 4 \citet{tml93} ; 5 \citet{lylg95}; 6 \citet{lnl+95}; 7 \citet{nll+95}; 8 \citet{cnst96}; 9 \citet{mld+96}; 10 \citet{bjb+97} ; 11 \citet{snt97}; 12 \citet{vdhm97} ; 13 \citet{kxl+98} ; 14 \citet{lml+98} ; 15 \citet{mhq98}; 16 \citet{tbms98} ; 17 \citet{kll+99} ; 18 \citet{mkkw00a}; 19 \citet{mhl+02}; 20 \citet{kbm+03}; 21 \citet{hfs+04}; 22 \citet{kjm05} ; 23 \citet{kjlb11}; 24 \citet{tpc+11}; 25 \citet{kkmj12}; 26 \citet{mhb+13}; 27 \citet{dhm+15}; 28 \citet{hwxh16}; 29 \citet{lmj+16}; 30 \citet{mdm+17}; 31 \citet{zwy+17}; 32 \citet{jk18}; 33 \citet{jvk+18}\\
\end{table}

\clearpage
\begin{table}[h]
\footnotesize{
\caption{The Spearman rank correlation $\rm R_s$ of spectral indices $\alpha$ versus $\rm log_{10}(x)$, where $x$ is one of the pulsar parameters $\rm P$, $\rm \dot{P}$, $\tau$, $\rm L_{1400}$, $\rm B_{surf}$ and $\rm {E}$ for different data sets. The sample size for each data set is listed in parentheses after the name.  $\rm R_s$ is the Spearman rank correlation coefficient, and p is the corresponding probability that the observed relationship has occurred by statistical chance.}\label{tb:spec-corr}
\begin{tabular}{lccccccccccccccc}
\hline
\hline
Sample/Parameter   &  & Spearman $\rm R_s(p)$ &   \\
\cline{2-4}
This work  & $\rm AP\ (27)$ & $\rm NP\ (21)$ & $\rm MSP\ (6)$ \\
\hline
$\rm P$       &$-0.29(0.18)$ &$-0.718(0.001)$&$-0.26(0.62)$ \\  
$\rm \dot{P}$ &$0.32(0.14)$  &$0.39(0.12)$ &$-0.37(0.47)$ \\     
$\rm \tau$    &$-0.44(0.04)$ &$-0.57(0.02)$&$0.14(0.79)$  \\ 
$\rm L_{1400}$&$0.30(0.16)$  &$0.21(0.41)$&$-0.03(0.96)$ \\     
$\rm B_{surf}$&$0.11(0.61)$  &$0.02(0.95)$&$-0.26(0.62)$ \\     
$\rm \dot{E}$ &$0.623(0.002)$ &$0.723(0.001)$&$0.37(0.47)$\\         
\hline
\citet{jvk+18}  & $\rm AP\ (276) $& $\rm NP\ (267) $& $\rm MSP\ (9)$ \\
\hline
$\rm \dot{P}$ & 0.25(3.8e-05) &0.28(4.9e-06)&$0.05(0.9)$ \\     
$\rm \tau$    &-0.35(1.9e-09)&-0.39(5.9e-11)&$-0.5(0.17)$ \\ 
$\rm \dot{E}$ &0.44(4e-14) &0.43(1.4e-13)&$0.33(0.38)$\\         
\hline
\citet{hwxh16}  &  &$\rm NP\ (572)$&  \\
\hline
$\rm P$       &--&$-0.21$&-- \\  
$\rm \dot{P}$ &--&$0.13$ &-- \\     
$\rm \tau$    &--&$-0.20$&--\\ 
$\rm B_{surf}$&--&$0.03$&--\\     
$\rm \dot{E}$ &--&$0.26$&--\\    
\hline
\citet{lylg95} &$\rm AP\ (343)$&$\rm NP\ (323)$ &$\rm MSP\ (20)$ \\
\hline
$\rm P$       &$-0.15(0.005)$& $-0.22$(6e-5)& $+0.51(0.020)$ \\  
$\rm \tau$    &$\rm -0.22$(6e-5)&$-0.19$(8e-4)&-- \\ 
\hline
\end{tabular}\\
\tablenotetext{}{Note: AP: all pulsars; NP: nomal pulsars; MSP: millisecond pulsars.}}
\end{table}
\clearpage

\begin{table}[h]
\footnotesize{
\caption{Power-law fit parameters for spectral indices
    $\alpha$ versus $\rm log_{10}(x)$.}\label{tb:spec-fit}
\begin{tabular}{lccccccccccccc}
\hline
\hline
Correlation &Slope & Intercept \\
\hline$\rm \alpha\ vs.\ log_{10}(P)$      &$-0.12\pm0.14$ &$-1.78\pm0.16$ \\  
$\rm \alpha\ vs.\ log_{10}\dot{P}$&$0.04\pm0.06$  &$-1.06\pm0.86$ \\     
$\rm \alpha\ vs.\ log_{10}\tau$   &$-0.12\pm0.07$ &$-0.86\pm0.48$ \\ 
$\rm \alpha\ vs.\ log_{10}L_{1400}$&$0.25\pm0.11$ &$-1.95\pm0.16$ \\     
$\rm \alpha\ vs.\ log_{10}B_{surf}$&$0.04\pm0.08$ &$-2.10\pm0.91$ \\     
$\rm \alpha\ vs.\ log_{10}\dot{E}$ &$0.29\pm0.07$ &$-11.33\pm2.23$ \\     
\hline
\end{tabular}
}
\end{table}

\clearpage

\begin{table}[h]
\caption{Power-law ($\nu^{\beta}$) indices for component separations
  of seven pulsars with decreasing separation at higher
  frequencies. $\sigma$ is the rms residual from the fit.}\label{tb:FW_ng}
\begin{tabular}{clcc}
\hline
\hline
PSR  & Freq & $\beta$ & $\sigma$ \\
 Name  & ~~(MHz)  &   & ($\degr$) \\
 \hline
 J0358+5413 & $\,\ $925$-$4820	& $-0.26\pm0.01$  & 0.55 \\
 J0528+2200 & $\,\ $408$-$4820	& $-0.12\pm0.02$  & 0.28 \\
 J1509+5531 & $\,\ $925$-$8600 & $-0.23\pm0.01$  & 0.32 \\
 J1600$-$3053 & $\,\ $730$-$4820 & $-0.21\pm0.03$  & 1.29 \\
 J1826$-$1334 & 1369$-$4820 & $-0.11\pm0.01$  & 0.60 \\
 J1937+2544 & $\,\ $653$-$4820 & $-0.11\pm0.02$  & 0.48 \\
 J2354+6155 & $\,\ $408$-$4820 & $-0.58\pm0.03$  & 0.43\\
\hline
\end{tabular}
\end{table}

\clearpage

\begin{table}[h]
\caption{As for Table~\ref{tb:FW_ng} for eleven pulsars with
  essentially no frequency dependence of component separation. The
  mean component separation, $\langle\Delta\phi\rangle$, is given in
  last column.}\label{tb:FW_fl}
\begin{tabular}{clccr}
\hline
\hline
PSR  & Freq & $\beta$ & $\sigma$ & $\langle\Delta\phi\rangle$\\
 Name  & ~~(MHz)  &   & ($\degr$) \\
 \hline
 J0147+5922 & 1642$-$8600& $\ \ 0.03\pm0.01$  & 0.43 & 5.9\\
 J0454+5543 & 1400$-$8600 & $\ \ 0.04\pm0.01$	& 0.12 & 13.7 \\
 J0820$-$1350 & $\,\ $653$-$4820	& $-0.05\pm0.01$  & 0.09  & 3.2\\
 J1559$-$4438 & 1382$-$4820		& $\ \ 0.001\pm0.001$ & 0.25 & 16.0\\ 
 J1740$-$3015 & 1369$-$8600		& $-0.070\pm0.001$  & 0.03  & 1.7\\
 J1825$-$0935 & $\,\ $408$-$4820	& $-0.023\pm0.001$  & 0.17 & 16.4\\
 J1833$-$0827 & $\,\ $925$-$4820 & $\ \ 0.06\pm0.03$ & 1.00 & 46.1\\
 J1909+1102 & $\,\ $653$-$4820 & $-0.021\pm0.002$  & 0.21 & 8.0\\
 J2022+2854 & $\,\ $410$-$4820 & $-0.072\pm0.001$  & 0.23 & 9.8\\
 J2257+5909 & $\,\ $925$-$4820 & $\ \ 0.02\pm0.03$  & 0.32 & 5.0\\
 J2330$-$2005 & $\,\ $410$-$4820 & $\ \ 0.06\pm0.01$  & 0.09 &3.0\\
\hline
\end{tabular}
\end{table}

\clearpage

\begin{table}[h]
\caption{As for Table~\ref{tb:FW_ng} for two pulsars with increasing separation at higher
  frequencies.}\label{tb:FW_ps}
\begin{tabular}{clcc}
\hline
\hline
PSR  & Freq & $\beta$ & $\sigma$ \\
 Name  & ~~(MHz)  &   & ($\degr$) \\
 \hline
 J1946+1805 & $\,\ $317$-$4820 & $0.22	\pm0.01$ & 0.55 \\
 J1954+2923 & 1330$-$4820 		& $0.10\pm0.02$ & 0.25 \\
\hline
\end{tabular}
\end{table}

\clearpage

\begin{table}[h]
\caption{Core widths and magnetic inclination angles for 23 pulsars.}\label{tb:core-width}

\begin{tabular}{l r r r r r r r r r r r}
\hline
\hline
PSR     & $\rm P$    &$\rm W_{50}^{core,5.0}$ &$\rm W_{50}^{cone,5.0}$ & Ref. & $\rm W_{50}^{core,1.0}$ & $\rm \alpha_B^{5.0}$	& $\rm \alpha_B^{8.6}$	& $\rm \alpha_B^{1.0}$\\
Name          & (s)	   &	(deg)          & (deg)    &   & (deg)			 & (deg)			& (deg)				& (deg)\\
\hline
J0332+5459    & 0.7145 & $3.22\pm0.02$  &$3.8\pm0.1$   & 1 & 5.5    & $57\pm0.6$		& --			    & 32\\
                  &        &                &$6.1\pm0.1$   &   &        &                   &                   &\\
                  &        &                &$8.5\pm0.4$   &   &        &                   &                   &\\
                  &        &                &$2.26\pm0.02$ &   &        &                   &                   &\\
J0454+5543    & 0.3407 & $4.3\pm0.2$    &$2.7\pm0.2$   & 2 & 7.2   	& $67_{-7}^{+10}$	& --			   & 36\\
                  &        &                &$4.1\pm0.3$   &   &        &                   &                   &\\
                  &        &                &$11.9\pm2.2$  &   &        &                   &                   &\\
J0534+2200i   & 0.0331 & $7.3\pm0.5$    & --           & 3 & --	    & 90	            & $57^{+6}_{-5}$    & --\\
J0738$-$4042  & 0.3749 & $11.7\pm1.0$   &$14.6\pm0.3$  & 2 & 14	    & $18.6\pm1.4$	    & --			    & 17\\
J0837$-$4135  & 0.7516 & $2.59\pm0.04$  &$4.1\pm0.5$   & 2 & 3.7    & 90			    & $63\pm3$		   & 50\\
                  &        &                &$2.6\pm0.3$   &   &        &                   &                   &\\
J0944$-$1354  & 0.5703 & $4.3\pm0.2$    &--            & 3 & 4.6    & $44\pm2$		    & --			    & 45\\
J1509+5531    & 0.7397 & $2.3\pm0.4$    &$3.6\pm0.4$   & 2 & 5	    & 90			    & --			    & 35\\
                  &        &                &$2.9\pm0.2$   &   &        &                   &                   &\\
J1644$-$4559  & 0.4551 & $3.61\pm0.01$  &$3.5\pm0.1$   & 2 & 6.7    & $70.2\pm0.5$	    & $40.2\pm0.3$	   & 33\\
                  &        &                &$2.3\pm0.1$   &   &        &                   &                   &\\
                  &        &                &$5.25\pm0.01$ &   &        &                   &                   &\\
                  &        &                &$9.6\pm0.2$   &   &        &                   &                   &\\
J1645$-$0317  & 0.3877 & $3.08\pm0.04$  &$4.6\pm0.3$   & 2 & 4.2    & 90				& $90_{-10}$	    & 70\\
                  &        &                &$5.0\pm0.5$   &   &        &                   &                   &\\
J1705$-$1906i & 0.2990 & $1.8\pm0.1$    &--            & 2 & 4.5    & 90				& $58^{+23}_{-15}$   & 85\\
J1739$-$2903m & 0.3229 & $5.45\pm0.05$  &--            & 2 & --	    & $47.8\pm0.6$	    & --			    & --\\
J1739$-$2903i & 0.3229 & $6.0\pm0.7$    &--            & 2 & --	   	& $43^{+7}_{-5}$	& --			   & --\\
J1740+1311    & 0.8031 & $8.4\pm0.7$    &$3.9\pm0.1$   & 3 & 4.1    & $17.7\pm1.4$	    & --			    & 42\\
                  &        &                &$3.3\pm0.6$   &   &        &                   &                   &\\
J1748$-$1300  & 0.3941 & $4.3\pm0.2$    &$5.1\pm0.1$   & 3 & 4	    & $59\pm5$		    & --			    & 77\\
                  &        &                &$5.0\pm0.1$   &   &        &                   &                   &\\
J1807$-$0847  & 0.1637 & $6.8\pm0.2$    &$7.8\pm0.2$   & 2 &$\sim7$ & $56\pm3$	        & $43^{+10}_{-6}$   & 60\\
                  &        &                &$7.0\pm0.3$   &            &                   &                   &\\
J1825$-$0935  & 0.7690 & $2.34\pm0.05$  &$5.7\pm0.5$   & 2 & 2.8    & 90		        & --			    & 86\\
                  &        &                &$1.6\pm0.1$   &   &        &                   &                   &\\
J1829$-$1751  & 0.3071 & $12.7\pm2.1$   &$2.5\pm0.1$   & 2 & 9	    & $19\pm4$		    & --			    & 29\\
                  &        &                &$2.69\pm0.03$ &   &        &                   &                   &\\
J1833$-$0827  & 0.0853 & $15.1\pm1.5$   &$17.4\pm3.8$  & 2 & --	    & $31\pm4$		    & --			    & --\\
                  &        &                &$12.8\pm1.7$  &   &        &                   &                   &\\
                  &        &                &$15.8\pm0.9$  &   &        &                   &                   &\\
1848$-$0123   & 0.6594 & $8.7\pm0.6$    &$4.2\pm0.1$   & 2 &$\sim5$ & $19\pm1$	        & $19^{+3}_{-2}$    & 39\\
                  &        &                &$5.6\pm0.2$   &            &                   &                   &\\
J1909+1102    & 0.2836 & $9.4\pm1.1$    &$4.7\pm0.5$   & 2 & 6.1   	& $27^{+4}_{-3}$	& --			   & 49\\
J1935+1616    & 0.3587 & $4.6\pm0.1$    &$3.4\pm0.3$   & 2 & 5.25   & $56\pm3$		    & $48\pm4$		   & 51\\
                  &        &                &$4.8\pm0.2$   &            &                   &                   &\\
J1955+5059    & 0.5189 & $3.0\pm0.1$    &--            & 2 & 5.8    & 90				& --			    & 36\\
J2048$-$1616  & 1.9615 & $3.9\pm0.1$    &$2.3\pm0.1$   & 4 & 4.0    & $25.0\pm0.8$	    & $53^{+29}_{-11}$   & 26\\
                  &        &                &$1.80\pm0.03$ &   &        &                   &                   &\\

\hline
\end{tabular}\\
\justifying
Note: Col.(2) gives the period of 24 pulsars; the corresponding core widths at 5.0 and 1.0~GHz $\rm W_{50}^{c,5.0}$ and $\rm W_{50}^{c,1.0}$ are listed in col.(3) and (5); col.(6)--(8) give the inclination angles estimated at 8.6, 5.0 and 1.0~GHz. Col.(4) gives the references for core widths at 5.0~GHz: 1 \citet{sld+17}; 2 This paper; 3 \citet{hkk98}; 4 \citet{kkwj98}\\
\end{table}

\clearpage

\begin{figure}[phtb]
\caption{Integrated pulse profiles at 5.0~GHz obtained with the TMRT
  for 71 pulsars. The $x$ axis is degrees of pulse longitude or phase
  and only the region around the pulse is shown. All these profiles
  are normalized by peak flux density, the pulse period is given for
  each pulsar and the red bar represents $\pm\sigma_b$, the rms
  baseline noise. Most profiles have 1024 phase bins per pulse period;
  six pulsars (PSRs J0538+2817, J1518+4904, J1643-1224, J1651-4246,
  J1703-3241, J1900-2600) have 512 phase bins per period and two (PSRs
  J1012+5307 and J1600-3053) have 256 phase bins per
  period.} \label{fg:psrprf} 
 \includegraphics[scale=0.9]{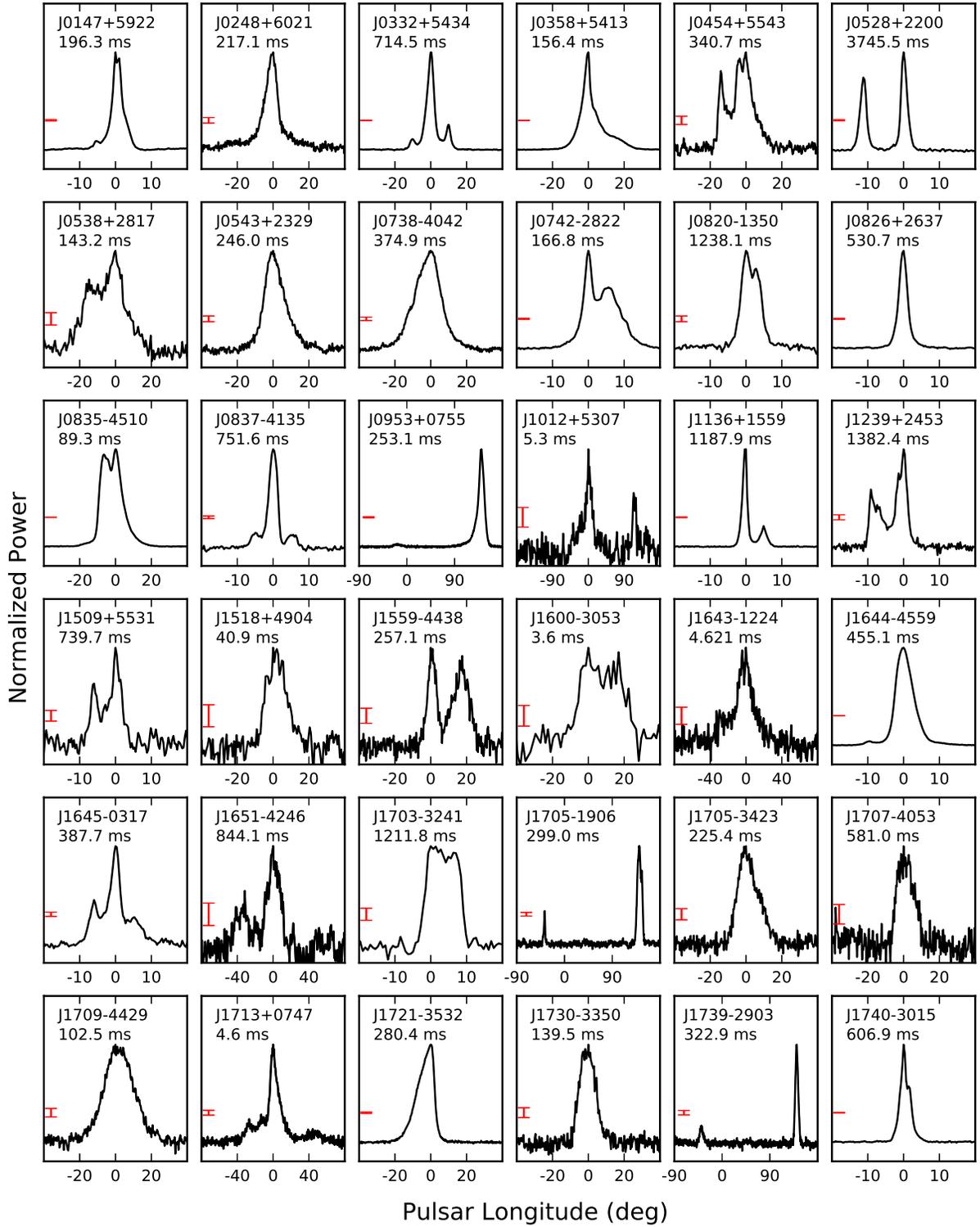}
\end{figure}
\addtocounter{figure}{-1}
\begin{figure}[phtb]
\caption{-continued}
 \includegraphics[scale=0.9]{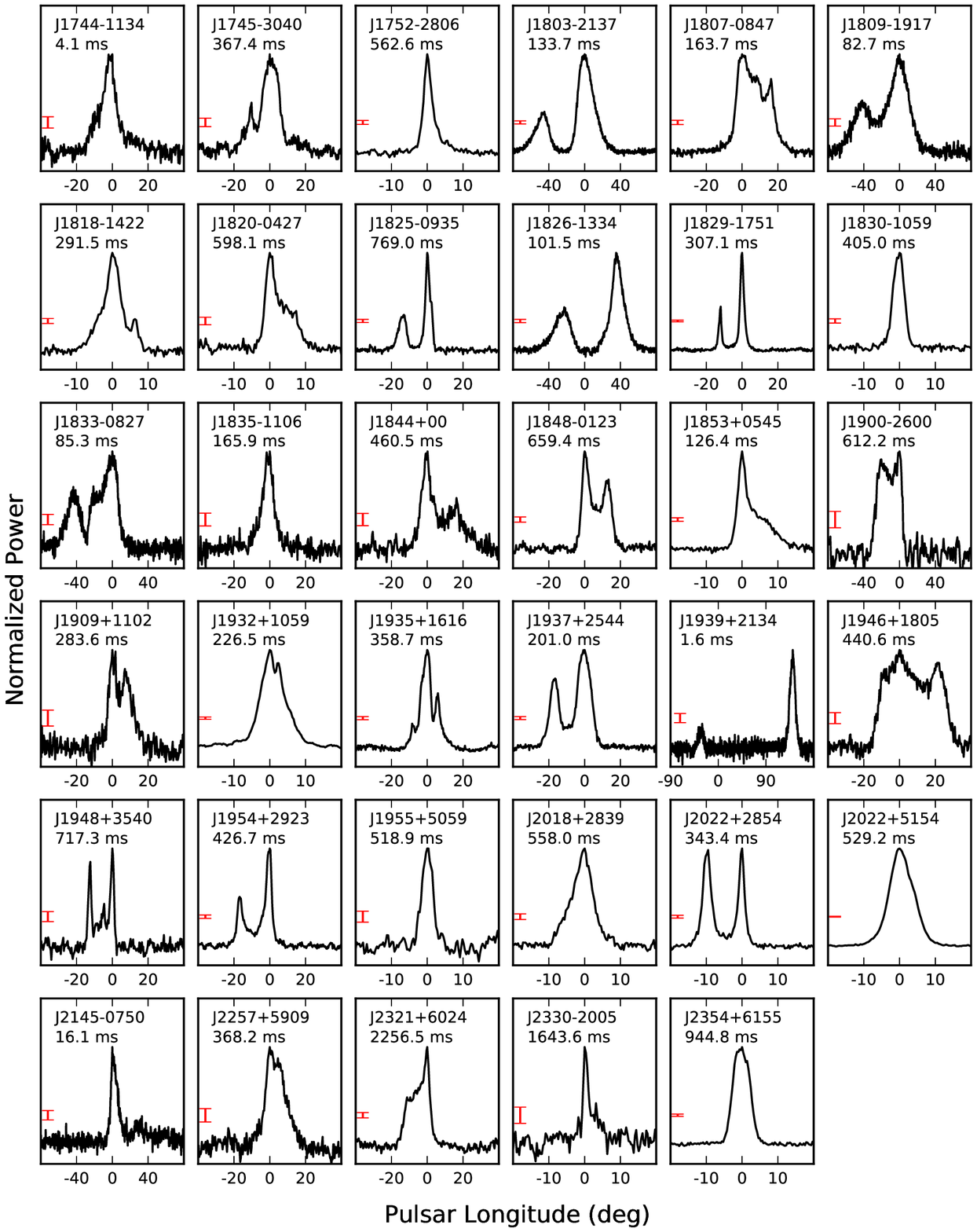}
\end{figure}

\clearpage
 
\begin{figure}[h]
\begin{center}
\begin{tabular}{cc}
\resizebox{0.53\hsize}{!}{\includegraphics[angle=0]{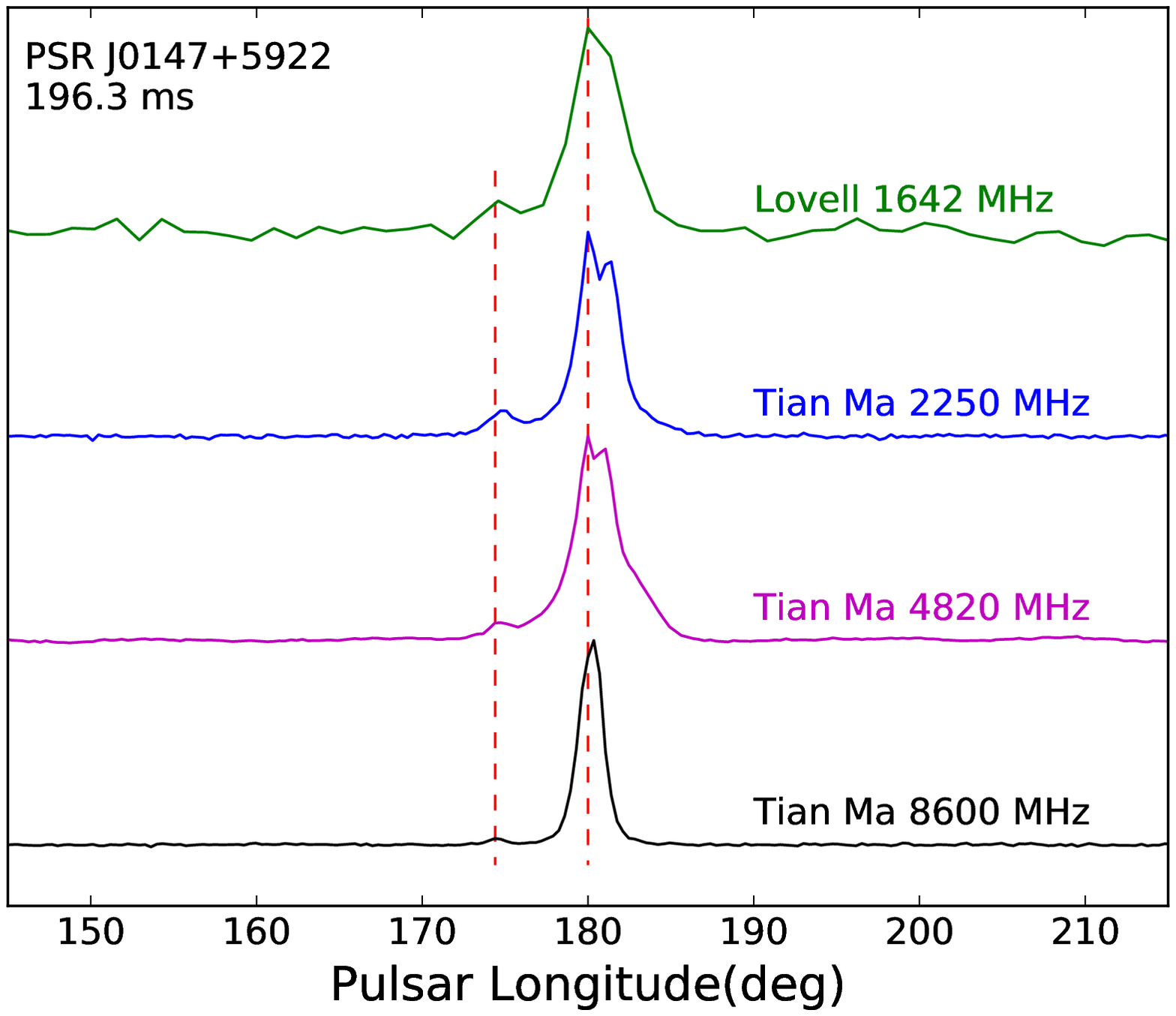}}&
\resizebox{0.53\hsize}{!}{\includegraphics[angle=0]{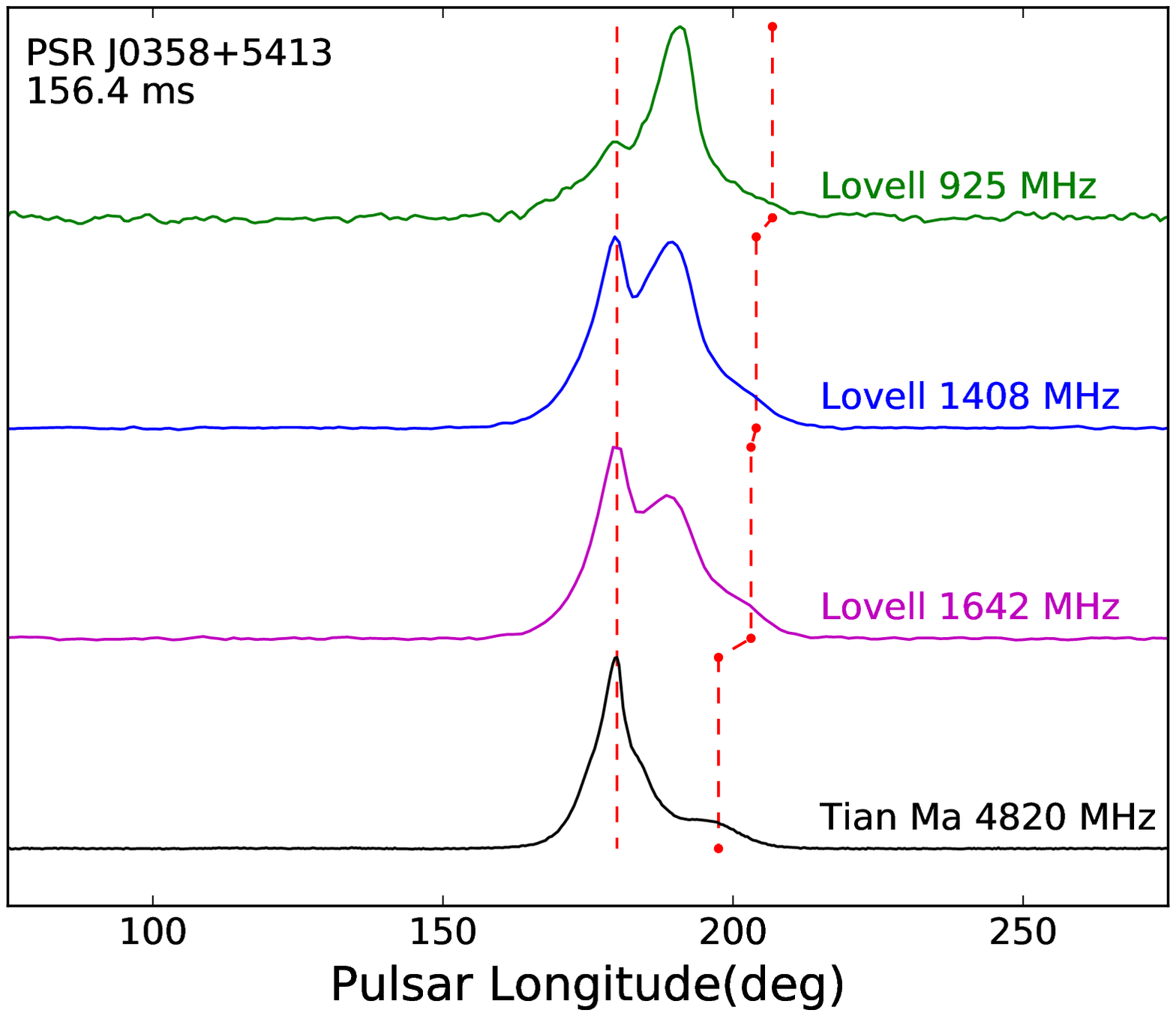}}\\
\resizebox{0.53\hsize}{!}{\includegraphics[angle=0]{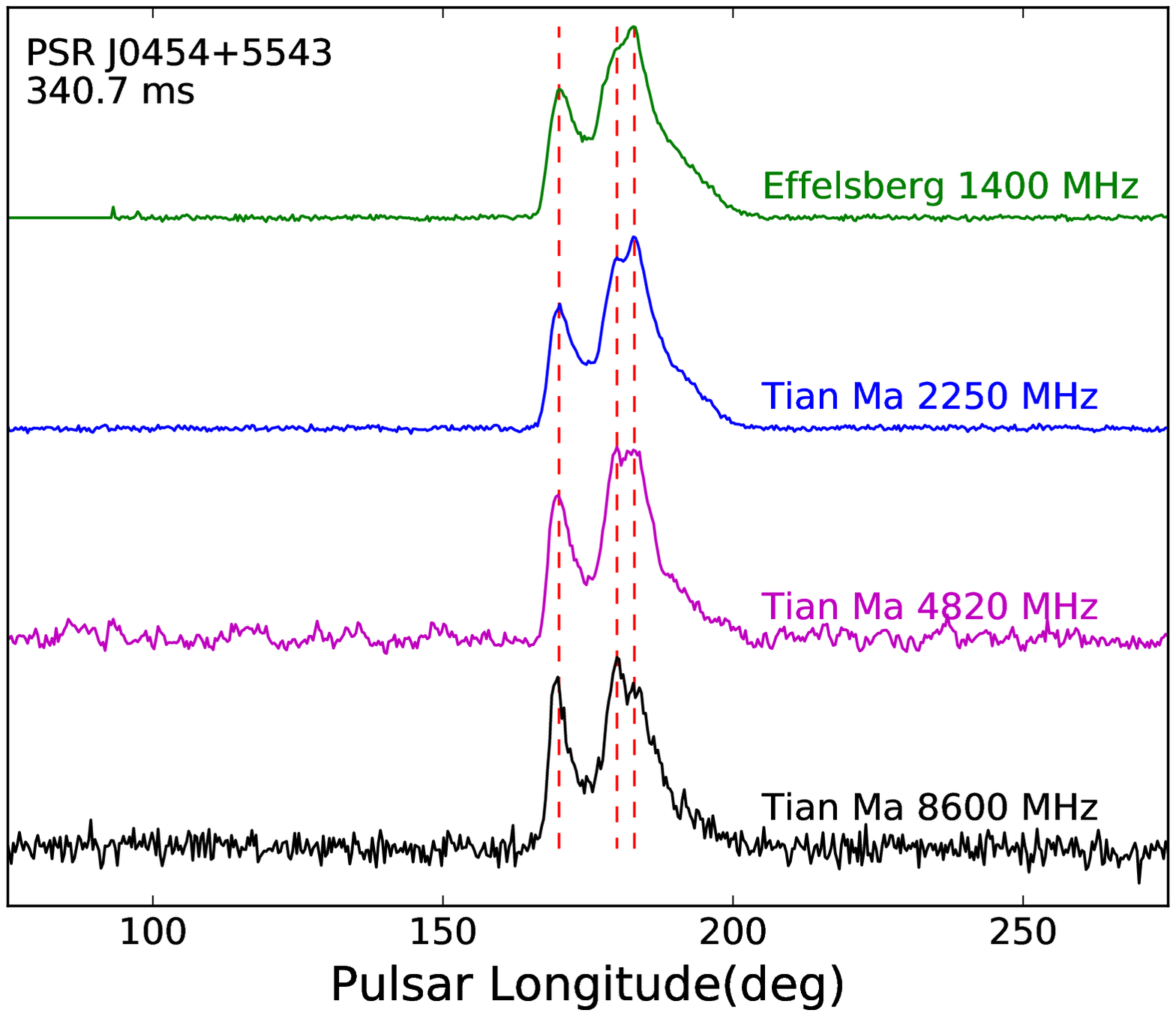}}&
\resizebox{0.53\hsize}{!}{\includegraphics[angle=0]{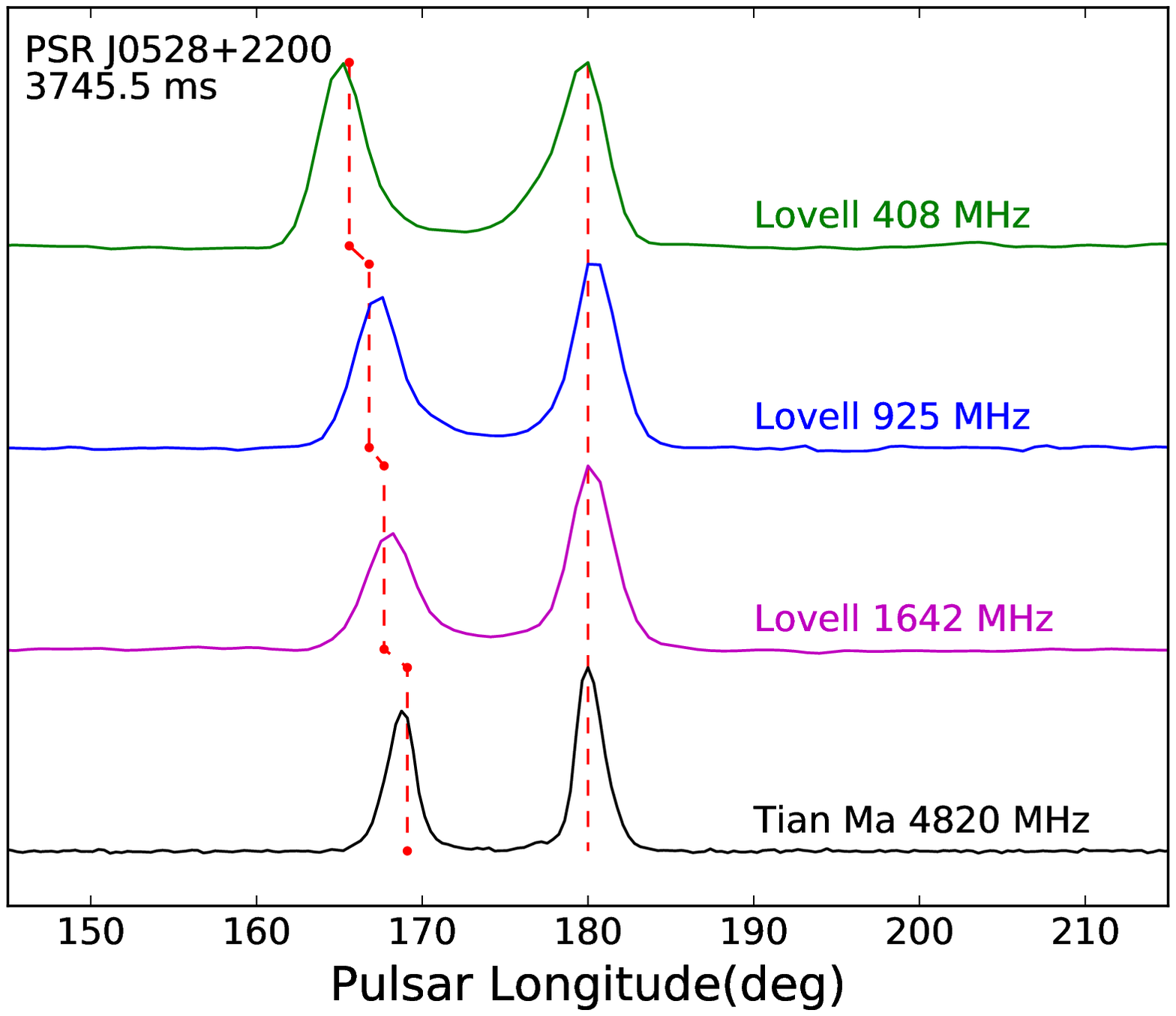}}\\
\resizebox{0.53\hsize}{!}{\includegraphics[angle=0]{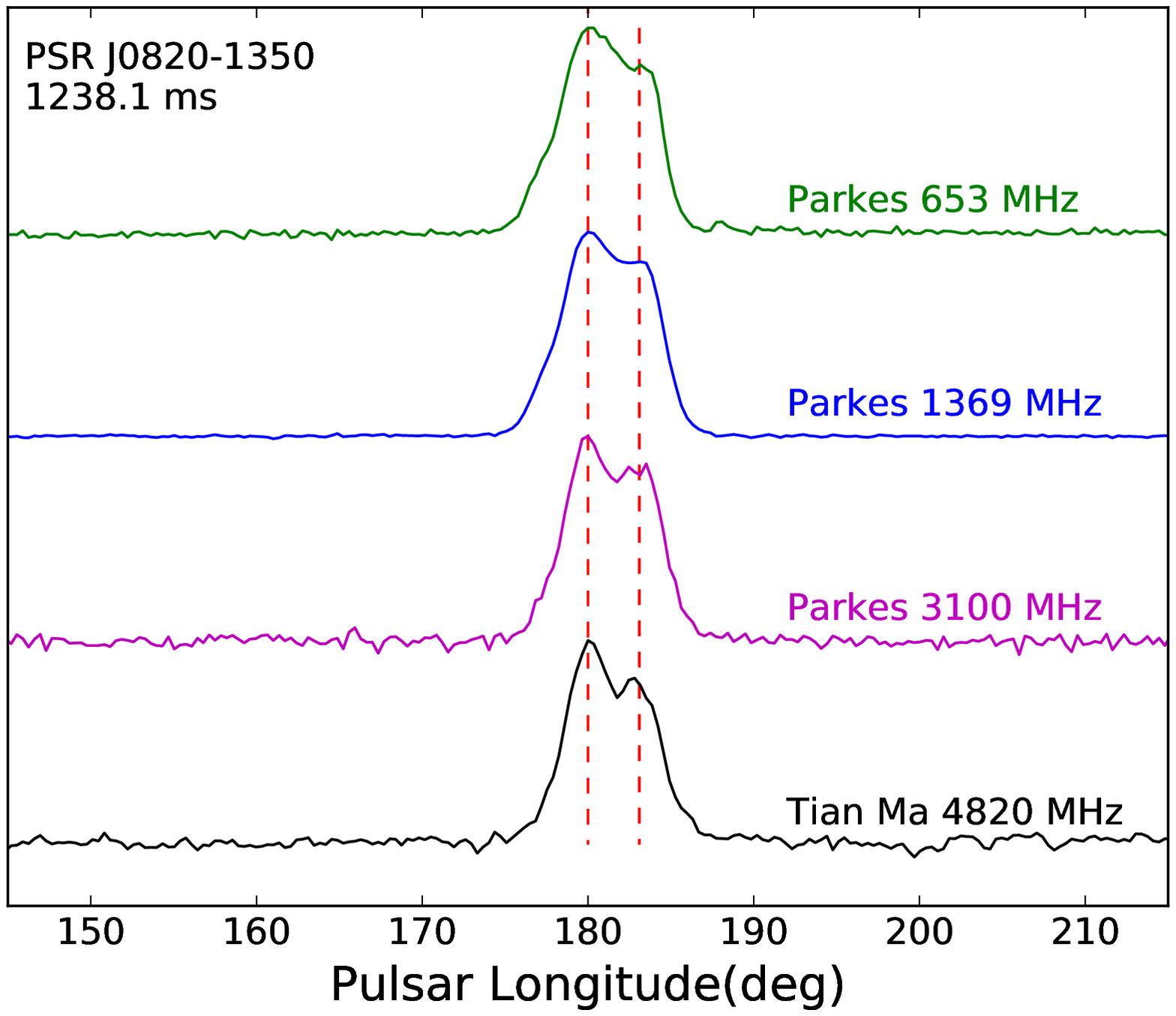}}&
\resizebox{0.53\hsize}{!}{\includegraphics[angle=0]{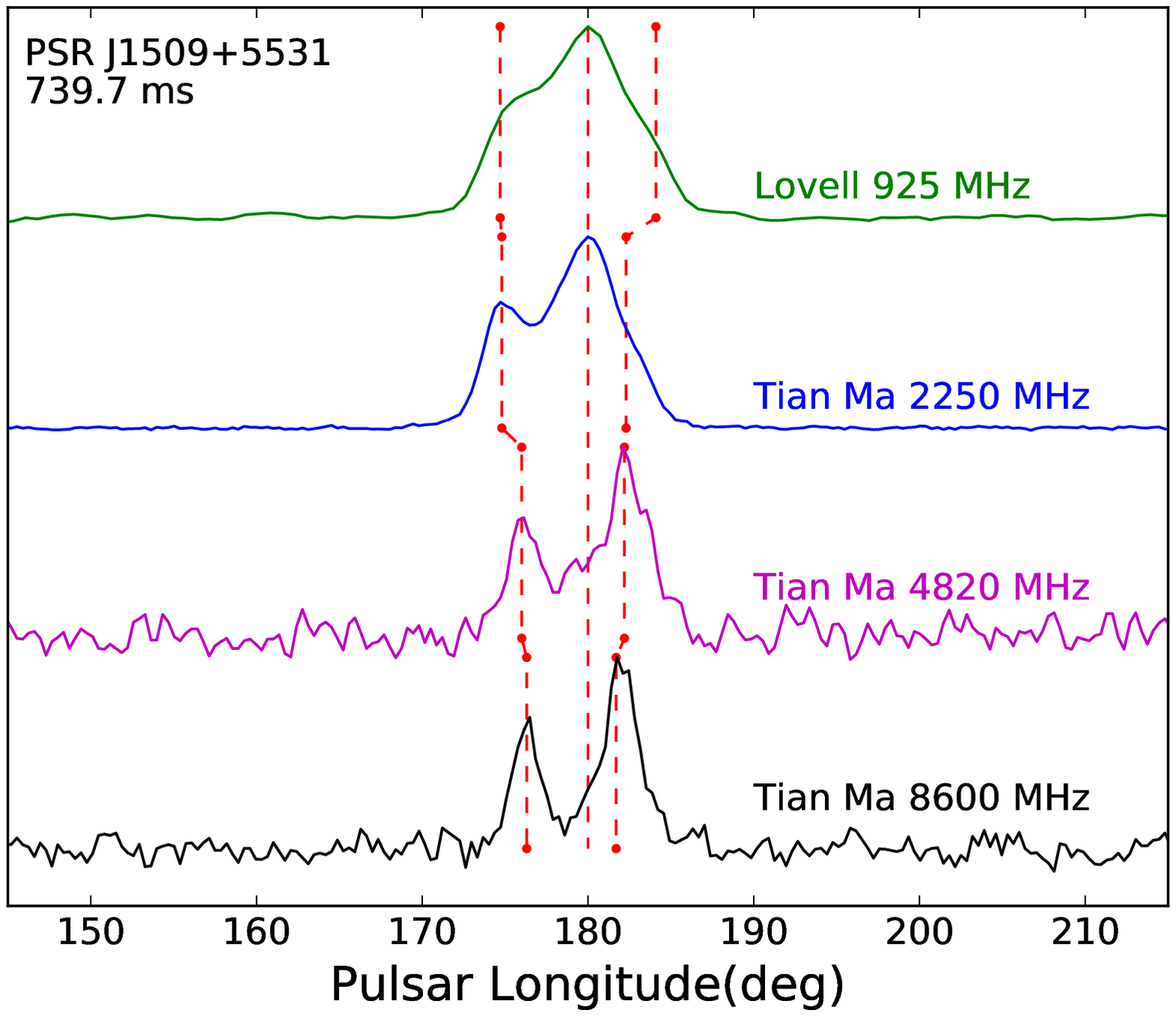}}\\

\end{tabular}
\end{center}
\caption{Integrated pulse profiles for 20 pulsars at 
four frequencies: 5.0~GHz from the TMRT and three other 
frequencies from the ATNF and EPN databases. All profiles are 
normalized and the corresponding components are connected by dashed 
lines.}\label{fg:mul-freq}
\end{figure}
\addtocounter{figure}{-1}
\begin{figure}[h]
\begin{center}
\begin{tabular}{cc}
\resizebox{0.53\hsize}{!}{\includegraphics[angle=0]{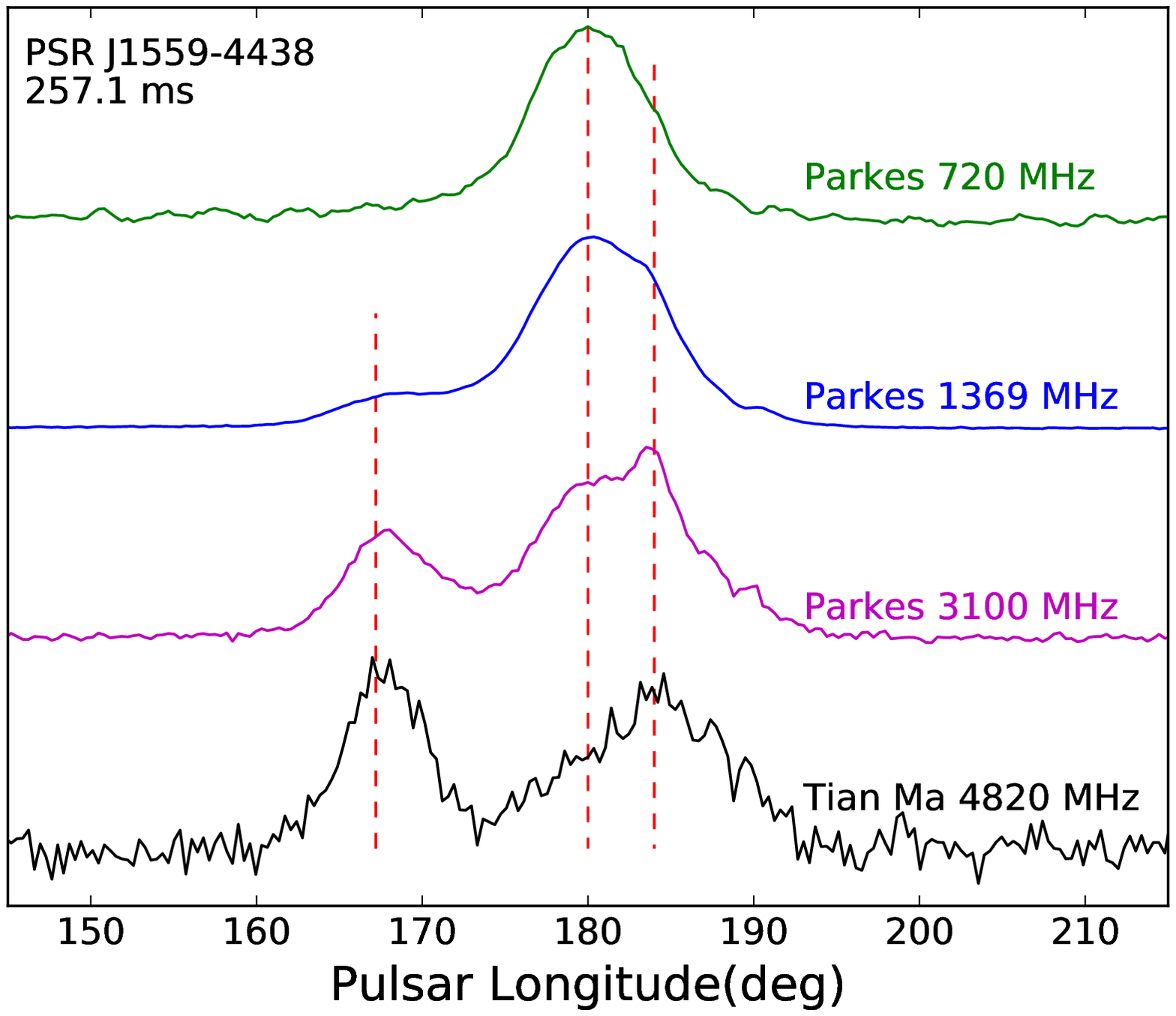}}&
\resizebox{0.53\hsize}{!}{\includegraphics[angle=0]{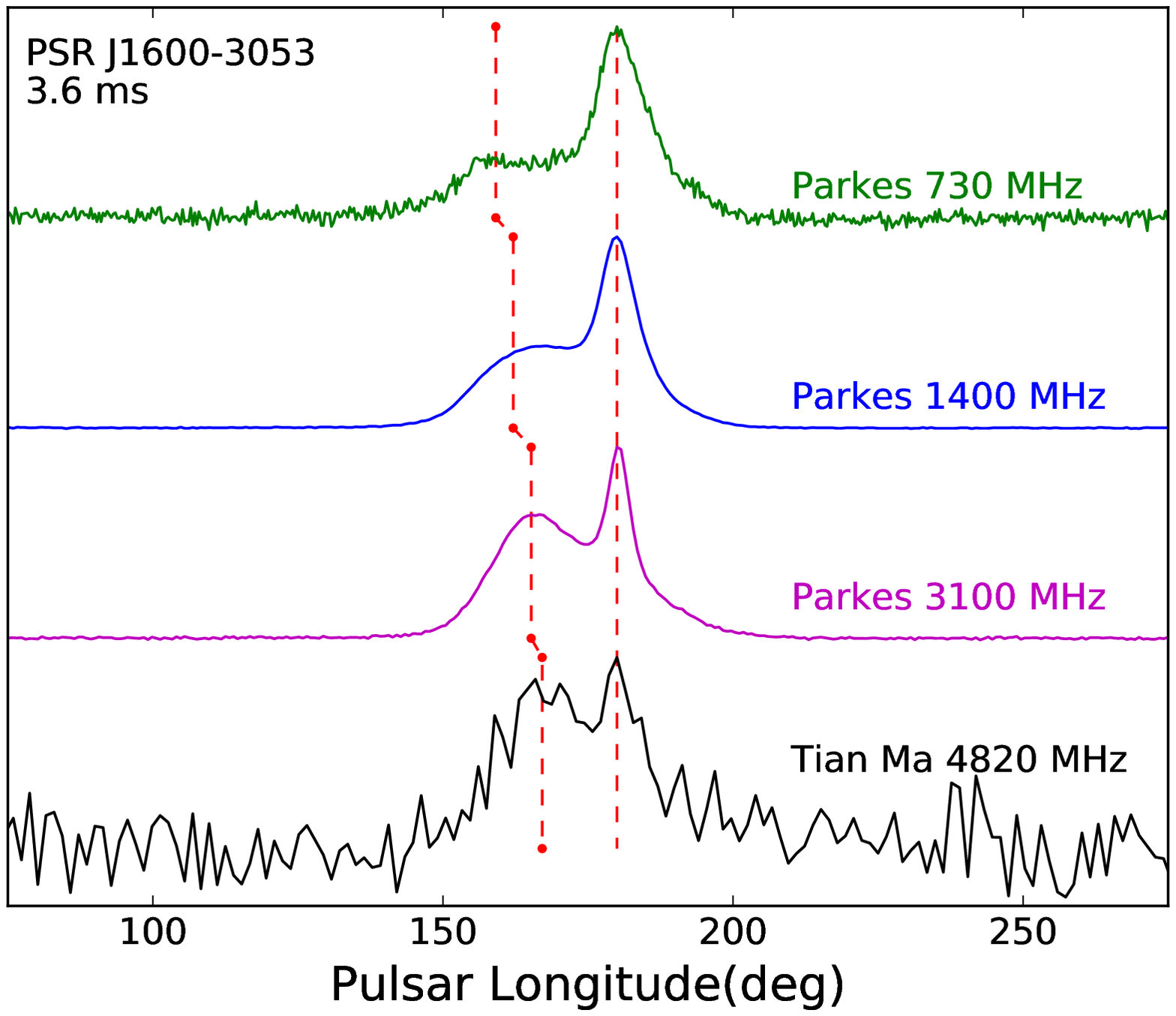}}\\
\resizebox{0.53\hsize}{!}{\includegraphics[angle=0]{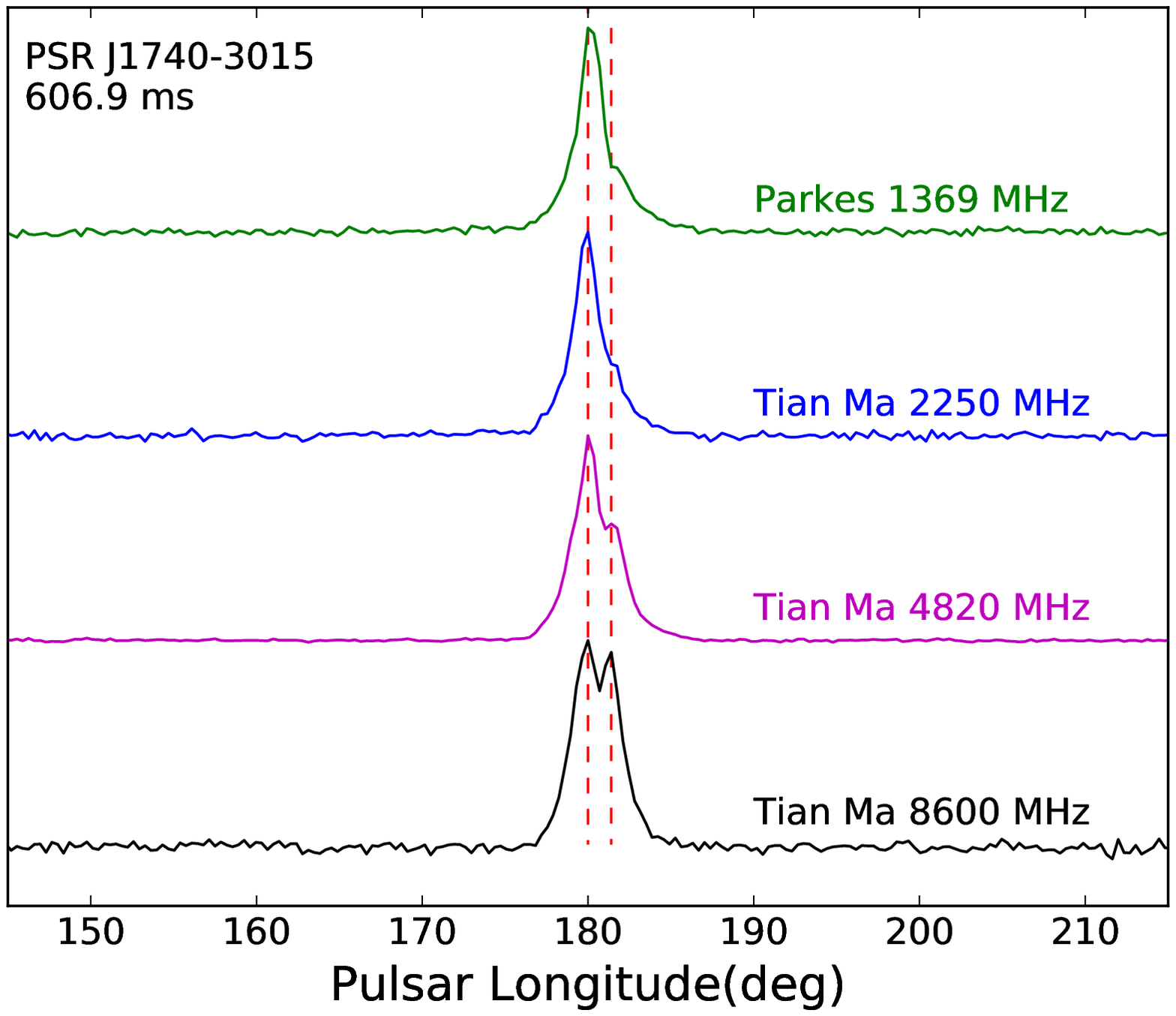}}&
\resizebox{0.53\hsize}{!}{\includegraphics[angle=0]{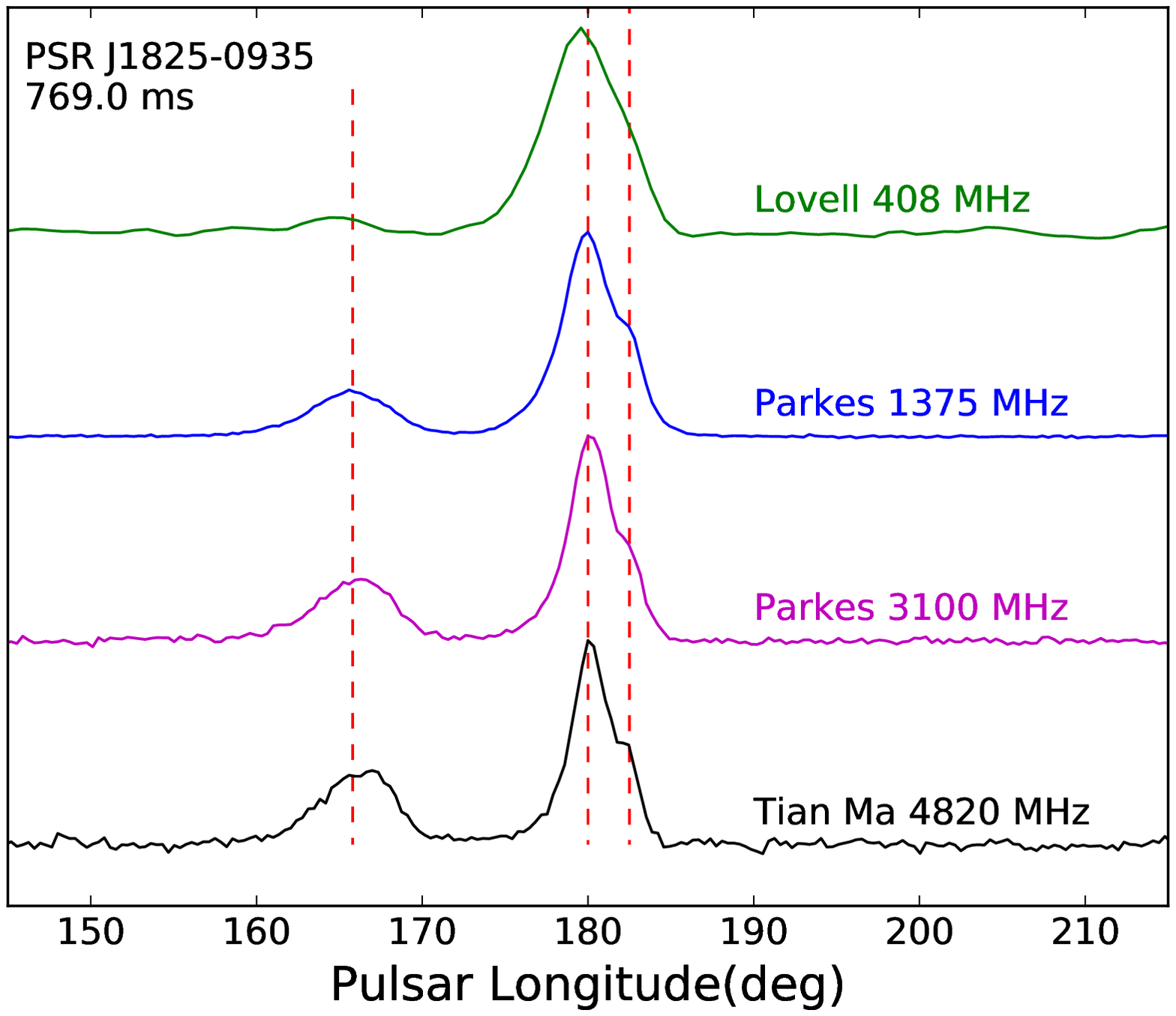}}\\
\resizebox{0.53\hsize}{!}{\includegraphics[angle=0]{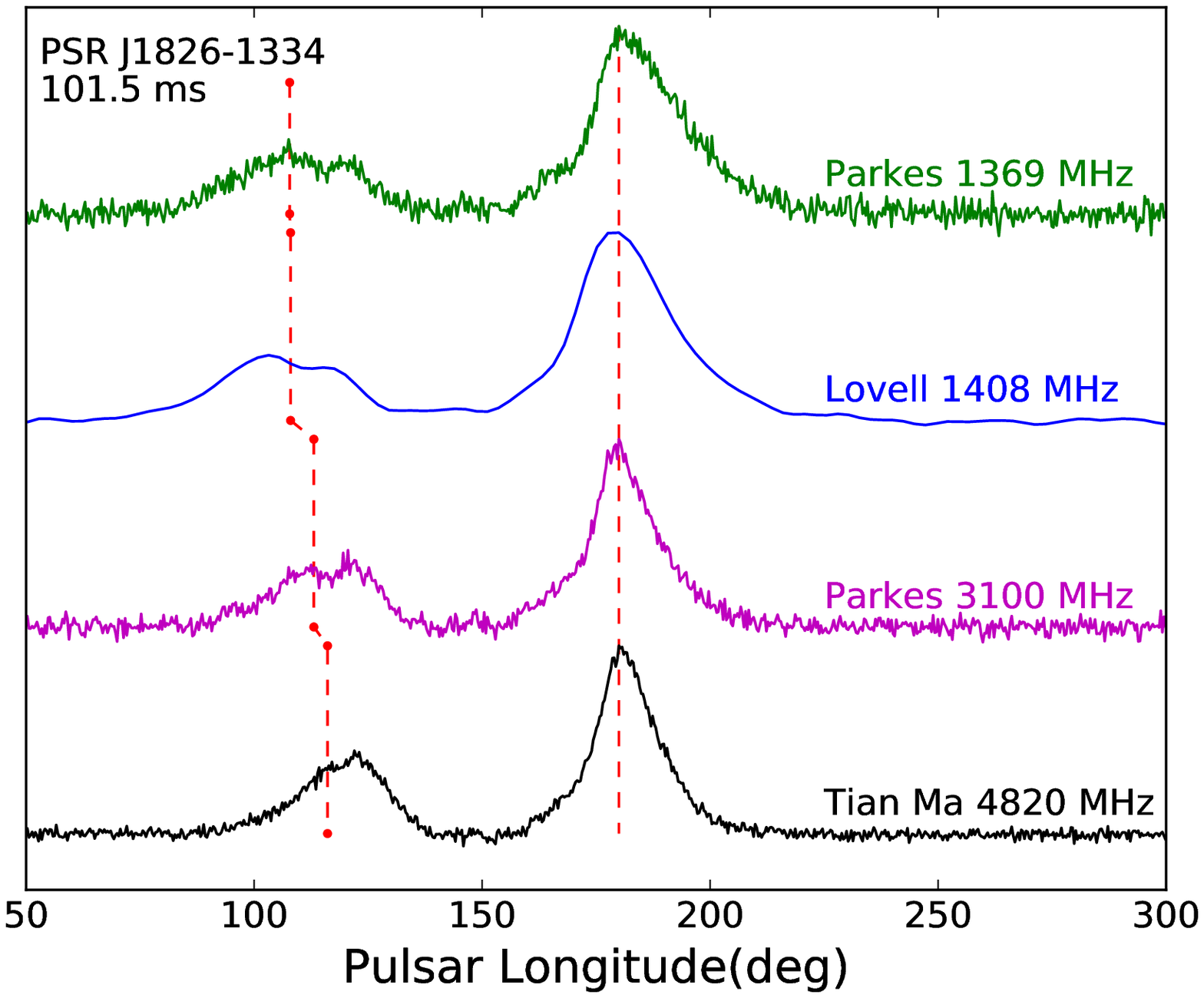}}&
\resizebox{0.53\hsize}{!}{\includegraphics[angle=0]{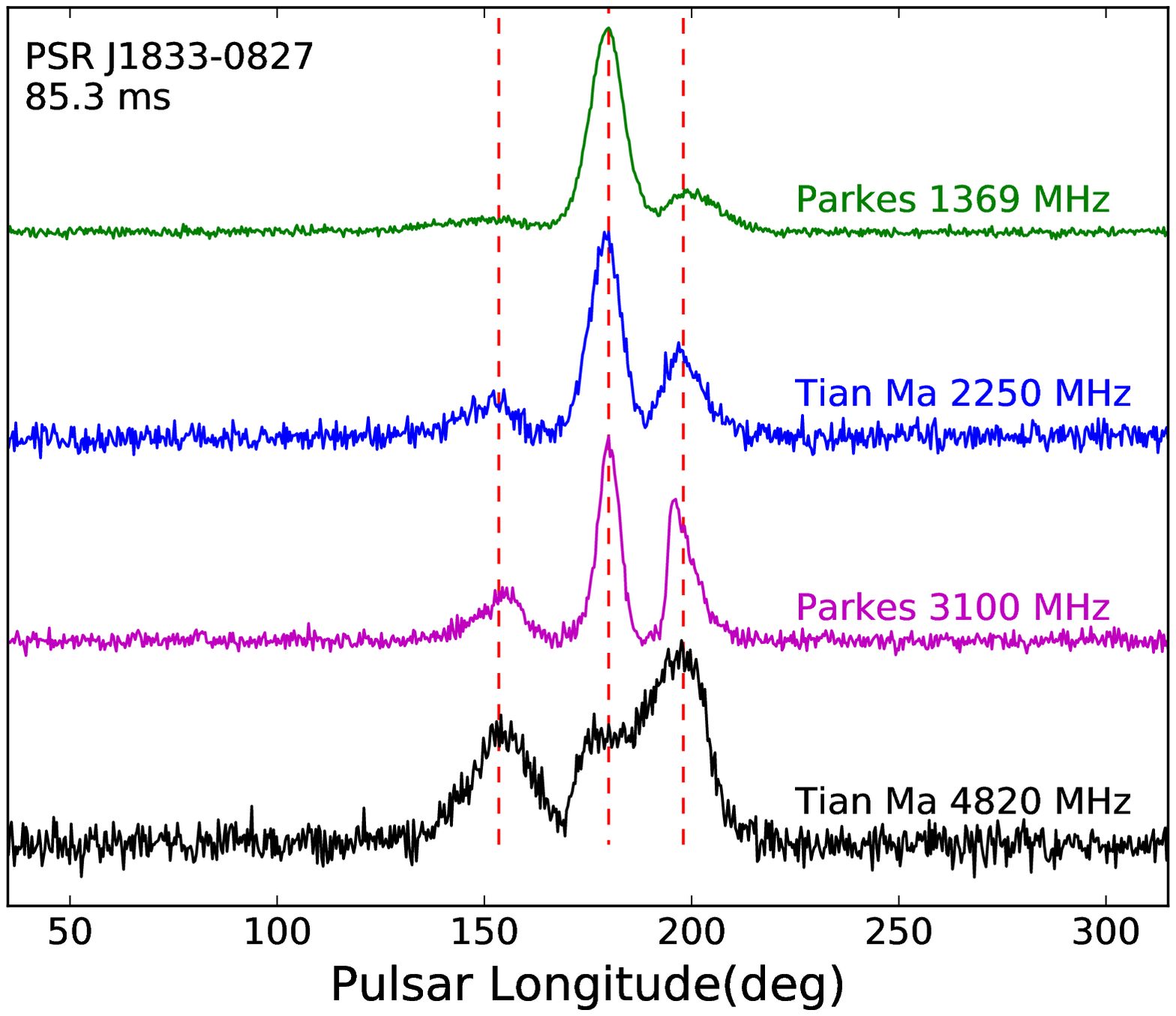}}\\

\end{tabular}
\end{center}
\caption{-continued}
\end{figure}
\addtocounter{figure}{-1}
\begin{figure}[h]
\begin{center}
\begin{tabular}{cc}

\resizebox{0.53\hsize}{!}{\includegraphics[angle=0]{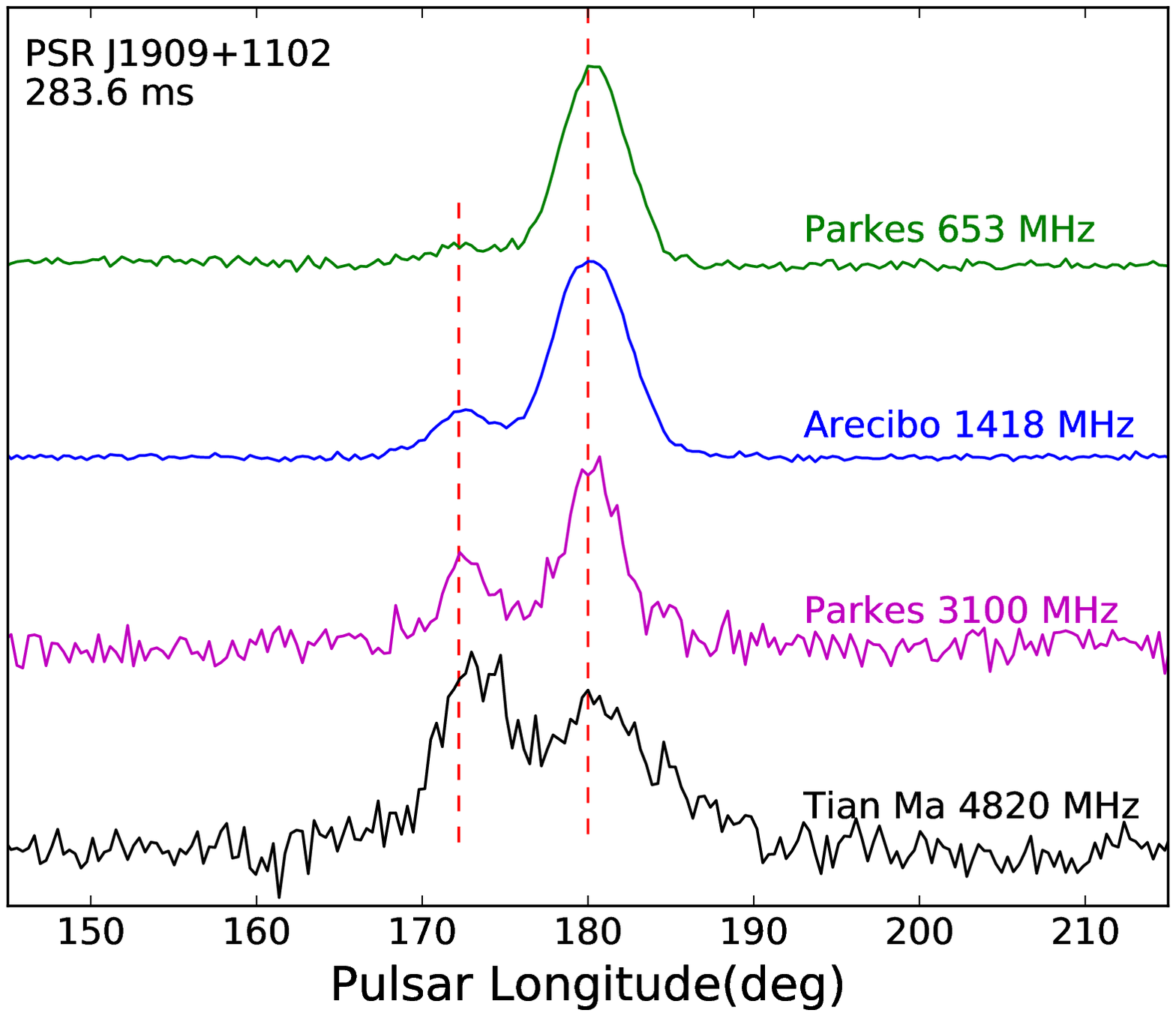}}&
\resizebox{0.53\hsize}{!}{\includegraphics[angle=0]{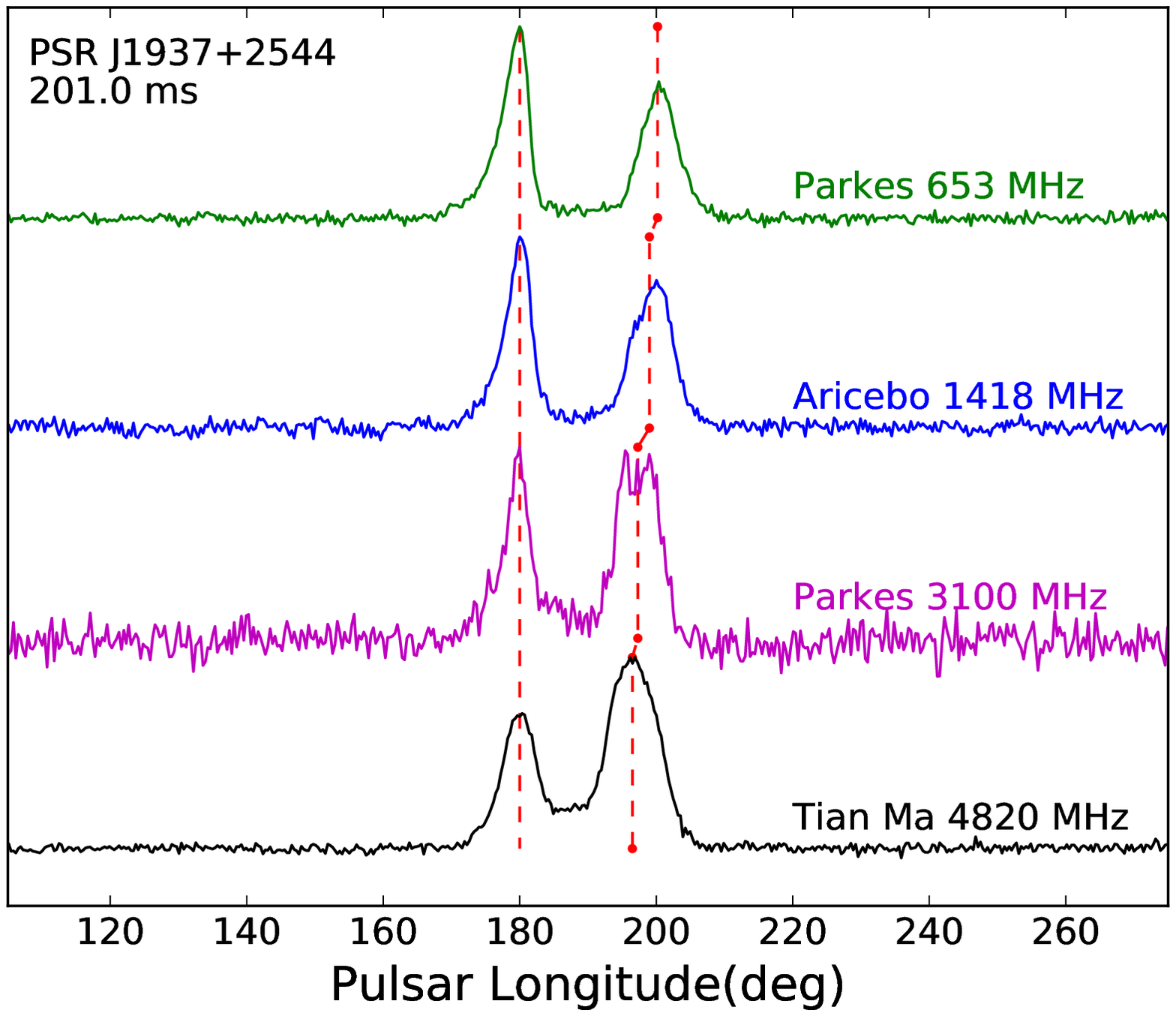}}\\
\resizebox{0.53\hsize}{!}{\includegraphics[angle=0]{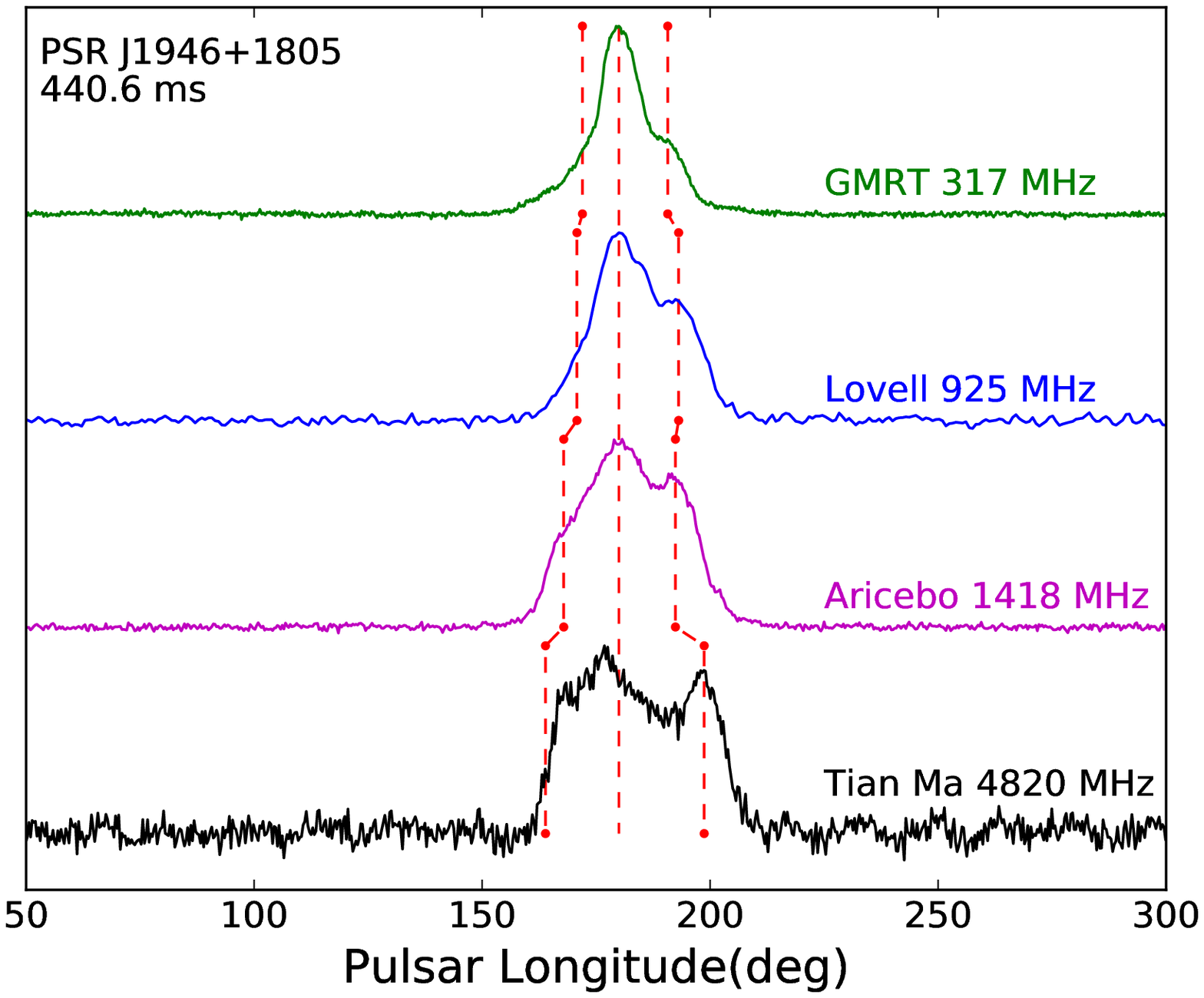}}&
\resizebox{0.53\hsize}{!}{\includegraphics[angle=0]{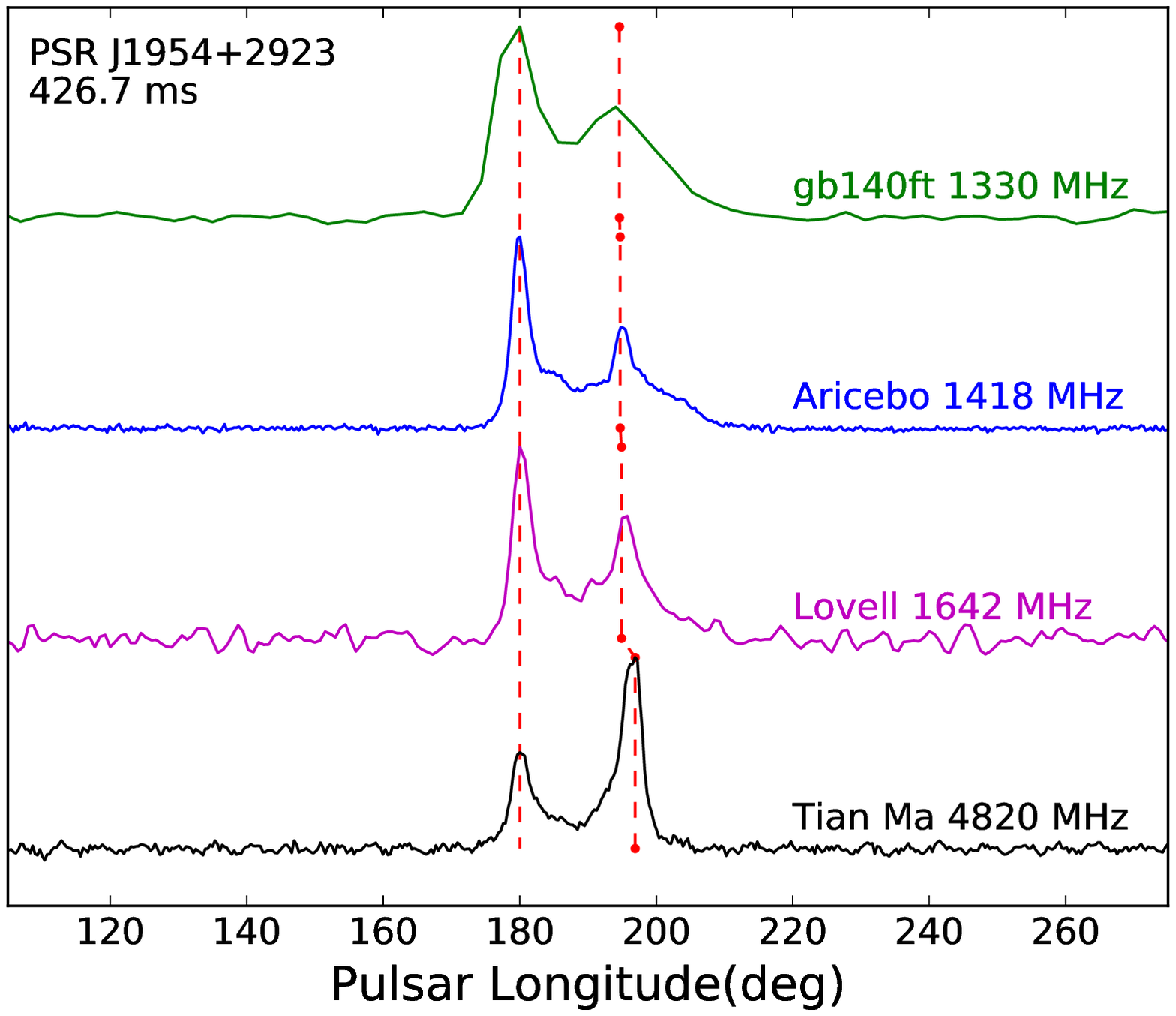}}\\
\resizebox{0.53\hsize}{!}{\includegraphics[angle=0]{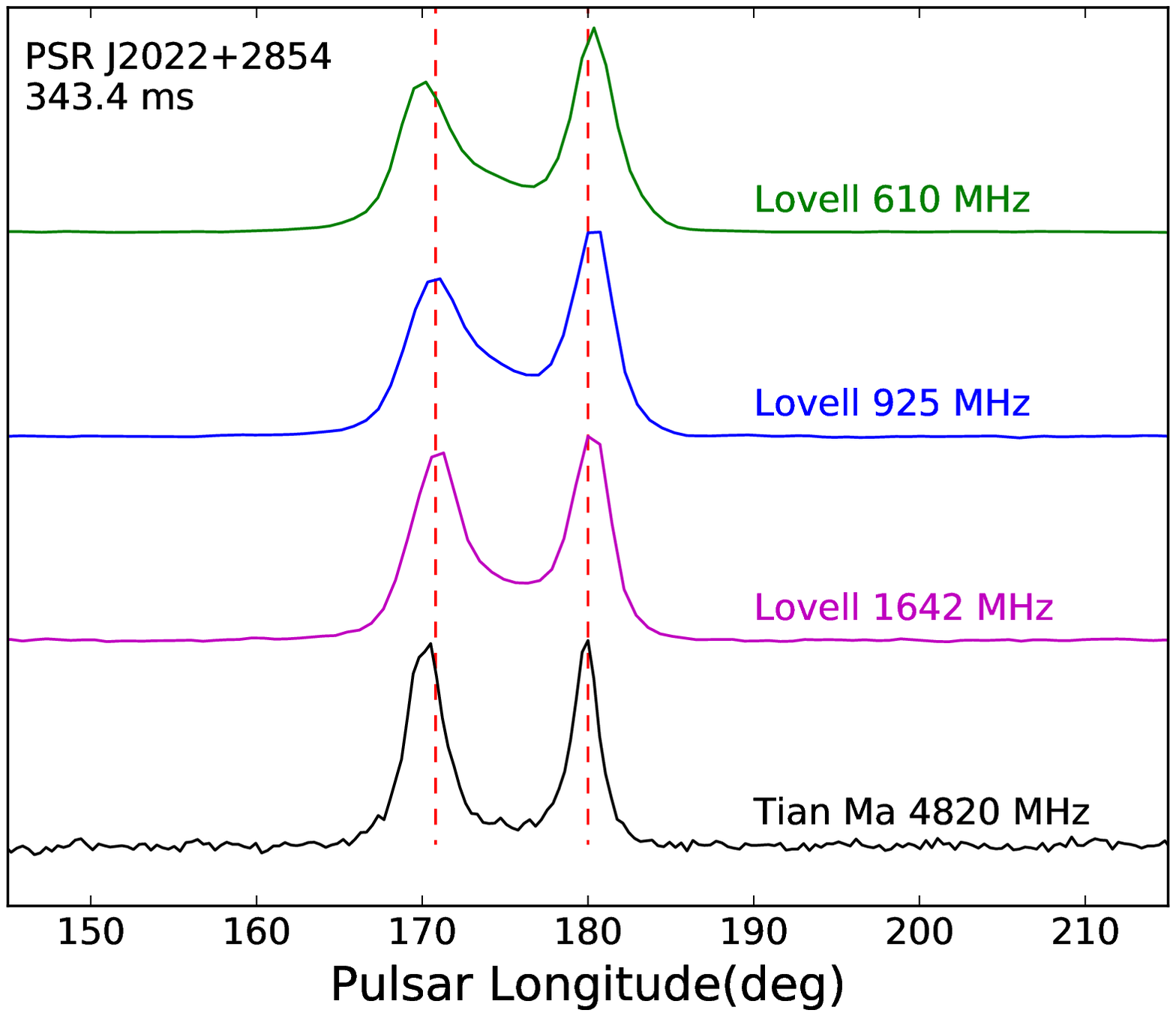}}&
\resizebox{0.53\hsize}{!}{\includegraphics[angle=0]{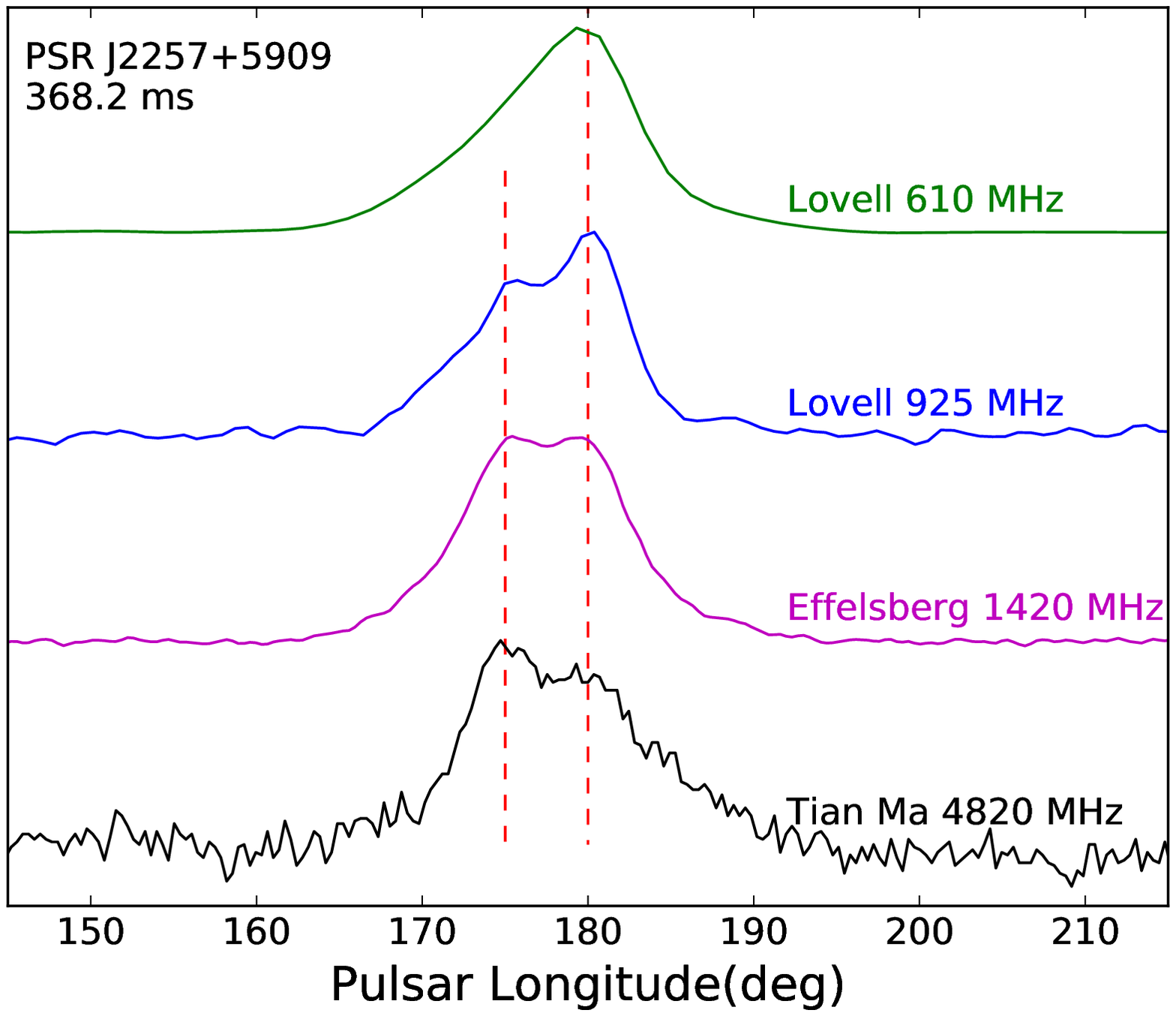}}\\
\end{tabular}
\end{center}
\caption{-continued.}
\end{figure}
\addtocounter{figure}{-1}
\begin{figure}[h]
\begin{center}
\begin{tabular}{cc}

\resizebox{0.53\hsize}{!}{\includegraphics[angle=0]{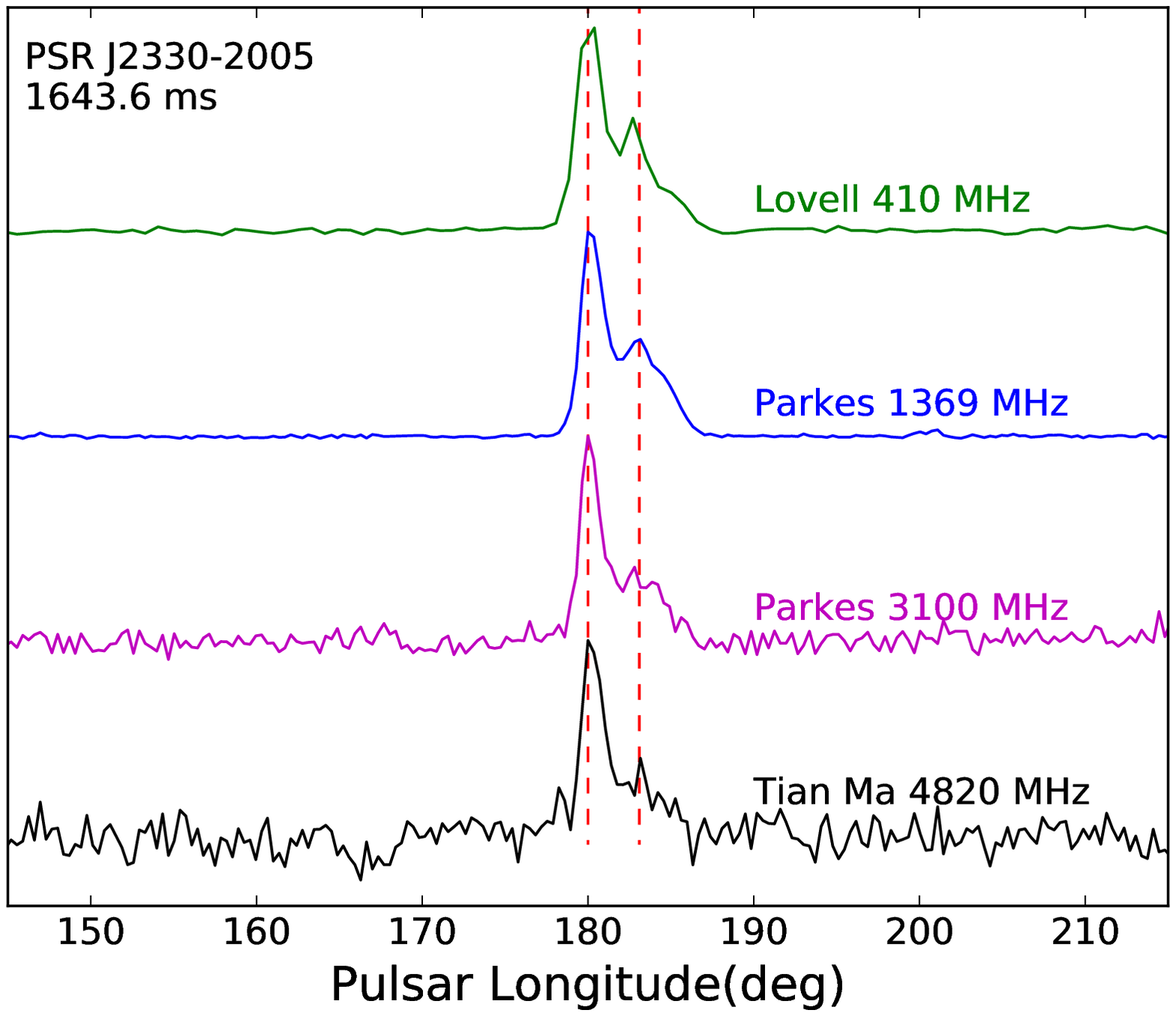}}&
\resizebox{0.53\hsize}{!}{\includegraphics[angle=0]{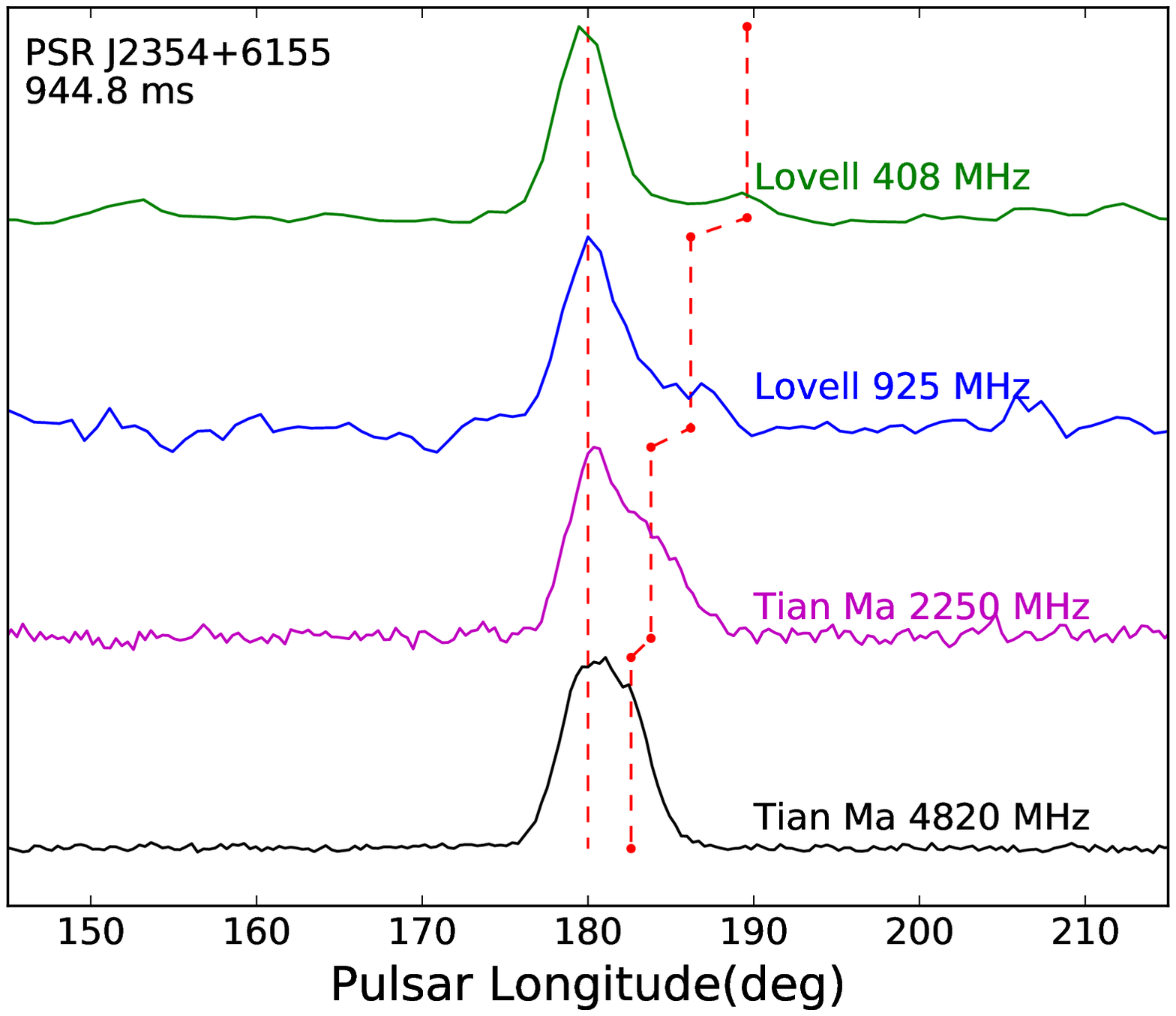}}\\
\end{tabular}
\end{center}
\caption{-continued.}
\end{figure}

 \clearpage

\begin{figure}[phtb]
\caption{Spectral plots for 27 pulsars along with power-law fits to
  the data. The red bars give the uncertainties of the flux
  densities at the different frequencies.} \label{fg:spec}
 \includegraphics[scale=0.9]{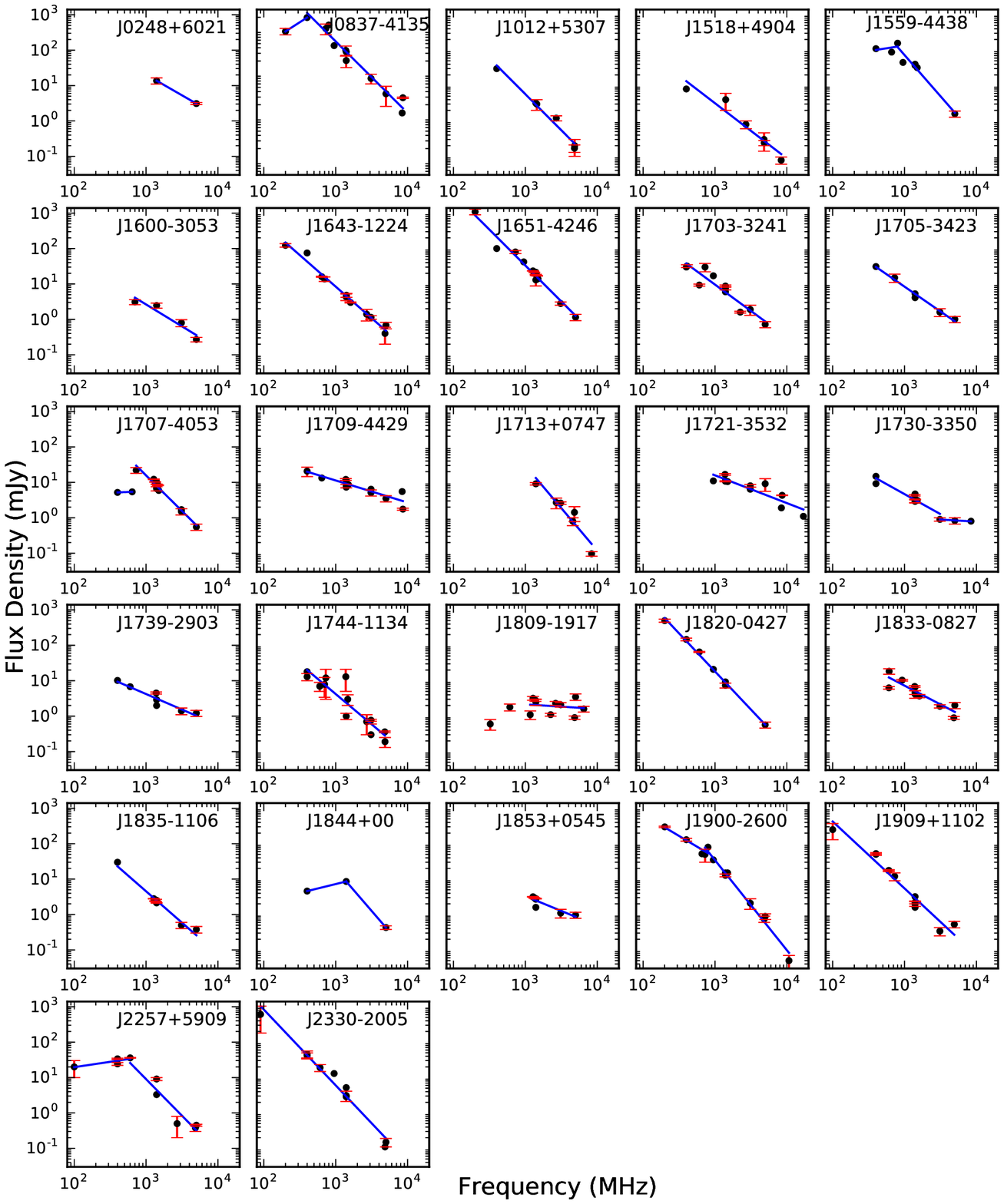}
\end{figure}

\begin{figure}[h]
\caption{Distributions of spectral index $\alpha$ versus pulsar period
  $\rm P$, period derivative $\rm \dot P$, characteristic age $\tau$,
  1400-MHz psuedo-luminosity $\rm L_{1400}$, surface magnetic field
  $\rm B_{surf}$ and spin-down luminosity $\rm \dot E$. In each
  sub-plot, the blue line is a weighted power-law fit to the points,
  and the dashed lines show the 1-$\sigma$ uncertainties in the
  power-law index. The parameters of the fit and the Spearman rank
  coefficient are given at the bottom of each
  sub-plot.} \label{fg:spec-corr}
\begin{center}
\begin{tabular}{ccc}
\resizebox{0.33\hsize}{!}{\includegraphics[angle=0]{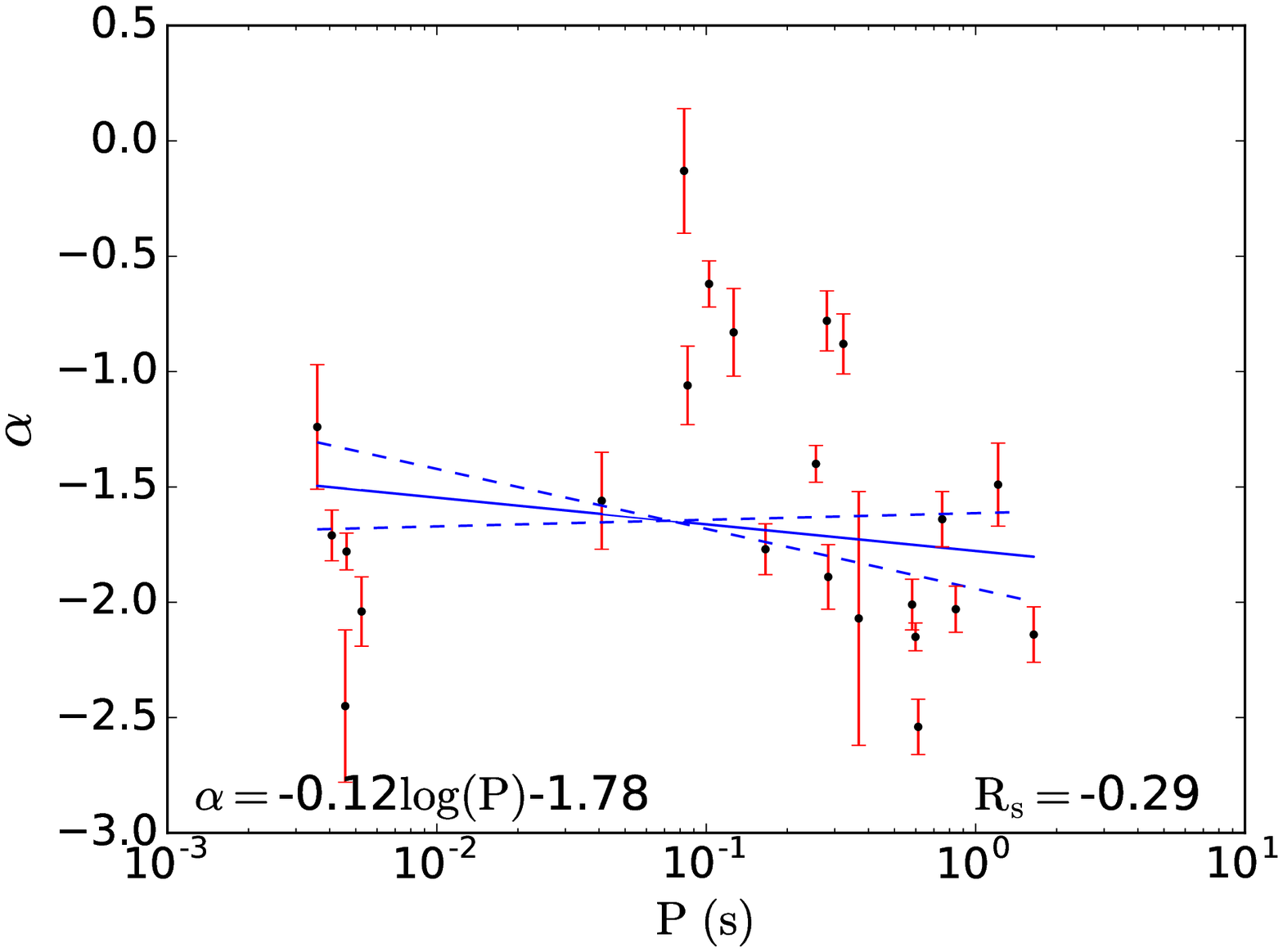}}&
\resizebox{0.33\hsize}{!}{\includegraphics[angle=0]{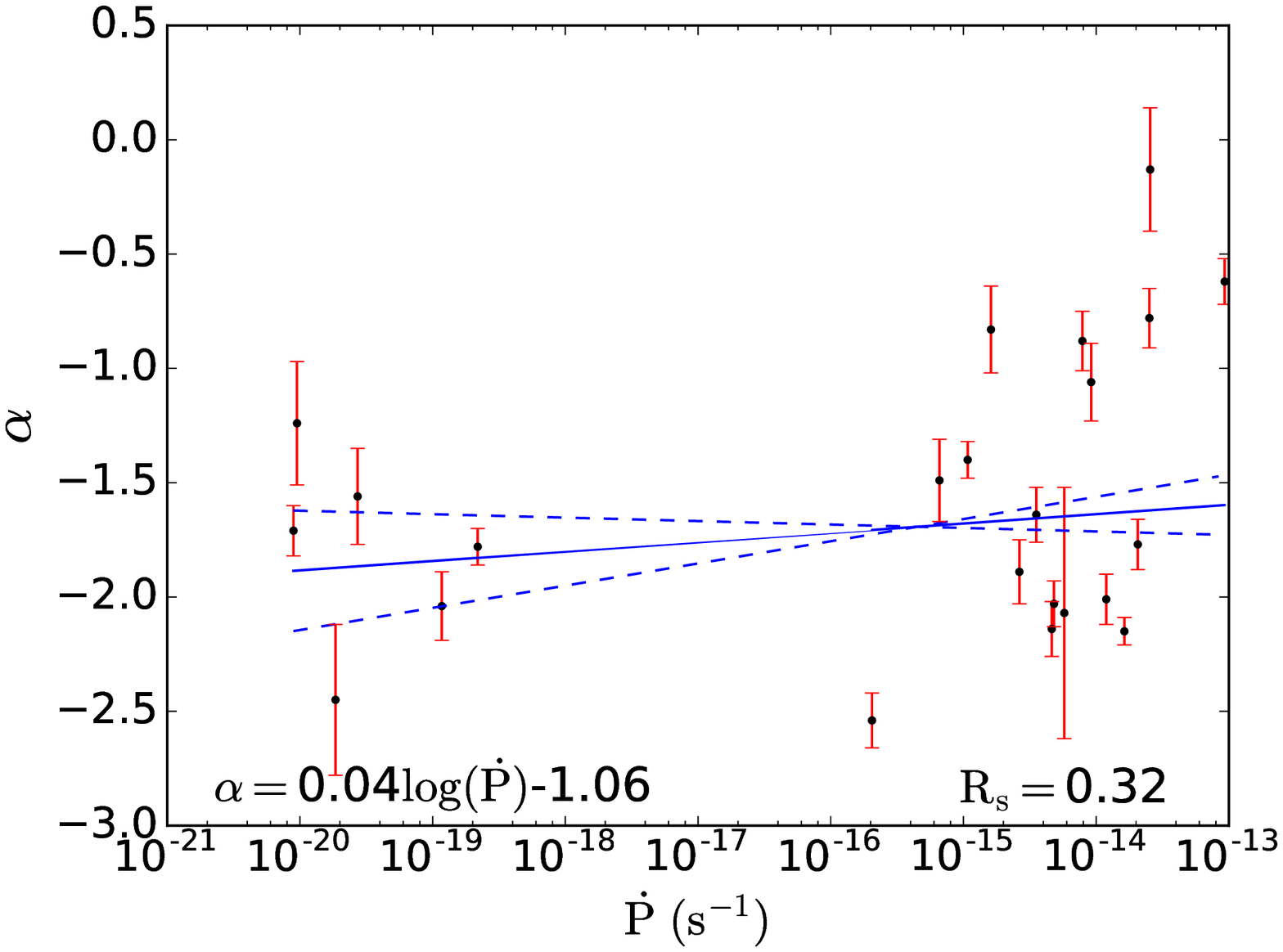}}&
\resizebox{0.33\hsize}{!}{\includegraphics[angle=0]{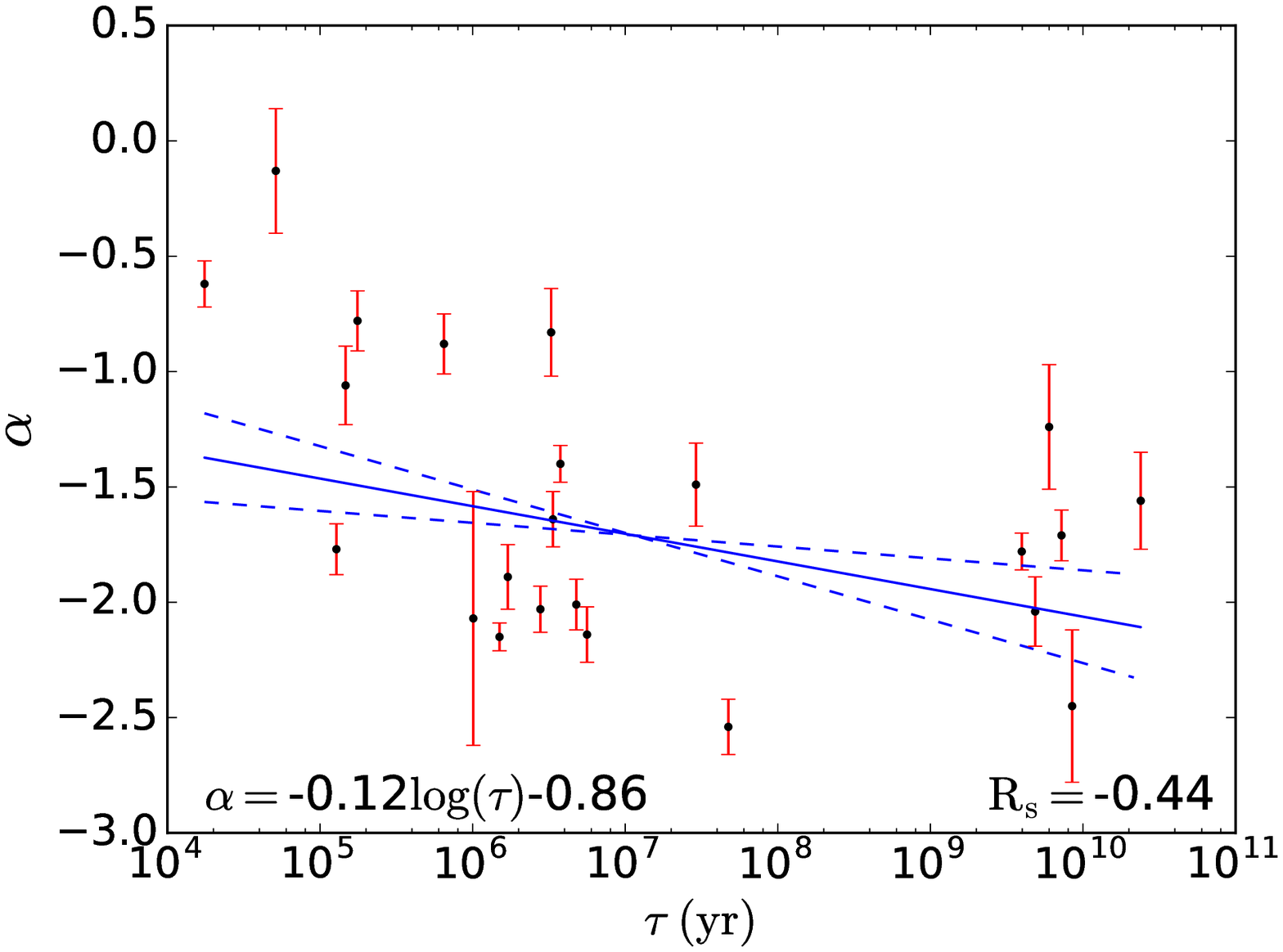}}\\
\resizebox{0.33\hsize}{!}{\includegraphics[angle=0]{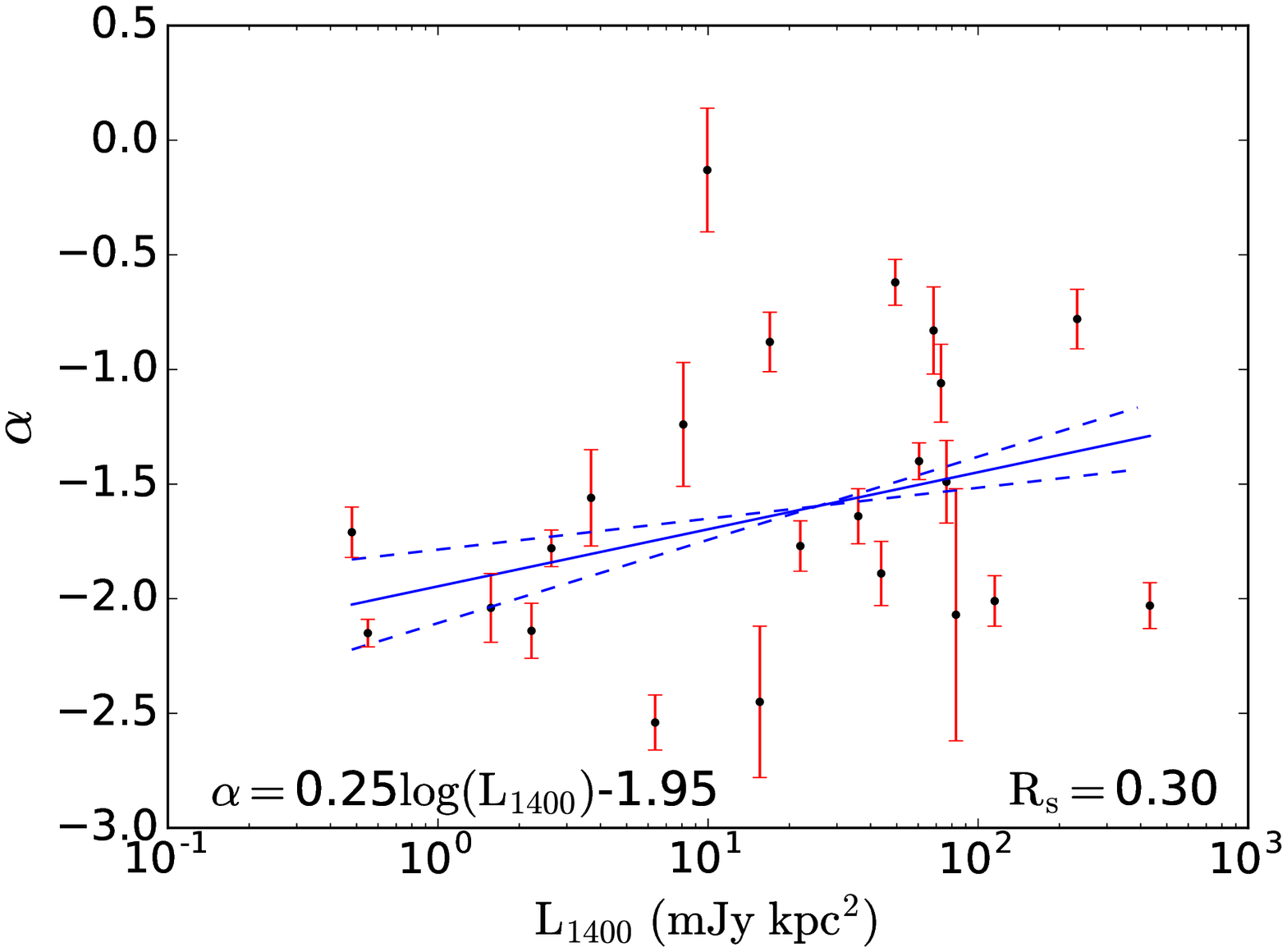}}&
\resizebox{0.33\hsize}{!}{\includegraphics[angle=0]{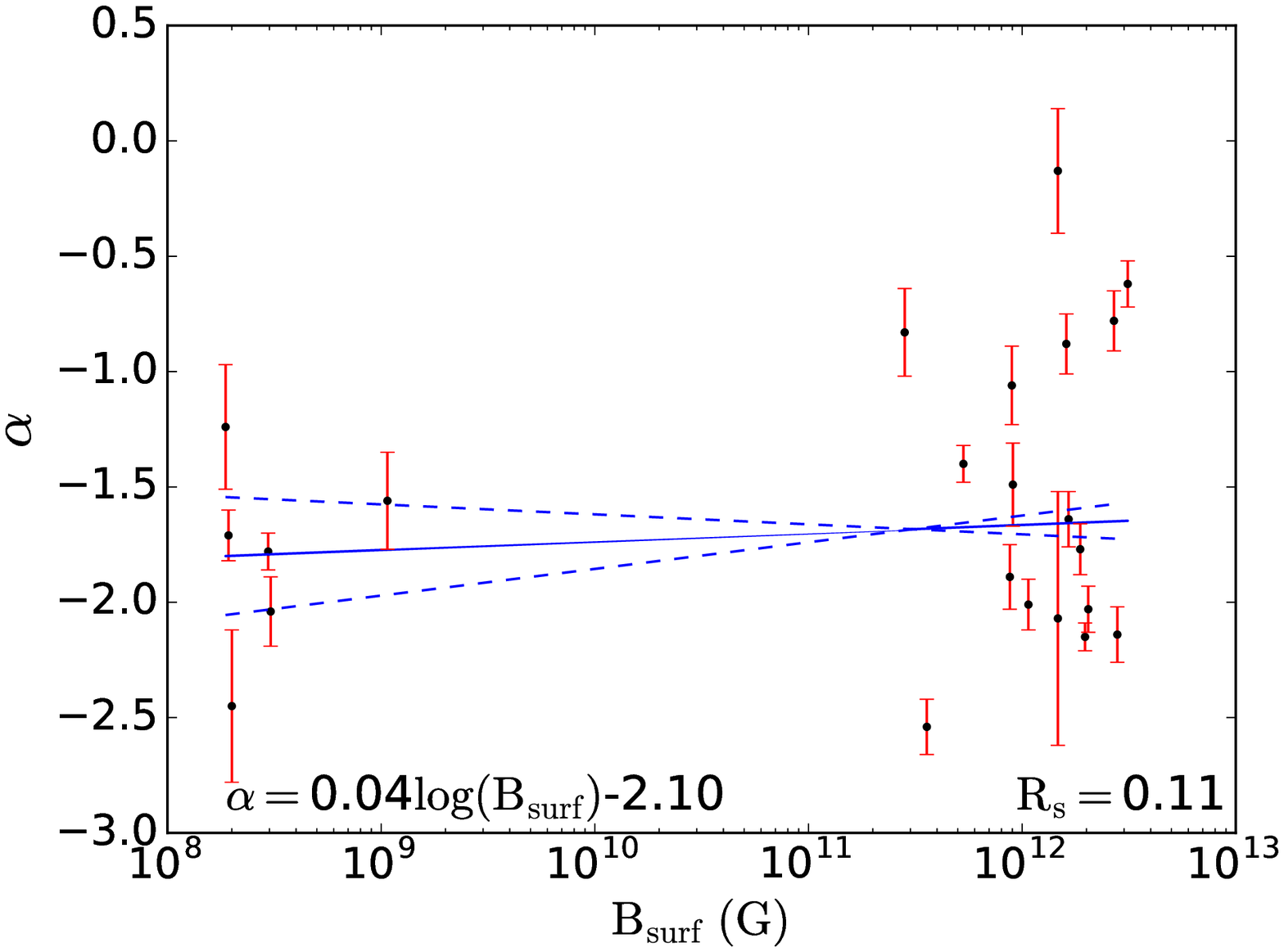}}&
\resizebox{0.33\hsize}{!}{\includegraphics[angle=0]{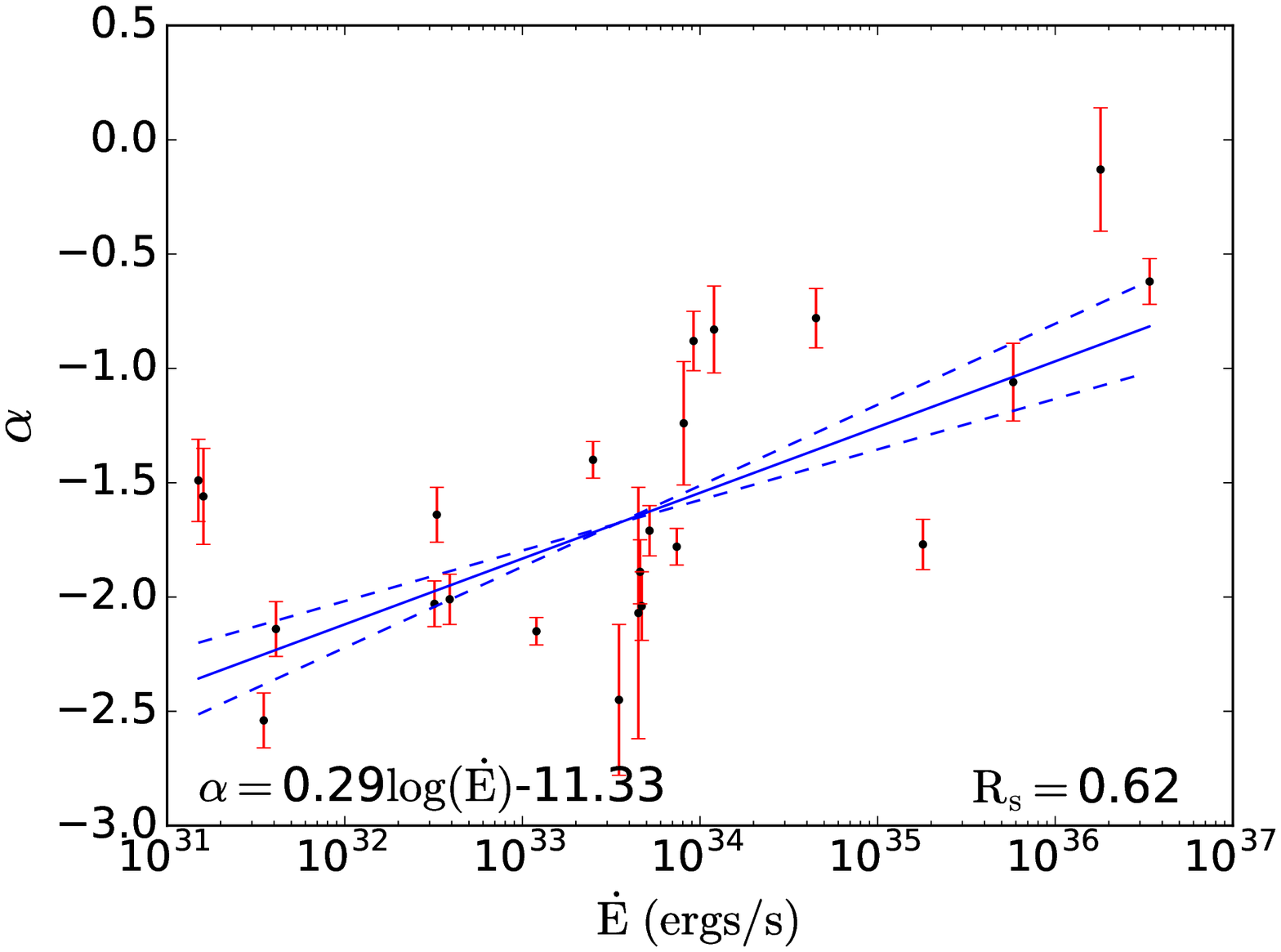}}\\
\end{tabular}
\end{center}
\end{figure}

\clearpage

\begin{figure}[h]
\caption{Left panel: integrated profiles for PSR J0332+5434
showing the two modes observed in TMRT observations at 5 GHz. Right panel: ratios of 
leading and trailing components to the central component ($\rm R_{12}$ and $\rm R_{32}$) 
for different observations separated by green dashed lines.}
\label{fg:J0332_md}
\begin{center}
\begin{tabular}{ccc}
\resizebox{0.5\hsize}{!}{\includegraphics[angle=0]{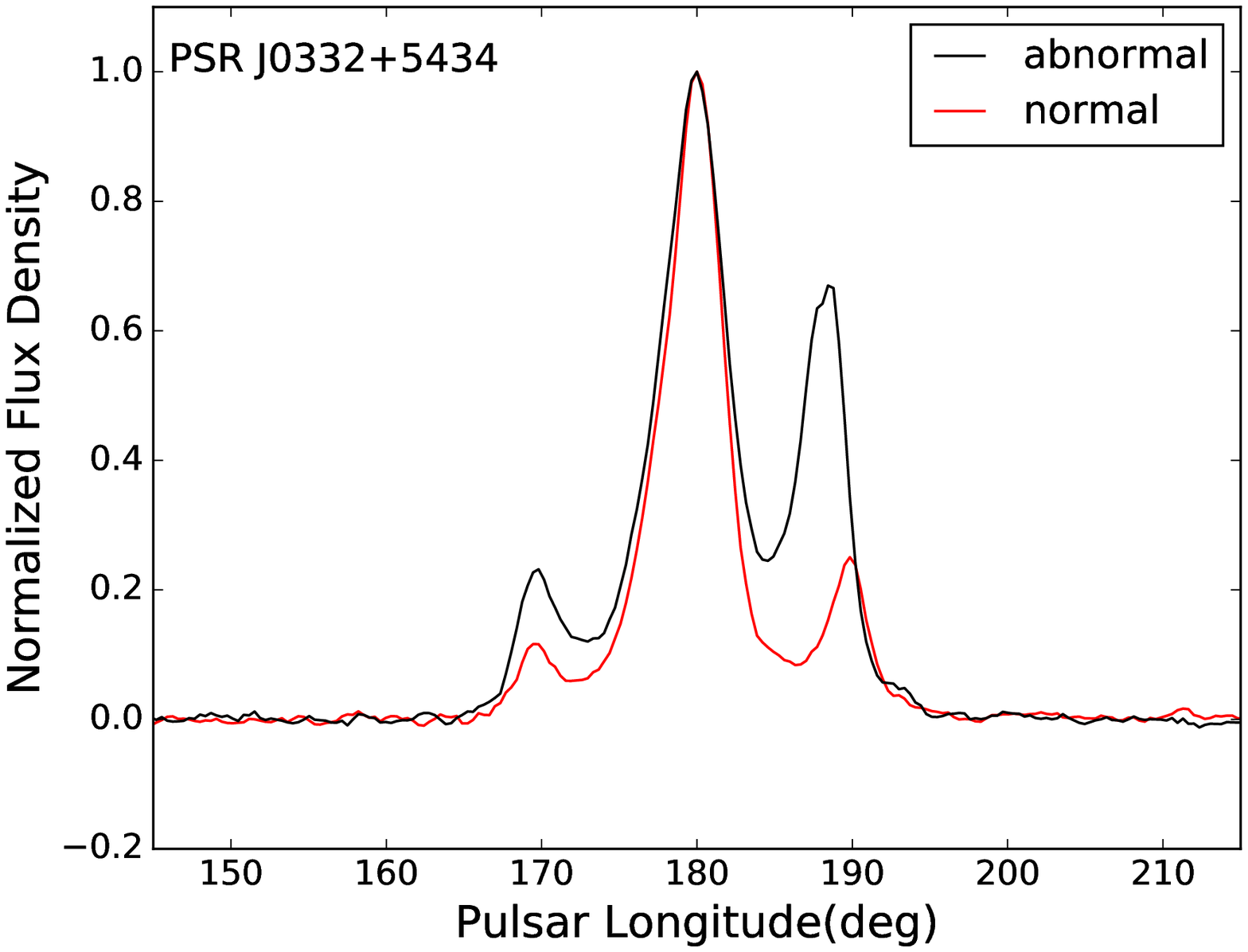}}&
\resizebox{0.5\hsize}{!}{\includegraphics[angle=0]{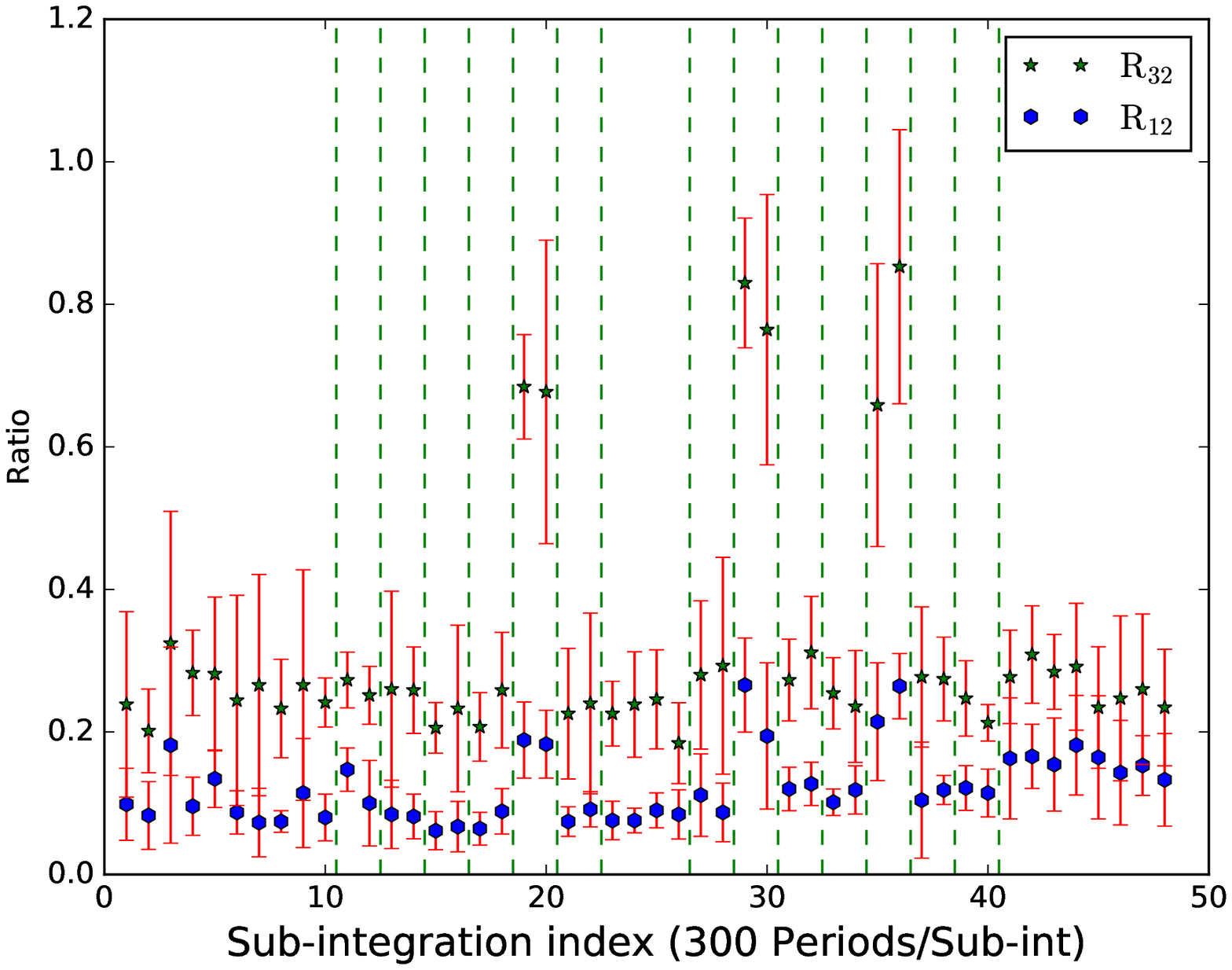}}\\
\end{tabular}
\end{center}
\end{figure} 

\clearpage

\begin{figure}[h]
\caption{Integrated profiles for PSR J0538+2817 showing the 
two modes at 5 GHz obtained in TMRT observations.}
\label{fg:J0538_md}
\begin{center}
\begin{tabular}{ccc}
\resizebox{0.5\hsize}{!}{\includegraphics[angle=0]{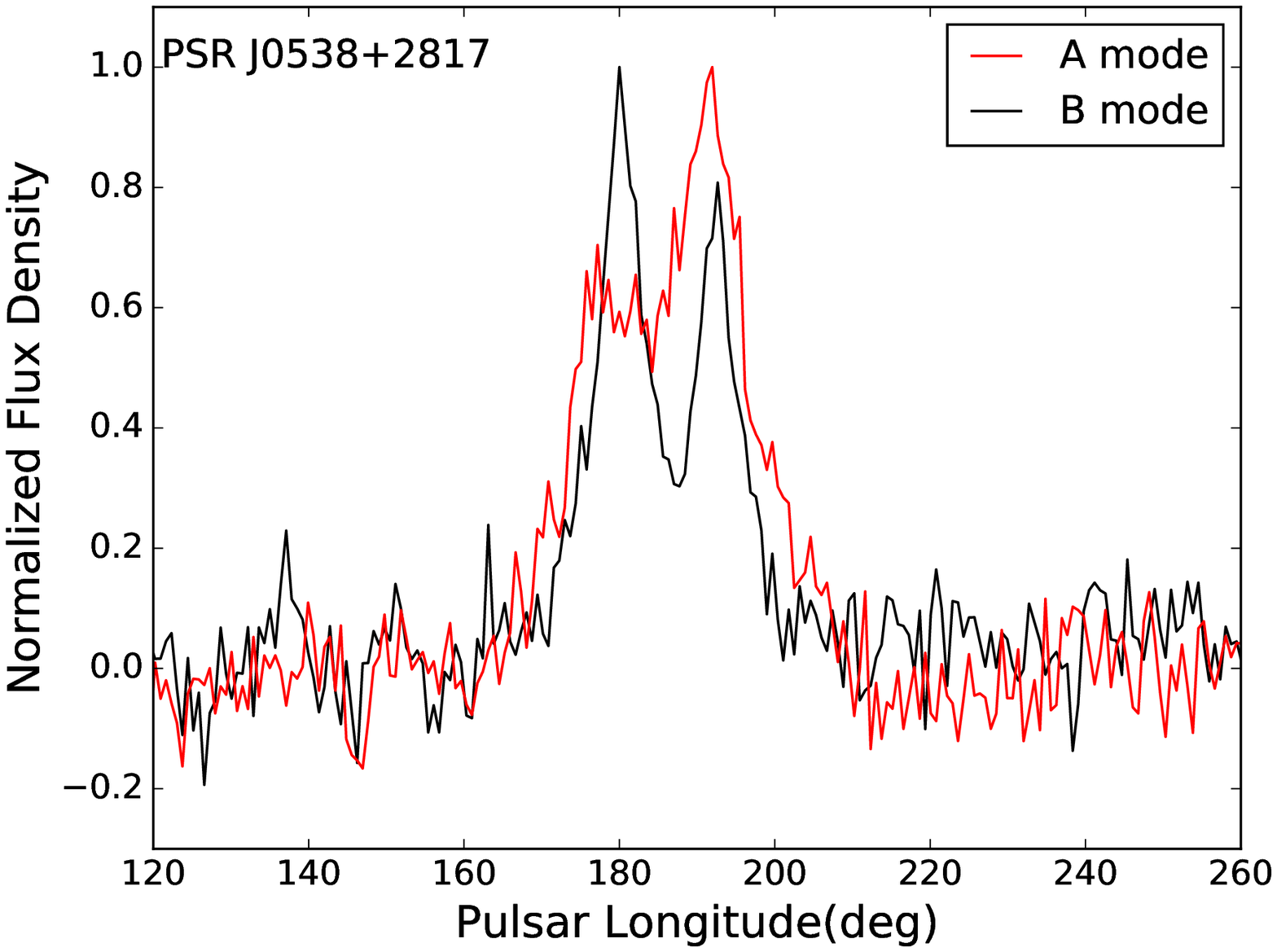}}\\
\end{tabular}
\end{center}
\end{figure} 

\clearpage

\begin{figure}[h]
\caption{The left panel is the mode changing of integrated profile for 
PSR J0742-2822, whose two modes were obtained in TMRT observations. The right 
panel is the $\rm R_{12}$ distribution against sub-integration numbers. The 
blue line shows the fit based on two gaussian components marked by dashed lines.}
\label{fg:J0742_md}
\begin{center}
\begin{tabular}{ccc}
\resizebox{0.5\hsize}{!}{\includegraphics[angle=0]{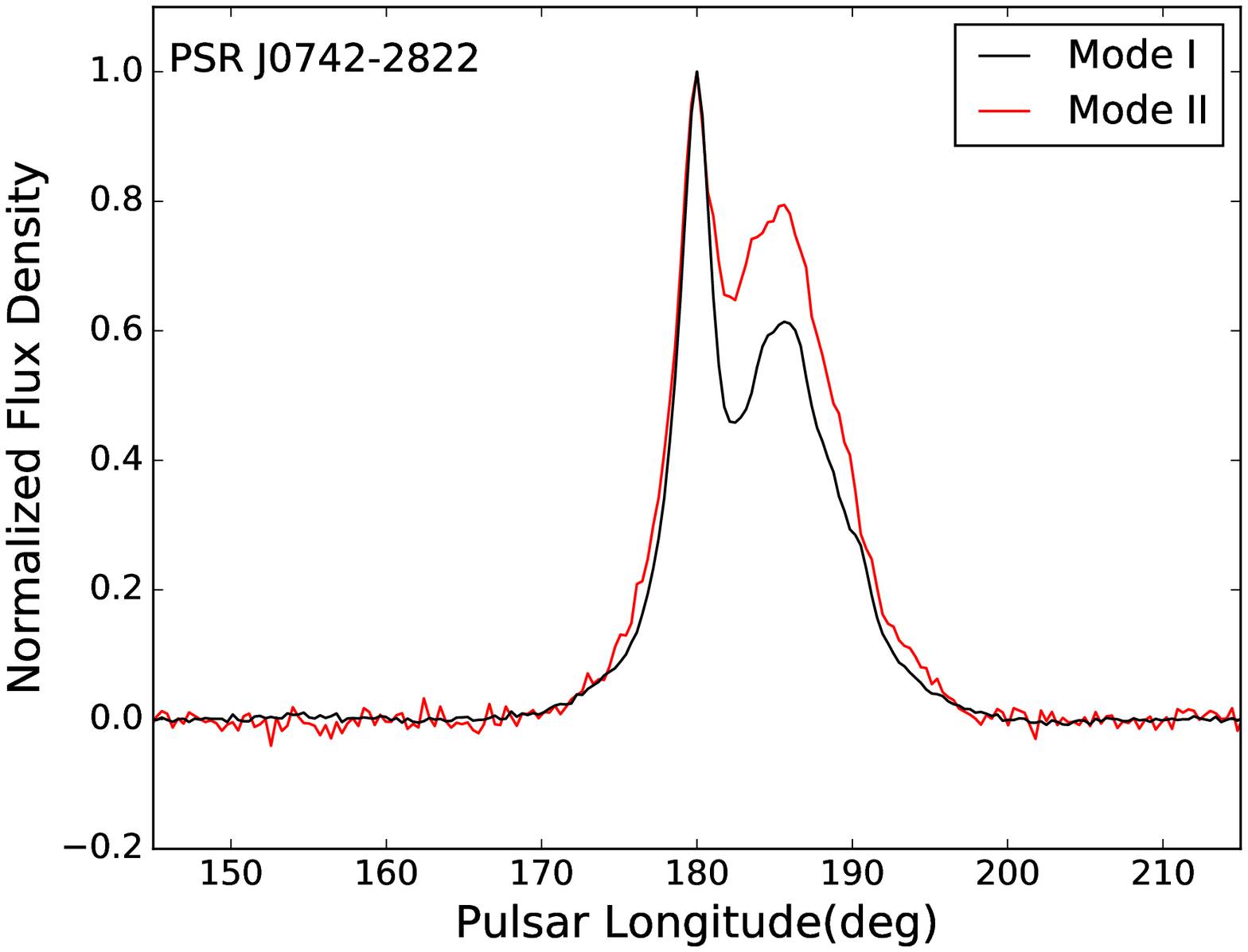}}&
\resizebox{0.5\hsize}{!}{\includegraphics[angle=0]{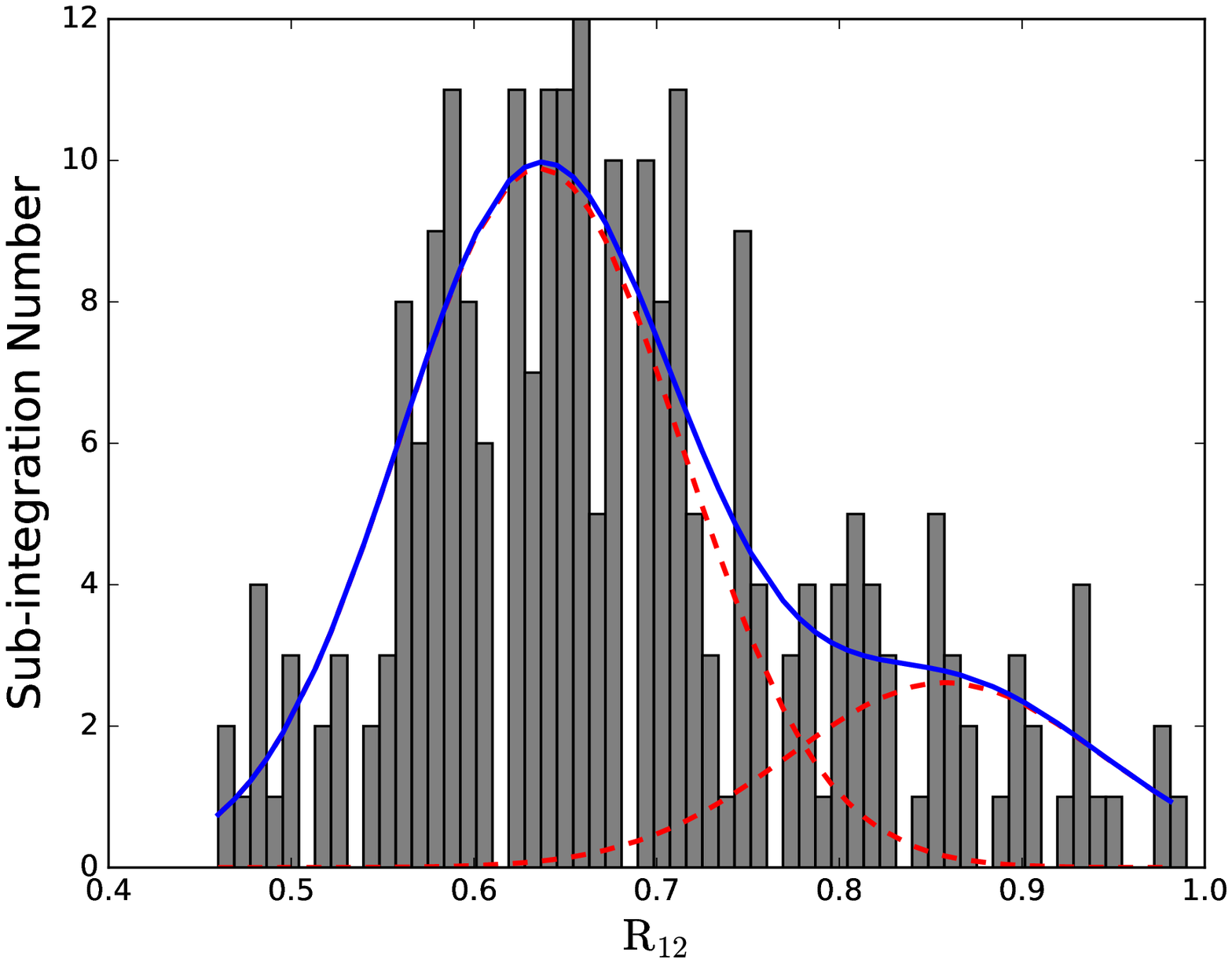}}\\
\end{tabular}
\end{center}
\end{figure} 

\clearpage

\begin{figure}[h]
\caption{Left panel: mode changing of integrated profile for PSR J1825-0925, 
whose two modes were obtained in TMRT observations. Right panel from top to
bottom: a zoom-in precursor and main pulse as a function of Sub-integration; the 
main pulse, precursor and interpulse signal-to-noise ratios of the detections 
in time bins of 30~s (about 40 periods). The x-axis in the bottom gives the
number of 40 periods. B and Q mode were distinguished by the exhibitition of 
precursor on/off switching (grey area). Different observations are splited by the 
red dash lines.}
\label{fg:J1825_md}
\begin{center}
\begin{tabular}{ccc}
\resizebox{0.5\hsize}{!}{\includegraphics[angle=0]{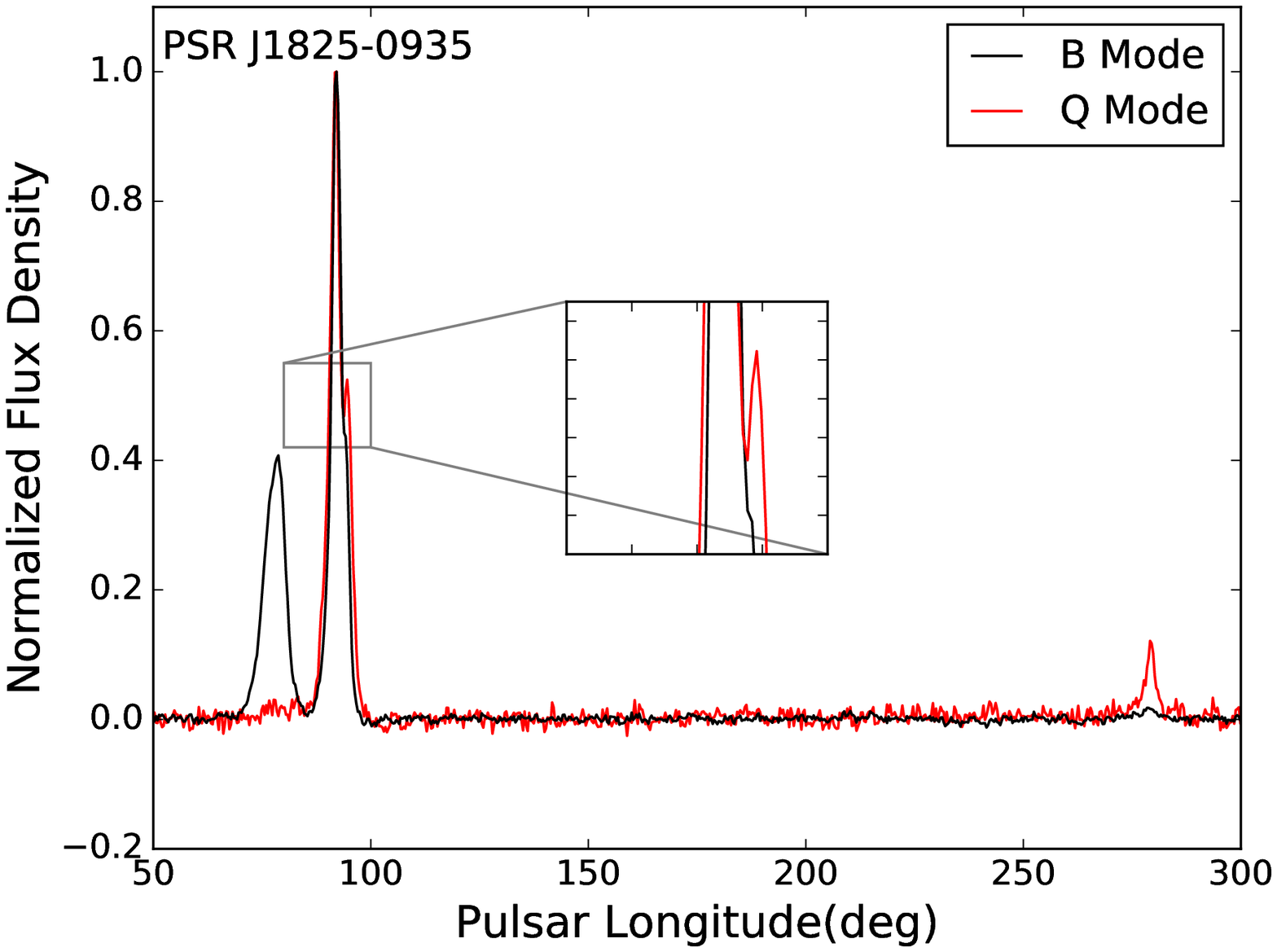}}&
\resizebox{0.5\hsize}{!}{\includegraphics[angle=0]{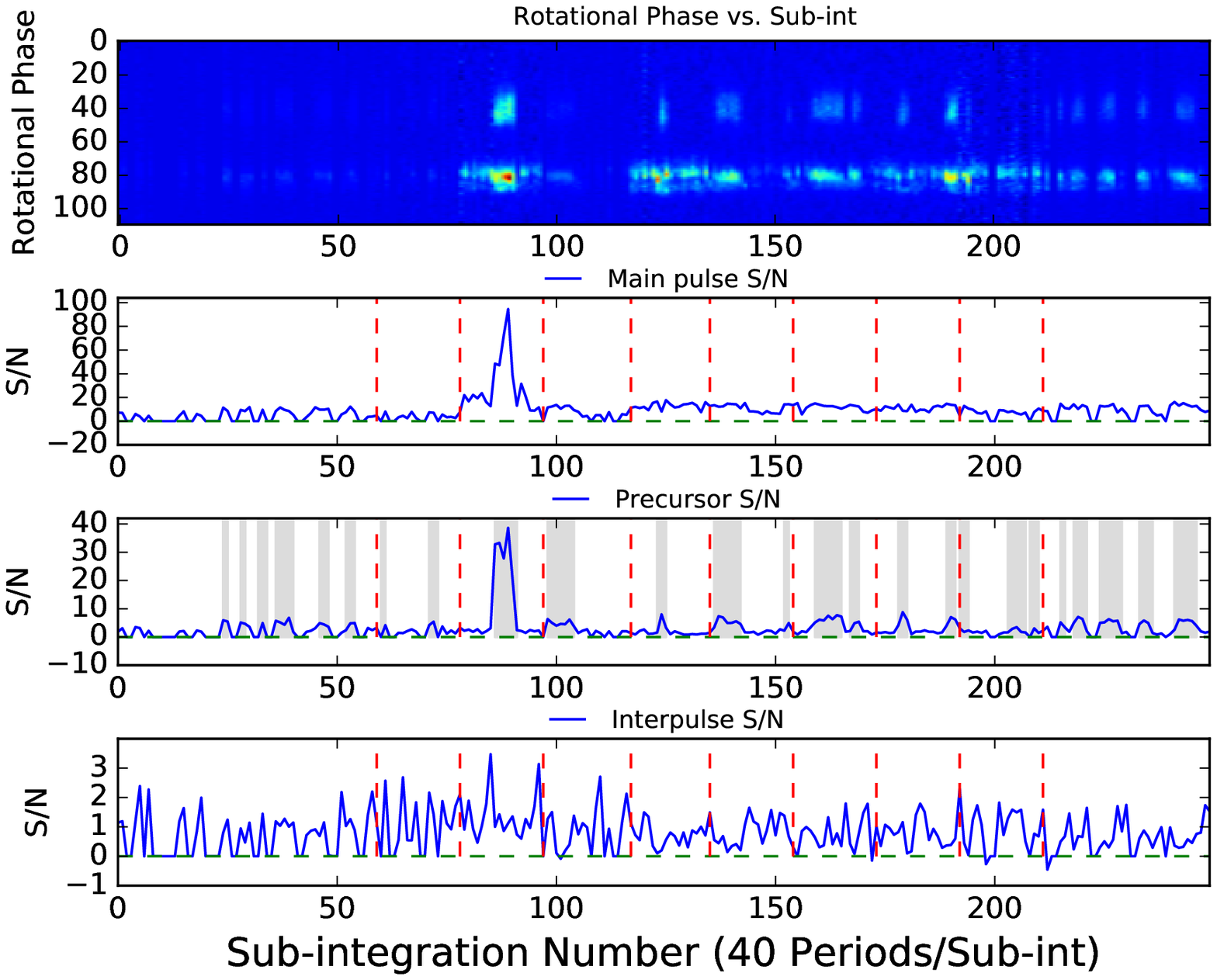}}\\
\end{tabular}
\end{center}
\end{figure} 

\clearpage

\begin{figure}[h]
\caption{The left panel is the mode changing of integrated profile for PSR J2022+2854, whose two modes were obtained in TMRT observations. The right panel is the $\rm R_{12}$ distribution against sub-integration numbers. The blue line shows the fit based on two gaussian components marked by dashed lines.}
\label{fg:J2022_md}
\begin{center}
\begin{tabular}{ccc}
\resizebox{0.5\hsize}{!}{\includegraphics[angle=0]{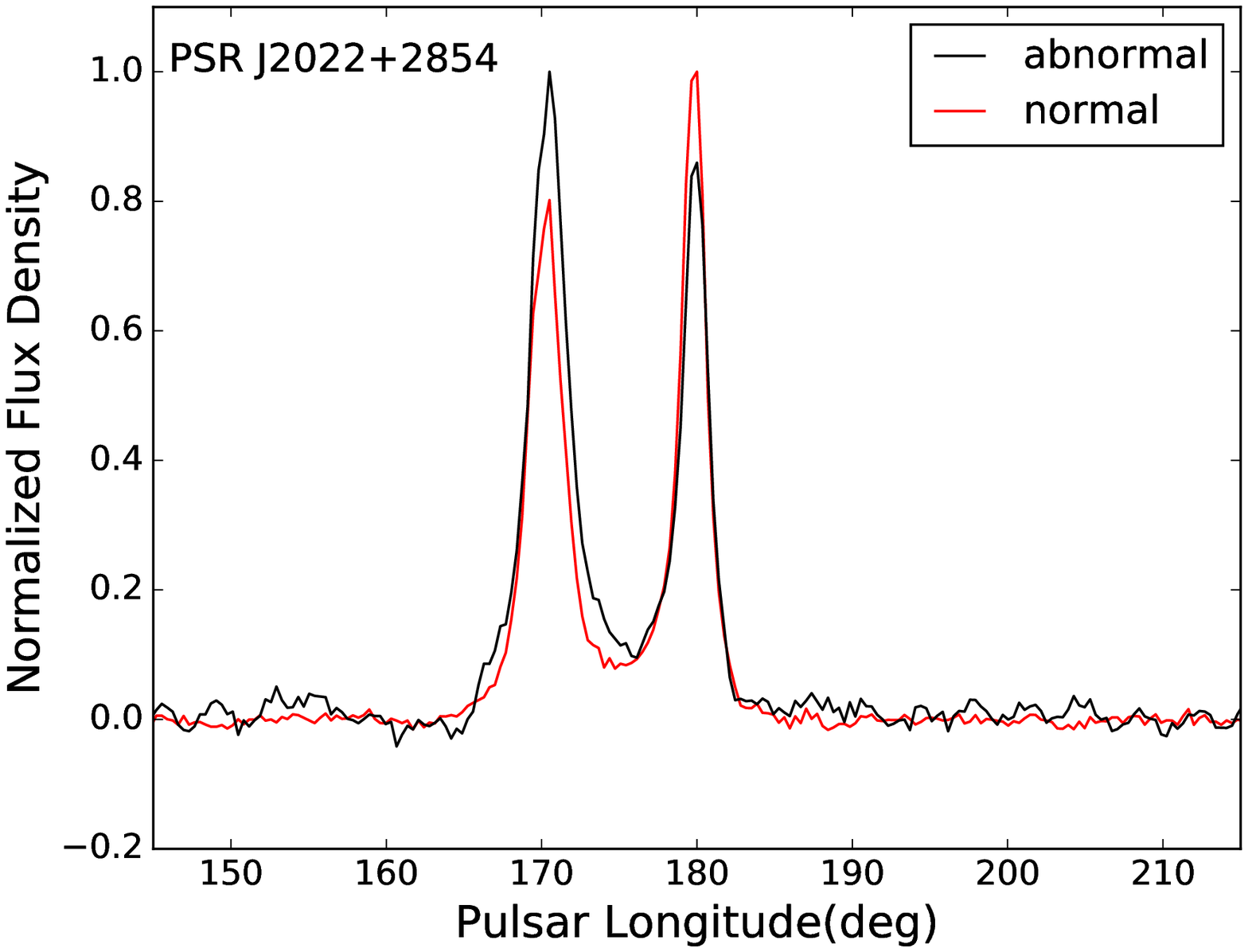}}&
\resizebox{0.5\hsize}{!}{\includegraphics[angle=0]{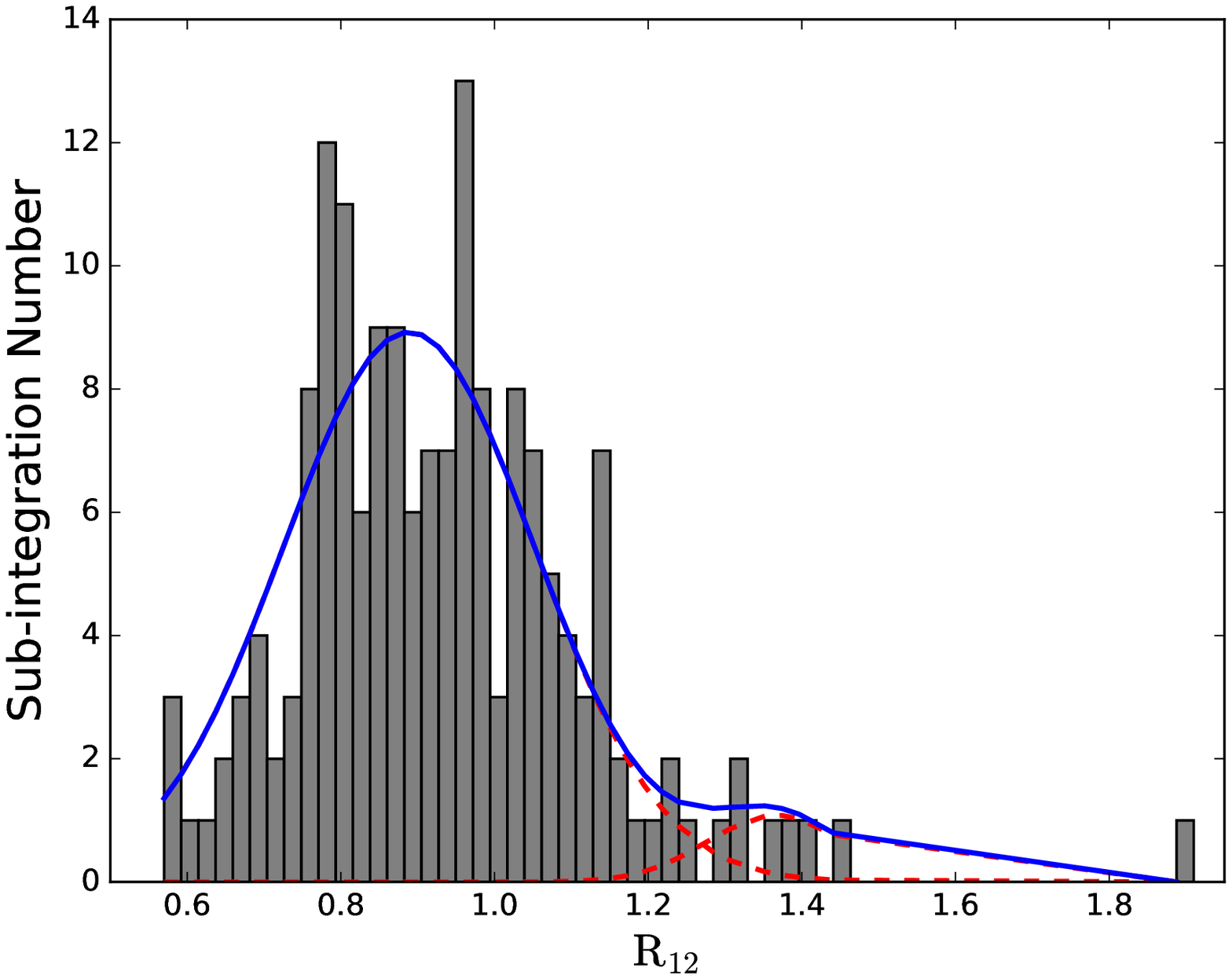}}\\
\end{tabular}
\end{center}
\end{figure} 

\clearpage

\begin{figure}[h]
\caption{a: Separation between the main pulse (MP) and interpulse (IP)
  as a function of frequency for six pulsars. b: Ratio of the peak
  flux density of the IP to the peak flux density of the MP as a
  function of frequency for the same six pulsars. Weighted power-law
  fits to the dependencies (in some cases segmented where the over-all
  dependence is non-linear) are shown.}
\label{fg:mp_ip}
\begin{center}
\begin{tabular}{cc}
\resizebox{0.53\hsize}{!}{\includegraphics[angle=0]{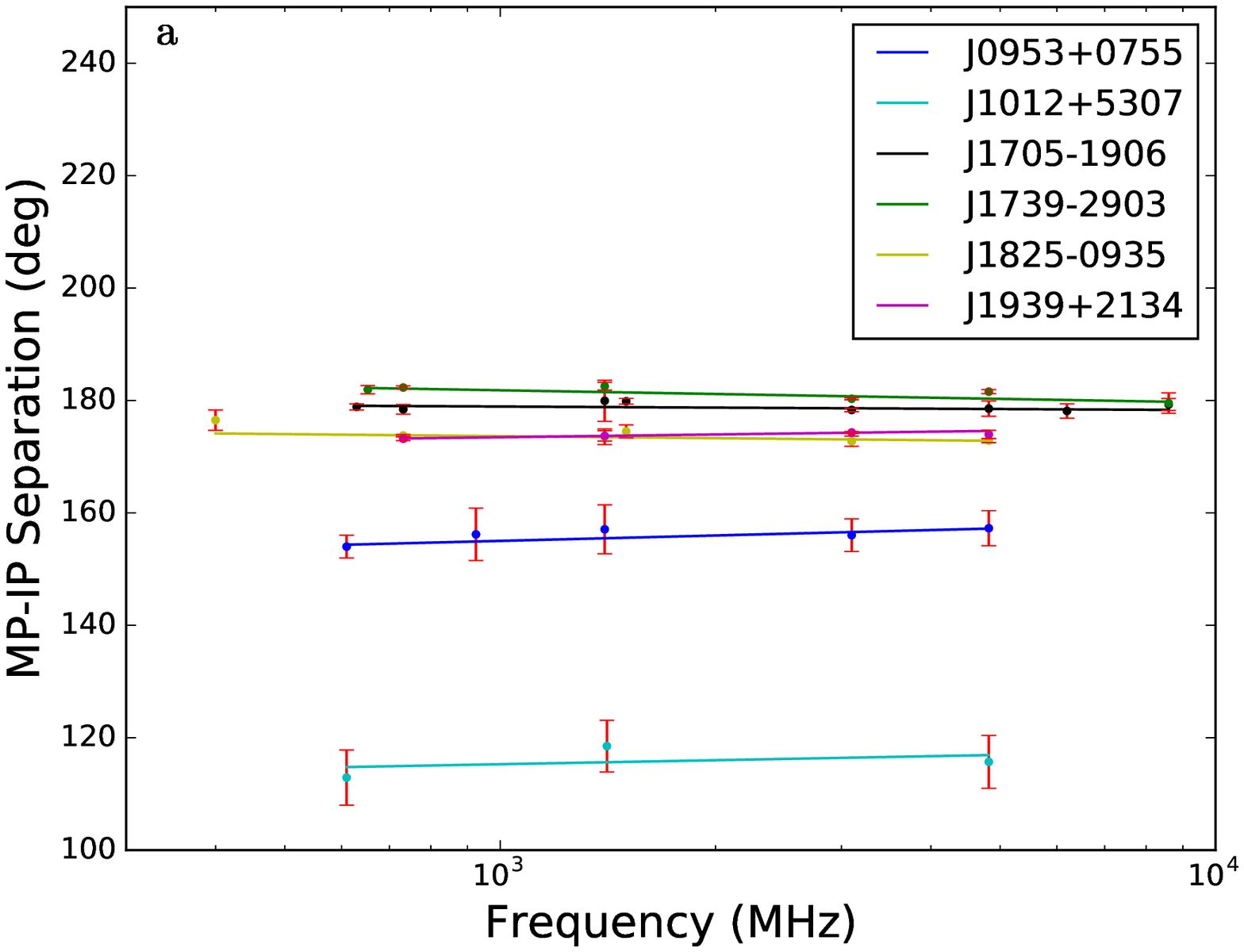}}&
\resizebox{0.53\hsize}{!}{\includegraphics[angle=0]{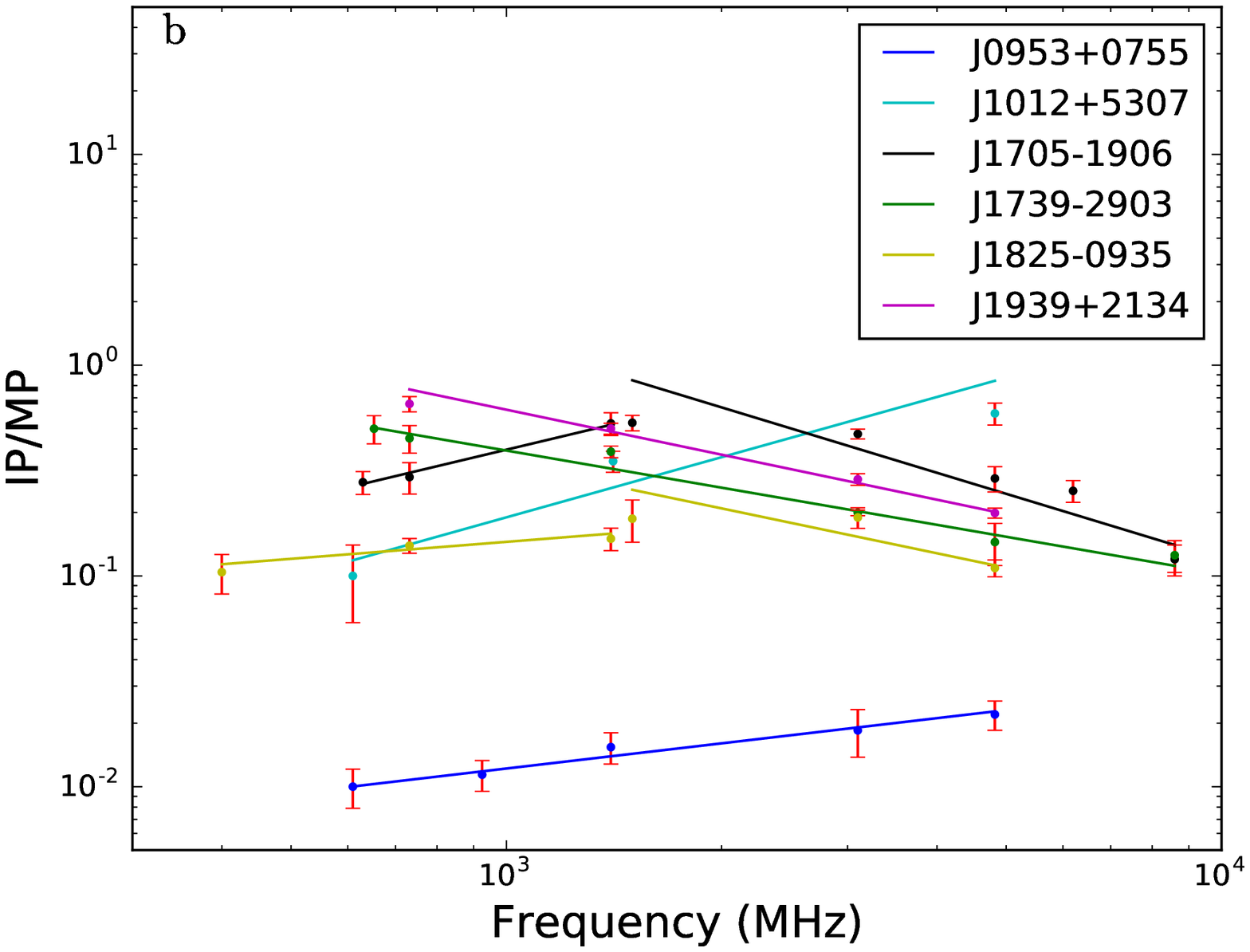}}\\
\end{tabular}
\end{center}
\end{figure} 

\clearpage

\begin{figure}[h]
\caption{Separation of profile outermost components as a function of
  frequency for seven pulsars where the separation decreases with
  increasing frequency (upper panel), for another eleven with nearly constant 
  component separation (central panel) and for the remaining two where   
  the separation increases with increasing frequency (lower panel).
  The lines are weighted power-law fits for each pulsar.}
\label{fg:sp_freq}
\begin{center}
\begin{tabular}{ccc}
\resizebox{0.53\hsize}{!}{\includegraphics[angle=0]{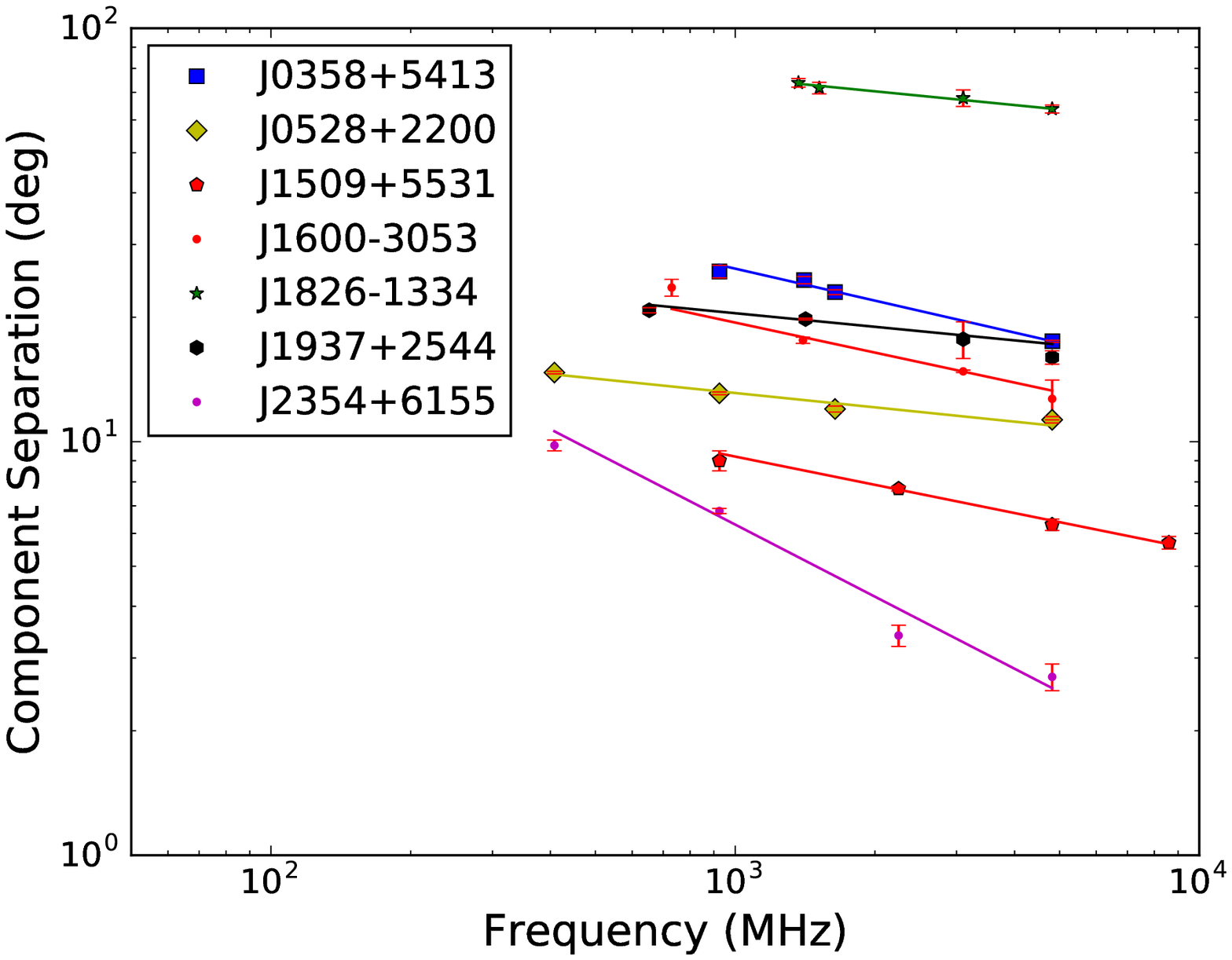}}\\
\resizebox{0.53\hsize}{!}{\includegraphics[angle=0]{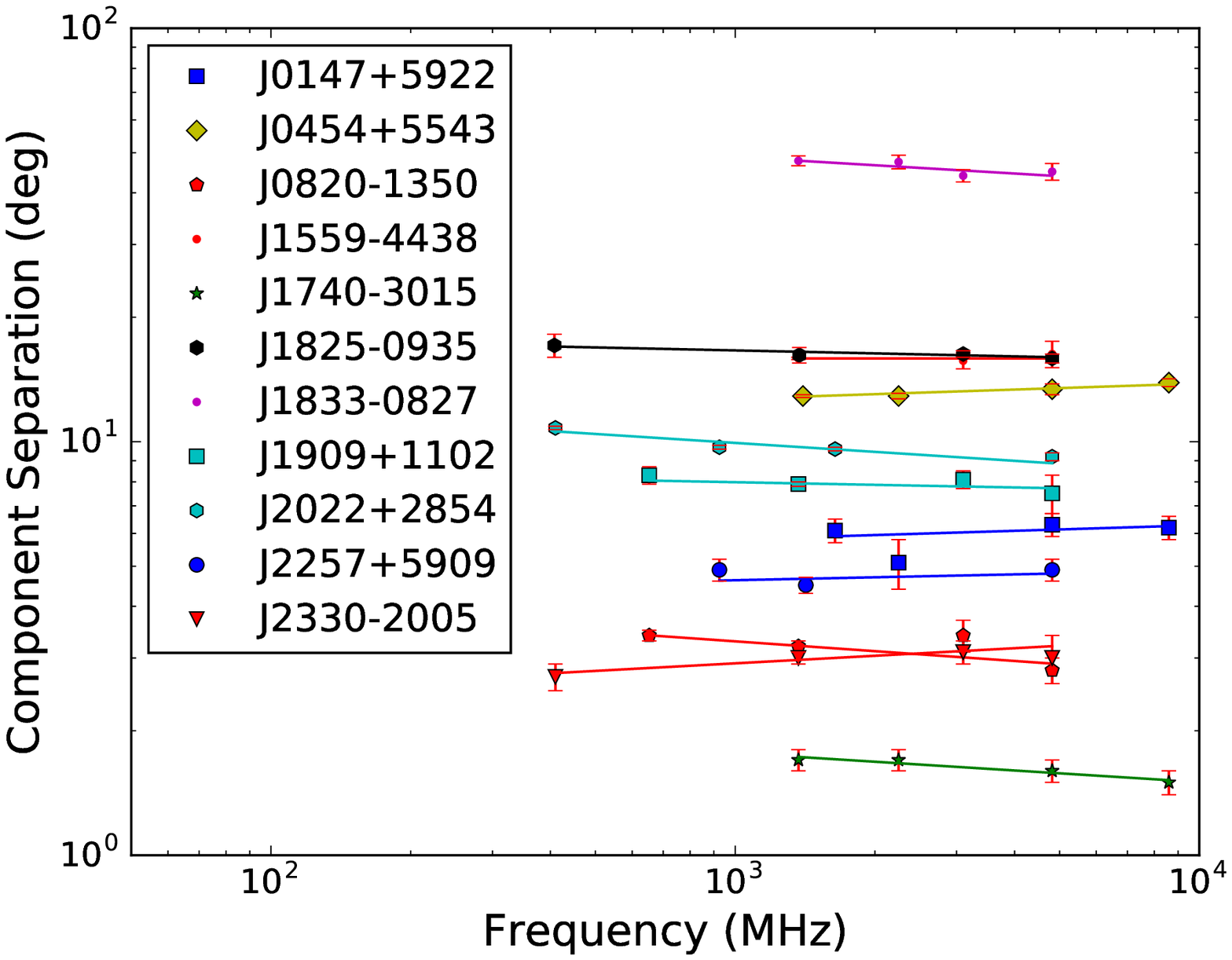}}\\
\resizebox{0.53\hsize}{!}{\includegraphics[angle=0]{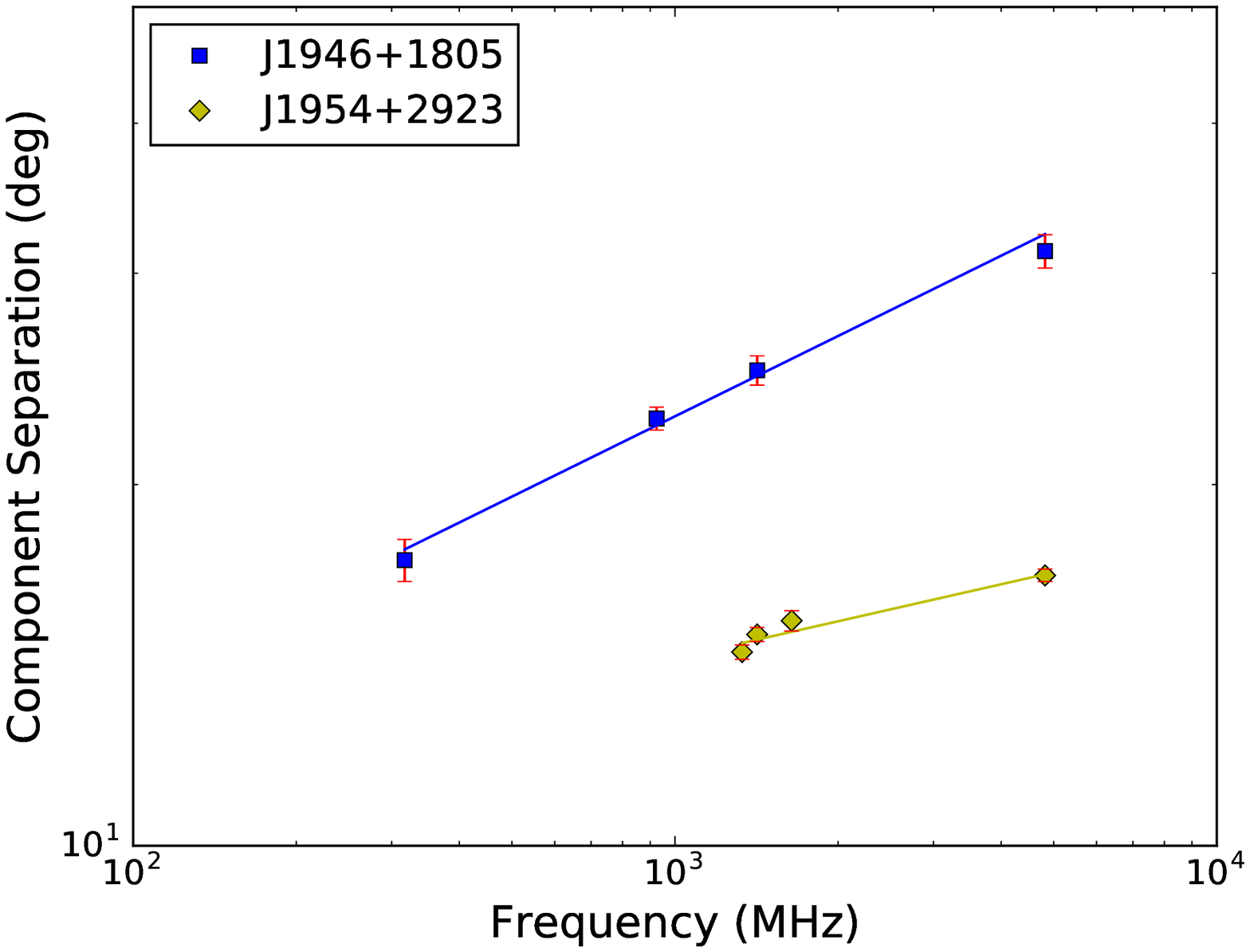}}\\
\end{tabular}
\end{center}
\end{figure} 

\clearpage

\begin{figure}[h]
\caption{Pulse profiles for PSR J1946+1805 at several frequencies with
  fitted Gaussian components. Related components are identified by
  colour and their sum is given by the red dashed line. The fit
  residual is given as a blue line. The peak of outmost components are
  marked by green dashed line.} 
\label{fg:J1946_G}
\begin{center}
\begin{tabular}{ccc}
\resizebox{0.53\hsize}{!}{\includegraphics[angle=0]{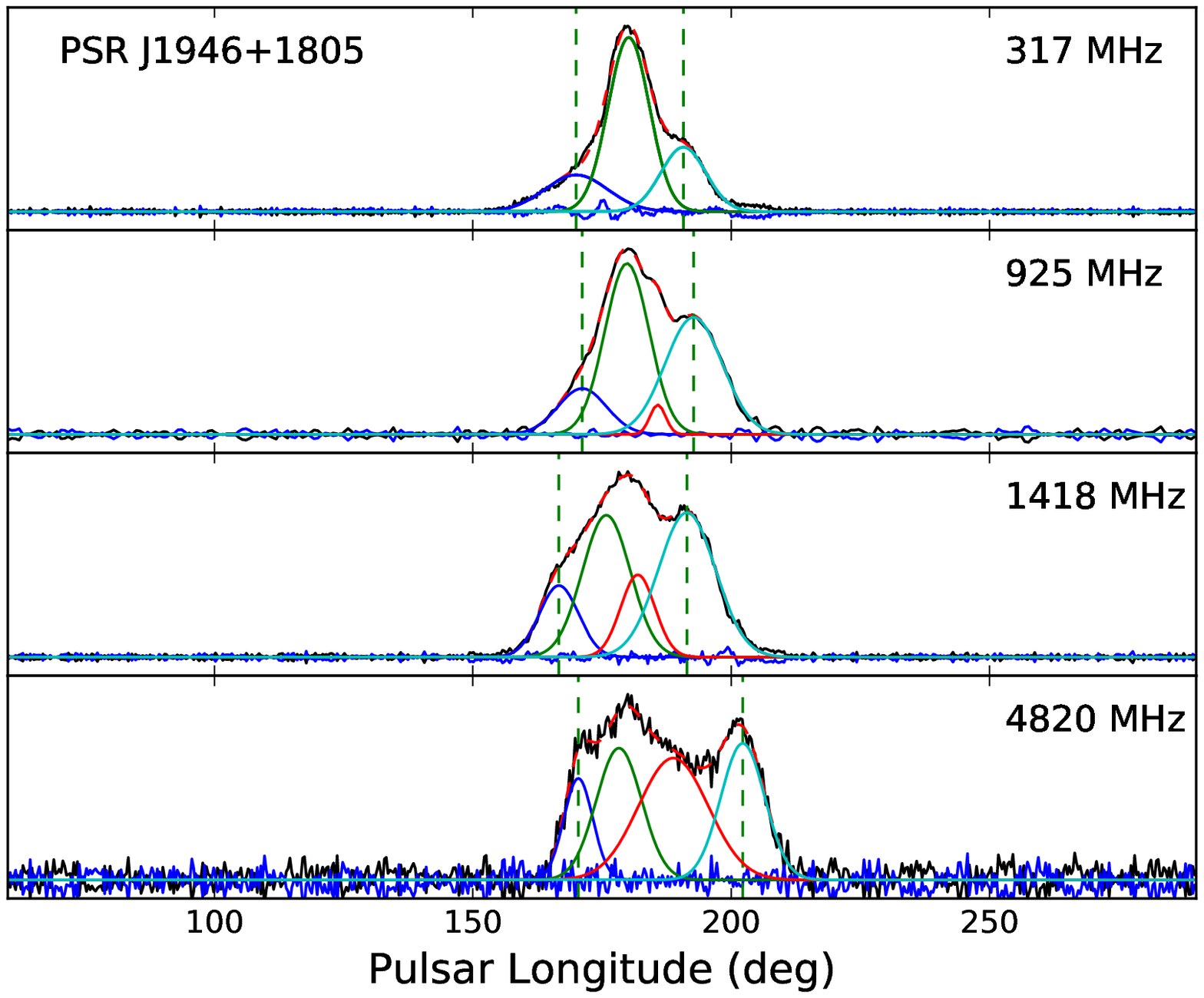}}\\
\end{tabular}
\end{center}
\end{figure}

\clearpage

\begin{figure}[h]
\caption{Pulse widths at 50\% of the peak flux density ($W_{50}$) and
  10\% of the peak flux density ($W_{10}$) at 5~GHz for 63 pulsars.
  (The eight millisecond pulsars in our sample were excluded as they
  form a distinct class.) The green lines are power-law fits with the
  same slope of $-0.37$ for $W_{50}$ and $W_{10}$, respectively.  The
  blue lines in both figures are the 5~GHz estimate of the lower
  boundary line using quantile regression, and the errors from our
  fits were marked by blue dashed lines. The red and black lines in the
  left panel are the lower boundaries of $W_{50}$ at 0.6 and 0.3~GHz,
  respectively \citep{sbm+18}. }
\label{fg:w-p}
\begin{center}
\begin{tabular}{cc}
\resizebox{0.53\hsize}{!}{\includegraphics[angle=0]{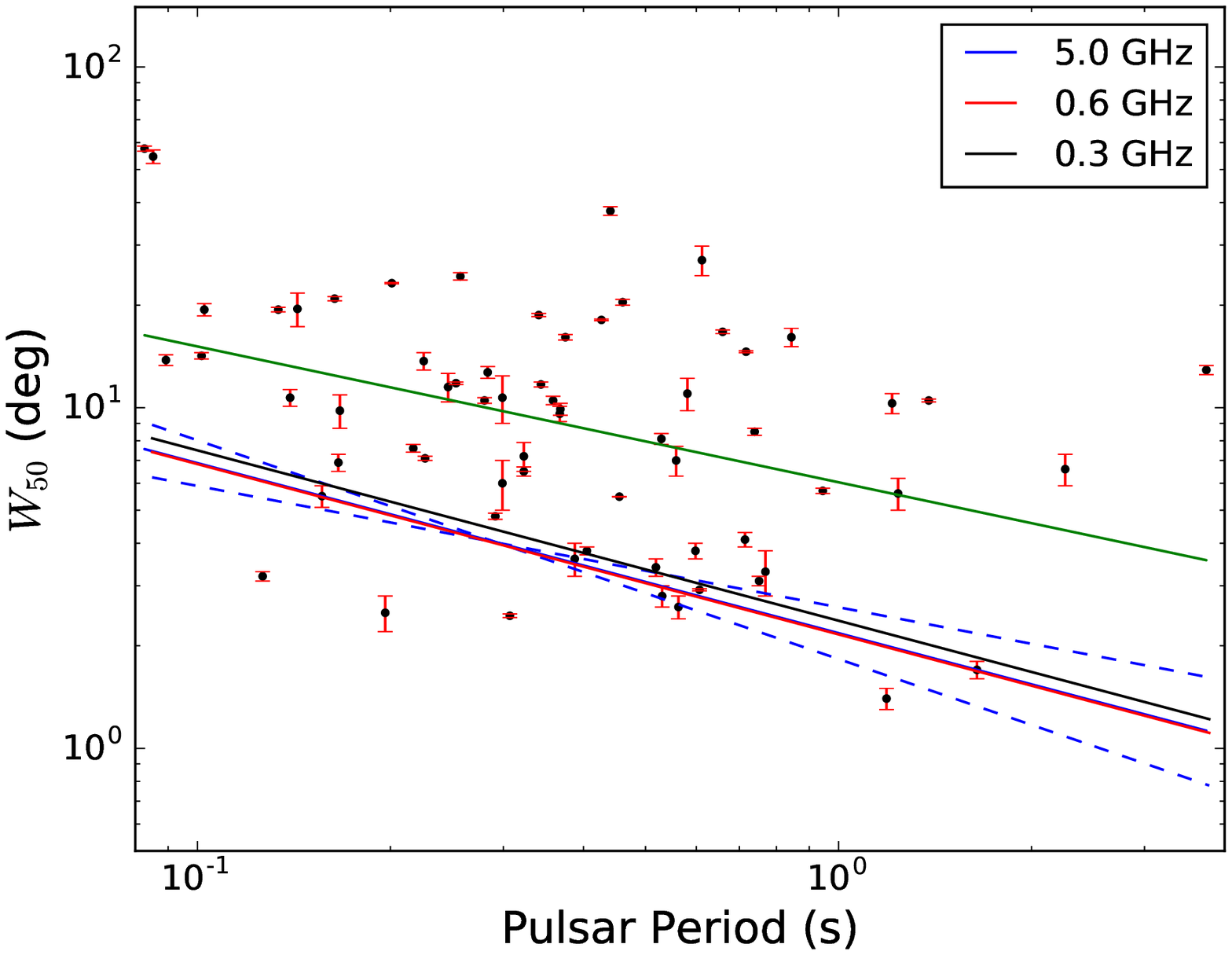}}&
\resizebox{0.53\hsize}{!}{\includegraphics[angle=0]{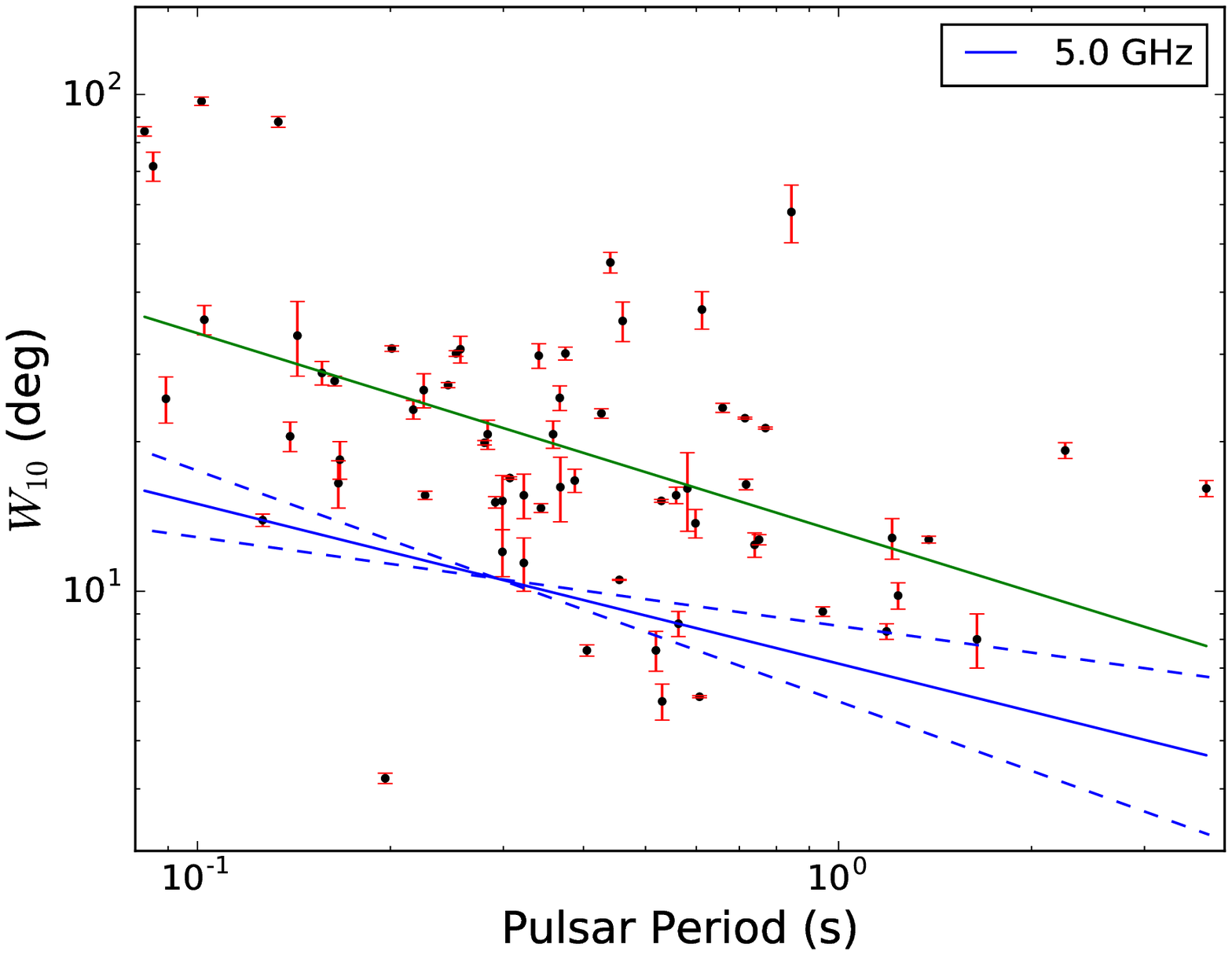}}\\
\end{tabular}
\end{center}
\end{figure} 

\clearpage

\begin{figure}[h]
\caption{Observed core- and conal-component half-power widths at 5.0
  GHz plotted against pulse period. The blue lines in both figures are
  the 5~GHz estimate of the lower boundary line using quantile
  regression, with the errors from our fits were marked by blue dashed
  lines. The lower-bound relations in the left panel are from
  \citet{ran90} from 1.0~GHz observations (red line), from
  \citet{zwy+17} from 8.6~GHz observations (black line) and from the
  scaled \citet{ran90} fit (green line). The red and black lines in
  the right panel are the lower boundary of conal-component half-power
  widths at 0.6 and 0.3~GHz respectively from \citet{sbm+18}.}
\label{fg:Wc-P}
\begin{center}
\begin{tabular}{cc}
\resizebox{0.53\hsize}{!}{\includegraphics[angle=0]{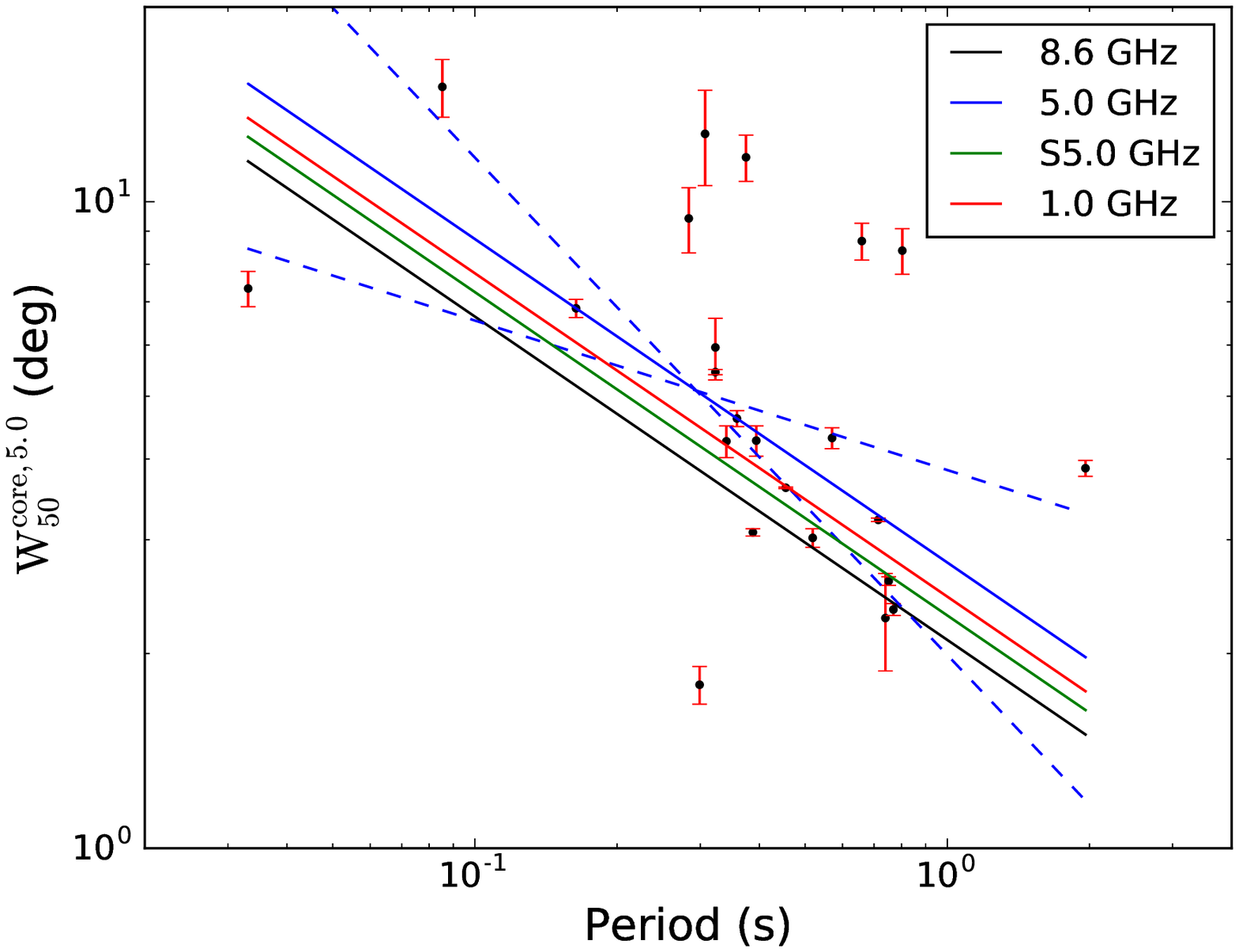}}&
\resizebox{0.53\hsize}{!}{\includegraphics[angle=0]{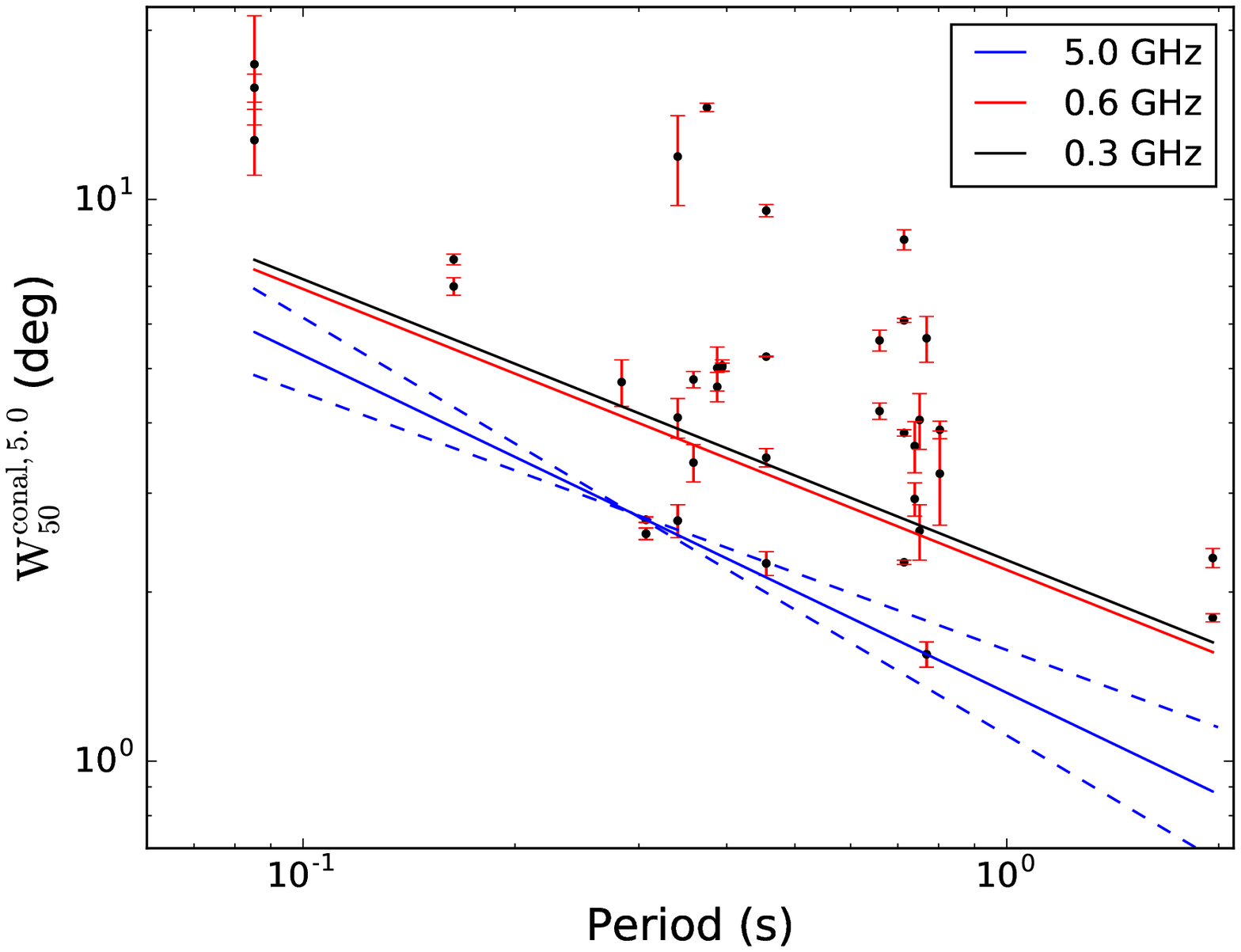}}\\
\end{tabular}
\end{center}
\end{figure}

\end{document}